\documentclass[11pt,3p,review,authoryear]{elsarticle}

\journal{ }
\usepackage{geometry, graphicx, amssymb, mathtools, enumerate, enumerate, amsthm,  natbib, booktabs, multirow, caption, float, placeins, indentfirst, subfigure, hyperref}
\usepackage{tikz, xcolor}

\usepackage{siunitx}
\newtheorem{theorem}{Theorem}
\newtheorem{lemma}{Lemma}
\newtheorem{definition}{Definition}
\newtheorem{remark}{Remark}
\newtheorem{assumption}{Assumption}
\newtheorem{proposition}{Proposition}

\newtheorem{algorithm}{Algorithm}

\hypersetup{
    colorlinks=true,
    linkcolor=blue,
    filecolor=magenta,
    urlcolor=cyan,
    }

\graphicspath{{figures/}}
\allowdisplaybreaks[4]
\begin{document}

\begin{frontmatter}

\title{A Semiparametric Real-Time GARCH Model with Nonstationary Variance}
\author{
Muyi Li\texorpdfstring{$^{a,b,c,d}$}{},
Yajia Chen\texorpdfstring{$^{b}$}{},
Yuqing Xu\texorpdfstring{$^{b}$}{},
Feiyu Jiang\texorpdfstring{$^{e}$}{}\\
\small{
\texorpdfstring{$^{a}$}{}  Wang Yanan Institute for Studies in Economics (WISE), Xiamen University, China}\\
\small{
\texorpdfstring{$^{b}$}{} Department of Statistics and Data Science, School of Economics, Xiamen University, China}\\
\small{
\texorpdfstring{$^{c}$}{} Key Laboratory of  Econometrics, Ministry of Education, Xiamen University, China}\\
\small{\texorpdfstring{$^{d}$}{} Fujian Key Laboratory of Statistical Science, Xiamen University, China}\\
\small{\texorpdfstring{$^{e}$}{} Department of Statistics and Data Science, School of Management, Fudan University, China}\\
}
\end{frontmatter}
\section*{Abstract}
We propose a semiparametric real-time GARCH (SRT-GARCH) model that combines a smooth deterministic long-run volatility component with a stationary short-run process driven by contemporaneous return innovations. The model accommodates gradual changes in unconditional variance while allowing short-run volatility to respond directly to current information rather than relying solely on lagged observations. We derive the variance normalization, conditional moments, and closed-form conditional density of the model. For estimation, we develop a two-step procedure that combines kernel smoothing with conditional Gaussian maximum likelihood. Under Gaussian innovations and mild regularity conditions, we establish consistency and asymptotic normality of the short-run parameter estimator. The analysis exploits geometric moment contraction to control serial dependence and shows that estimating the long-run component adds a positive semidefinite rank-one adjustment to the oracle covariance matrix obtained when the long-run component is known.  The resulting asymptotic covariance matrix admits a plug-in estimator that requires neither lag truncation nor additional tuning parameters. Monte Carlo simulations show good finite-sample performance of the estimator and the associated inference procedure. Applications to FTSE~100 returns and U.S. natural gas futures show that allowing for a time-varying long-run component improves forecasting relative to the parametric RT-GARCH benchmark, while the additional real-time feedback mechanism provides more pronounced gains in the natural gas application than in the FTSE application. The proposed model also delivers competitive one-step-ahead Value-at-Risk forecasts.

\noindent\textit{Keywords:}  variance decomposition; real-time GARCH; semiparametric estimation; volatility forecasting; Value-at-Risk

\noindent\textit{JEL classification:} C12; C14; C58

\section{Introduction}

Financial market volatility fluctuates over multiple horizons. Daily return innovations contain immediate information about current volatility, while structural changes in macroeconomic and market conditions can generate slow-moving variation in unconditional variance \citep{hong2017testing}. Capturing both sources of variation is important for volatility measurement, forecasting, and risk management. In the seminal generalized autoregressive conditional heteroskedasticity (GARCH) frameworks of \citet{engle1982autoregressive} and \citet{bollerslev1986generalized}, short-run volatility clustering is modeled through a conditional variance driven by lagged returns and past volatility. As emphasized by \citet{politis2007model}, however, conditioning exclusively on past information may overlook the contemporaneous signal contained in the current return. A related insight underlies the realized-volatility literature, which demonstrates the value of within-period information for tracking rapid changes in volatility \citep{andersen2003modeling,hansen2012realized}.

To incorporate contemporaneous information directly into daily return models, \citet{smetanina2017real} introduced the real-time GARCH (RT-GARCH) model, in which the current innovation enters both the return equation and the volatility process. Unlike stochastic volatility models, which introduce a separate latent volatility shock, RT-GARCH allows the volatility scale to respond immediately to the return innovation while retaining a tractable closed-form likelihood. \citet{smetanina2021asymptotic} subsequently established the large-sample theory for maximum likelihood estimation of the stationary RT-GARCH model. More recent extensions have incorporated joint nowcasting and forecasting, low-frequency macroeconomic covariates, heavy-tailed conditional distributions, realized volatility measures, and intraday range information; see, for example, \cite{ding2023simple,wu2023real,kruse2023improving,wu2025realized,xu2025realtime}. Despite their flexibility in describing short-run volatility dynamics, these models maintain a constant unconditional variance and may therefore be misspecified over long spans of financial data.

A complementary literature addresses nonstationarity in unconditional volatility. \citet{mikosch2004nonstationarities} show that unmodeled changes in baseline volatility can generate spurious integrated-GARCH behavior and distort estimates of volatility persistence when constant unconditional variance is imposed. This observation has motivated models with smooth deterministic trends or structural breaks \citep{stuaricua2005nonstationarities,engle2008spline,chen2016detecting,truquet2017parameter}, as well as locally stationary ARCH processes with time-varying parameters \citep{dahlhaus2006statistical}. Within this literature, multiplicative component models provide an especially convenient framework by decomposing returns into a smooth deterministic scale and a stationary GARCH process \citep{feng2004simultaneously,hafner2010efficient,jiang2021adaptive,chronopoulos2022choosing}. Related work studies adaptive estimation, specification testing, outlier robustness, and simultaneous inference in time-varying volatility models \citep{karmakar2022simultaneous,camposmartins2024modeling,ahlgren2024new,amado2025outlier}. Existing multiplicative GARCH specifications, however, continue to rely on lagged return shocks and therefore do not exploit contemporaneous return information in the short-run volatility recursion.

In this paper, we bridge these two strands of the literature by proposing a semiparametric real-time GARCH (SRT-GARCH) model. Returns are decomposed into a smooth deterministic scale function $\tau(t/T)$ and a stationary RT-GARCH process. A second-moment normalization of the short-run component identifies $\tau(t/T)$ directly as the time-varying unconditional variance of returns. The proposed specification nests the semiparametric GARCH (S-GARCH) model of \citet{jiang2021adaptive} when the contemporaneous feedback parameter is zero and a variance-targeted version of the RT-GARCH model of \citet{smetanina2017real} when the long-run scale is constant. We characterize the conditional moments and conditional density of the model. The long-run component determines the overall volatility, whereas the real-time feedback parameter introduces additional short-run distributional flexibility, including time-dependent conditional kurtosis and, under Gaussian innovations, exponential-type conditional heavy tails.

Combining an unknown long-run volatility component with contemporaneous short-run feedback creates nontrivial challenges for estimation and inference. The asymptotic theory cannot be obtained by simply combining existing results for S-GARCH and RT-GARCH models. First, because short-run volatility depends contemporaneously on the current innovation, recovering the latent volatility scale from observed returns requires inversion of a nonlinear quadratic equation at each time point. The resulting observed-data volatility recursion and its parameter derivatives therefore follow coupled nonlinear dynamics rather than the linear recursions familiar from conventional GARCH models. To control temporal dependence without imposing high-level $\alpha$- or $\beta$-mixing assumptions, we exploit a random-coefficient representation of the true short-run state to establish the geometric moment contraction property \citep{wu2005nonlinear,shao2007asymptotic}. The resulting coupling bounds are then extended to the nonlinear observed-data volatility process and its higher-order parameter derivatives.

Second, because the deterministic variance function $\tau(\cdot)$ is unknown, we estimate the model in two steps. We first estimate $\tau(\cdot)$ nonparametrically from squared returns using a boundary-reflected Nadaraya--Watson kernel estimator $\widehat{\tau}_t$ for $\tau_t=\tau(t/T)$. We then estimate the short-run parameters by Gaussian conditional maximum likelihood based on the standardized returns
$\widehat{u}_t=r_t/\sqrt{\widehat{\tau}_t}$, where $\{r_t\}$ is the observed return process, and $u_t$ denotes the corresponding oracle RT-GARCH process. 
Since $\widehat{u}_t^2$ enters both the contemporaneous quadratic root and all subsequent volatility states, the first-step estimation error propagates nonlinearly through the likelihood recursion and has a non-vanishing first-order effect on the limiting distribution of the second-step estimator. By projecting this infinite-horizon perturbation onto the first-step innovations, we establish consistency and asymptotic normality of the two-step estimator. Under the variance normalization, we further show that the second-step score is contemporaneously orthogonal to the martingale innovation driving the first-step error and that the infinite-series sensitivity vector admits a closed-form contemporaneous representation. These two structural properties reduce the contribution of first-step estimation to a positive semidefinite rank-one adjustment of the oracle covariance matrix and yield a tuning-free plug-in asymptotic covariance estimator that requires neither lag truncation nor additional tuning parameters.

We also derive closed-form expressions for multi-step-ahead volatility forecasts and one-step-ahead Value-at-Risk (VaR) forecasts. Monte Carlo experiments indicate that the finite-sample behavior of the estimator and the associated inference procedure is broadly consistent with the asymptotic theory. We apply the model to daily FTSE~100 returns and U.S. natural gas futures. In both applications, covariance stationarity is rejected, providing evidence consistent with a time-varying long-run volatility component. The forecasting results show that incorporating a time-varying long-run component improves substantially on the RT-GARCH benchmark, while the additional real-time feedback mechanism delivers more pronounced gains in the natural gas application than in the FTSE application. The VaR results are also broadly supportive of the proposed specification.

The remainder of the paper is organized as follows. Section~\ref{sec:model} introduces the SRT-GARCH model and characterizes its moment and distributional properties. Section~\ref{sec:estimation} develops the two-step estimation procedure, establishes the large-sample theory, and derives the plug-in covariance estimator. Section~\ref{sec:volatility-forecasting} presents the volatility forecasting framework and its multi-step extension. Sections~\ref{sec:simulations} and~\ref{sec:applications} report the Monte Carlo experiments and empirical applications, respectively. Section~\ref{sec:conclusion} concludes. All technical proofs are provided in the Appendix.

\section{The SRT-GARCH(1,1) Model}
\label{sec:model}

For an observed return series $\{r_t\}_{t=1}^T$, we consider the following semiparametric real-time GARCH
(SRT-GARCH) model:
\begin{equation}
\left\{
\begin{aligned}
r_t &= \sqrt{\tau_t}\,u_t,
\qquad \tau_t=\tau(t/T),\\
u_t &= \sqrt{h_t}\,\epsilon_t,\\
h_t &= \omega+\alpha u_{t-1}^2+\beta h_{t-1}
      +\varphi\epsilon_t^2,
\end{aligned}
\right.
\qquad t=1,\ldots,T,
\label{model}
\end{equation}
where $\tau:[0,1]\rightarrow(0,\infty)$ is an unknown,
smoothly time-varying deterministic function.

In model \eqref{model}, $\tau_t$ captures slow-moving changes in the unconditional variance, while $u_t$ governs stationary short-run volatility dynamics.
The innovations $\{\epsilon_t\}$  are assumed to be independent and identically distributed (i.i.d.)  with zero mean and unit variance. The short-run parameters satisfy $\omega>0$ and $\alpha,\beta,\varphi\geq0$. Here, the coefficient $\varphi$ measures the immediate feedback of the contemporaneous return shock on the short-run scale: setting $\varphi=0$ recovers the semiparametric GARCH (S-GARCH) model of \citet{jiang2021adaptive}, whereas restricting $\tau_t$ to be constant yields a variance-normalized version of the real-time GARCH (RT-GARCH) model of \citet{smetanina2017real}.

Unlike stochastic volatility models, model \eqref{model} does not introduce a separate latent volatility innovation. Because the same shock $\epsilon_t$ drives both the return equation and the short-run volatility update, $h_t$ and $\epsilon_t$ can be uniquely recovered from the observed return $r_t$ given past information via a nonlinear transformation. This property yields a closed-form conditional density and makes conditional maximum likelihood estimation   feasible.

\begin{remark}
Let $\mathcal F_t=\sigma(\epsilon_s:s\leq t)$ denote the $\sigma$-field generated by the innovations up to time $t$. Because the contemporaneous innovation $\epsilon_t$ enters the
recursion for $h_t$, the quantity $h_t$ is not measurable with
respect to the usual past information set $\mathcal{F}_{t-1}$ as in the standard GARCH models. It should therefore be
interpreted as a real-time short-run scale rather than as the
conventional one-step-ahead conditional variance.
\end{remark}

We begin with the regularity conditions that ensure the short-run process is well defined and the long-run component is identifiable. Let $\theta=(\alpha,\beta,\varphi)^\top$ and denote the true parameter as 
$\theta_0=(\alpha_0,\beta_0,\varphi_0)^\top$.

\begin{assumption}\label{Asm1}
\begin{enumerate}[(i)]
\item
The parameter space $\Theta$ is a compact subset of
\[
\left\{
(\alpha,\beta,\varphi)^\top\in[0,\infty)^3:
\alpha+\beta\leq\rho<1,\;
\omega(\theta)>0
\right\},
\]
where $\rho\in(0,1)$ is a fixed constant, and
\begin{equation}
\omega(\theta)
=
1-\alpha-\beta-\beta\varphi
+\varphi(\beta-1)\mu_4,
\qquad
\mu_4=\mathbb{E}(\epsilon_t^4).
\label{omega-normalization}
\end{equation}

\item
The true parameter satisfies $\theta_0\in\Theta$.

\item
The innovations $\{\epsilon_t\}$ are independently and identically
distributed, symmetric about zero, and satisfy
$\mathbb{E}(\epsilon_t)=0$, $\mathbb{E}(\epsilon_t^2)=1$, and
$\mu_4=\mathbb{E}(\epsilon_t^4)<\infty$.
\end{enumerate}
\end{assumption}

\subsection{\texorpdfstring{Normalization and
Second-Order Properties}{Normalization and Second-Order Properties}}
\label{subsec:normalization}

Assumption~\ref{Asm1}(i) defines a compact parameter space. The nonnegativity restrictions together with
$\omega(\theta)>0$, ensure the positivity of the real-time scale
$h_t$. The persistence condition $\alpha+\beta\leq\rho<1$ guarantees that the short-run dynamic  admits a strictly stationary and ergodic solution with finite second-order moments, as formalized in the following proposition.

\begin{proposition}
\label{prop:stationary-short-run}
Under Assumption~\ref{Asm1}, the short-run recursion has a unique
causal, strictly stationary, and ergodic solution satisfying
$\mathbb{E}(h_t)<\infty$ and $\mathbb{E}(u_t^2)<\infty$.
\end{proposition}
To separately identify the nonparametric scale $\tau_t$ and the short-run process $u_t$ in \eqref{model}, Assumption \ref{Asm1} also helps normalize the unconditional variance of $u_t$ to one so
that $\tau_t$ can be interpreted as the unconditional  variance.
\begin{proposition} 
\label{prop:normalization-proof}
Under Assumption~\ref{Asm1}, we have 
$
\mathbb{E}(u_t^2)=1,
$ and  $
\mathbb{E}(h_t)=1-\varphi(\mu_4-1).$
Moreover,
$\mathbb{E}(u_t\mid\mathcal F_{t-1})=0$ and
$\mathbb{E}(u_tu_{t-j})=0$ for every $j\geq1$, i.e. 
$\{u_t\}$ is a martingale difference sequence with unit unconditional variance.
\end{proposition}

Unlike standard GARCH models, where $\mathbb{E}(u_t^2)=\mathbb{E}(h_t)$, the normalization in \eqref{omega-normalization} must account for the contemporaneous covariance between $h_t$ and $\epsilon_t^2$. Writing $m_u=\mathbb{E}(u_t^2)$ and $m_h=\mathbb{E}(h_t)$ and taking unconditional expectations in the short-run equations $h_t = \omega + \alpha u_{t-1}^2 + \beta h_{t-1} + \varphi\epsilon_t^2$ and $u_t^2 = (\omega + \alpha u_{t-1}^2 + \beta h_{t-1})\epsilon_t^2 + \varphi\epsilon_t^4$ yields that 
\begin{align*}
m_h&=\omega+\alpha m_u+\beta m_h+\varphi,\\
m_u&=\omega+\alpha m_u+\beta m_h+\varphi\mu_4.
\end{align*}
Together with \eqref{omega-normalization} and
$\alpha+\beta<1$,  we obtain that $m_u=1$ and
$m_h=1-\varphi(\mu_4-1)$. Consequently,
$\mathbb{E}(r_t^2)=\tau_t$, so the long-run component has the direct
interpretation of unconditional return variance. When $\varphi=0$,
the normalization reduces to $\omega=1-\alpha-\beta$ as in \citet{francq2011merits} and \cite{jiang2021adaptive}; under standard Gaussian
innovations, it becomes
$\omega=1-\alpha-\beta-3\varphi+2\beta\varphi$.


\subsection{Conditional Moments}\label{sub_Model property}

While the deterministic function $\tau_t$ represents the overall level of return variance, the real-time feedback parameter $\varphi_0$ can change the shape of the conditional return distribution. Following \citet{smetanina2017real}, we characterize their different roles through the second and fourth conditional moments of $r_t$.

Let $b_{t-1}=\omega_0+\alpha_0u_{t-1}^2+\beta_0h_{t-1}$ denote the short-run scale, so that $h_t=b_{t-1}+\varphi_0\epsilon_t^2$ and $u_t^2=b_{t-1}\epsilon_t^2+\varphi_0\epsilon_t^4$. Write $\mu_j=\mathbb{E}(\epsilon_t^j)$ for $j=2,4,6,8$, with $\mu_2=1$.

\begin{proposition}\label{pro_moment}
Suppose that Assumption~\ref{Asm1} holds and
$\mathbb{E}(\epsilon_t^8)<\infty$. Then
\begin{align}
\mathbb{E}\!\left(r_t^2\mid\mathcal{F}_{t-1}\right)
&=
\tau_t
\left(
b_{t-1}+\varphi_0\mu_4
\right),
\label{conditional-second-moment}
\\
\mathbb{E}\!\left(r_t^4\mid\mathcal{F}_{t-1}\right)
&=
\tau_t^2
\left(
\mu_4 b_{t-1}^2
+2\varphi_0\mu_6 b_{t-1}
+\varphi_0^2\mu_8
\right).
\notag
\end{align}
Consequently,
\begin{equation}
\frac{
\mathbb{E}\!\left(r_t^4\mid\mathcal{F}_{t-1}\right)
}{
\left[
\mathbb{E}\!\left(r_t^2\mid\mathcal{F}_{t-1}\right)
\right]^2
}
=
\frac{
\mu_4 b_{t-1}^2
+2\varphi_0\mu_6 b_{t-1}
+\varphi_0^2\mu_8
}{
\left(
b_{t-1}+\varphi_0\mu_4
\right)^2
}.
\label{standardized-fourth-moment}
\end{equation}
\end{proposition}
Proposition~\ref{pro_moment} clarifies the different roles of the long-run and short-run components. The deterministic part $\tau_t$ proportionally scales the conditional variance \eqref{conditional-second-moment} and fourth moment, but cancels out of the standardized fourth moment \eqref{standardized-fourth-moment}. By contrast, when $\varphi_0>0$, the conditional kurtosis in \eqref{standardized-fourth-moment} varies endogenously with the  short-run component $b_{t-1}$. Thus, $\tau_t$ governs low-frequency shifts in the variance level, whereas contemporaneous feedback generates time-varying conditional tail heaviness.

\subsection{Conditional Density}
\label{sub:conditional-density}
Conditional on $\mathcal{F}_{t-1}$, the map
$\epsilon_t\mapsto
u_t=\epsilon_t\sqrt{b_{t-1}+\varphi_0\epsilon_t^2}$ is strictly
increasing and hence invertible for every $\varphi_0\ge0$.  Squaring both sides gives the quadratic equation $h_t^2 - b_{t-1}h_t - \varphi_0 u_t^2 = 0$ in $h_t = b_{t-1} + \varphi_0\epsilon_t^2 > 0$. Therefore, solving for the positive root at $u=r/\sqrt{\tau_t}$ yields the implied short-run scale 
\[
h_t(r)
=
\frac{
b_{t-1}
+\sqrt{b_{t-1}^2+4\varphi_0r^2/\tau_t}
}{2}.
\]
By the change of variables, we obtain the following conditional return density. 

\begin{proposition}\label{pro:conditional-density}
Suppose that Assumption~\ref{Asm1} holds and that the innovation
$\epsilon_t$ has density $f_\epsilon$. Then the conditional density
of $r_t$ given $\mathcal{F}_{t-1}$ is
\begin{equation}
f_t(r\mid\mathcal{F}_{t-1})
=
\begin{cases}
\displaystyle
\frac{\sqrt{h_t(r)}}
{\sqrt{\tau_t}\,
 \sqrt{b_{t-1}^2+4\varphi_0r^2/\tau_t}}
f_\epsilon\!\left(
\frac{r}{\sqrt{\tau_t h_t(r)}}
\right),
& \varphi_0>0, \\[3ex]
\displaystyle
\frac{1}{\sqrt{\tau_t b_{t-1}}}
f_\epsilon\!\left(
\frac{r}{\sqrt{\tau_t b_{t-1}}}
\right),
& \varphi_0=0.
\end{cases}
\label{eq:conditional-density}
\end{equation}
\end{proposition}
Comparing the two cases in \eqref{eq:conditional-density} showcases how real-time feedback can change the conditional return distribution. When $\varphi_0=0$, the Jacobian factor reduces to $(\tau_t b_{t-1})^{-1/2}$, recovering the S-GARCH \citep{jiang2021adaptive} conditional density with constant conditional kurtosis $\mu_4$. 
When $\varphi_0>0$, the nonlinear transformation from $\epsilon_t$ to $r_t$ introduces a state-dependent Jacobian term
that depends on both the predetermined state \(b_{t-1}\) and the current return \(r_t\)
and stretches large realizations of $\epsilon_t$. Indeed, the proof in the Appendix shows that  when $f_\epsilon$ is standard Gaussian, any $\varphi_0>0$ transforms the conditional tails of $r_t$ from Gaussian-type ($\log f_t(r\mid\mathcal{F}_{t-1}) \asymp -r^2$) to exponential-type ($\log f_t(r\mid\mathcal{F}_{t-1}) \asymp -|r|$). This  helps the model capture the heavy-tailed phenomenon commonly observed in financial returns.

\subsection{Distributional Implications of the Real-Time Component}\label{sub:distributional}
To illustrate how the contemporaneous feedback term $\varphi\epsilon_t^2$ affects the shape of the return distribution, we compare the SRT-GARCH model with its nested S-GARCH counterpart under three different long-run volatility patterns. Specifically,
\[
\begin{aligned}
\text{DGP 1 (SRT-GARCH 1):}\quad
&\tau_1(x)=1+0.5\sin(2\pi x),
&&(\alpha,\beta,\varphi)=(0.20,0.60,0.05),\\
\text{DGP 2 (SRT-GARCH 2):}\quad
&\tau_2(x)=1+4(x-x^2),
&&(\alpha,\beta,\varphi)=(0.20,0.50,0.03),\\
\text{DGP 3 (SRT-GARCH 3):}\quad
&\tau_3(x)=1+4x^2,
&&(\alpha,\beta,\varphi)=(0.15,0.65,0.04).
\end{aligned}
\]
DGPs 1–3 capture  periodic variation, a temporary rise followed by a decline, and a
monotone nonlinear increase in unconditional variance, respectively. For each $j=1,2,3$, S-GARCH~$j$ shares the same long-run component $\tau_j(x)$ and parameters $(\alpha,\beta)$ as DGP~$j$, but sets $\varphi=0$. In all specifications, the intercept $\omega$ is determined by the normalization in \eqref{omega-normalization} under standard Gaussian innovations. 

\begin{figure}[htb]
\centering
\includegraphics[width=1.0\textwidth]{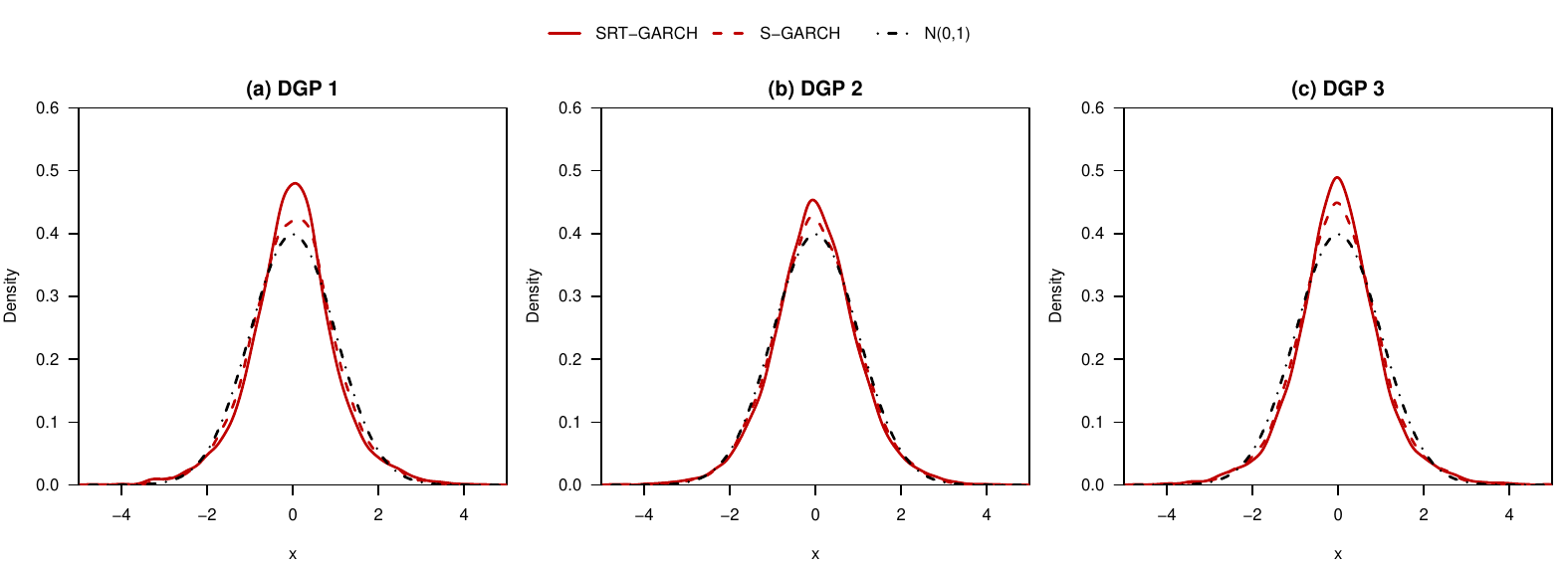}
\caption{Kernel density estimates for the sample-standardized simulated returns with $T=10{,}000$.}
\label{fig:heavy_tail}
\end{figure}

Figure~\ref{fig:heavy_tail} plots the kernel density estimates of the simulated return series with $T=10{,}000$ observations alongside the standard normal reference density. Across the three long-run specifications,  the SRT-GARCH return distributions exhibit a sharper peak around zero and noticeably heavier tails than their nested S-GARCH counterparts. Thus, incorporating real-time innovation feedback with a time-varying unconditional variance generates more pronounced  leptokurtosis without requiring a heavy-tailed innovation distribution.  



\section{Two-Step Estimation Procedure}
\label{sec:estimation}
\label{sub_Two-step estimation}
The normalization in Section~\ref{sec:model} separates long-run scale from short-run dynamics and suggests a natural two-step estimator.  
In the first step, we recover the smooth unconditional variance function $\tau(\cdot)$ nonparametrically from the squared returns $r_t^2$. In the second step, we standardize the observed returns as $\widehat{u}_t=r_t/\sqrt{\widehat{\tau}_t}$ by the estimated scale $\sqrt{\widehat\tau_t}$ and estimate the short-run parameter  $\theta=(\alpha,\beta,\varphi)^\top\in\Theta$, whose true value is denoted by $\theta_0=(\alpha_0,\beta_0,\varphi_0)^\top$, via Gaussian conditional maximum likelihood estimation. The parameter space allows zero values so that
the nested S-GARCH specification $(\varphi=0)$ remains admissible.  
While our consistency result in Theorem~\ref{thm:consistency} permits $\theta_0$ to lie on the boundary when the parameters are identified, deriving the asymptotic distribution in Theorem~\ref{thm:asymptotic-normality} requires $\theta_0$ to be an interior point of $\Theta$. The boundary case leads to nonstandard asymptotics and is highly non-trivial. We refer to \citet{francq2009testing} for a detailed treatment of boundary inference and testing in GARCH type models.

The two-step procedure follows the broad semiparametric strategy of
\citet{jiang2021adaptive}, but the presence of contemporaneous feedback $\varphi\epsilon_t^2$ changes both the second-step likelihood recursion and its first-order asymptotic behavior in two fundamental ways.  First, given the observations, the implied short-run scale $h_t(\theta)$ is defined through a nonlinear quadratic root rather than a linear recursion as in standard GARCH models, so the observed-data volatility process and its parameter derivatives follow coupled nonlinear functions.  Second, because the generated standardized returns $\widehat u_t=r_t/\sqrt{\widehat\tau_t}$ enter both the current scale and the recursive state at every period, the first-step kernel estimation error can propagate through the entire volatility process and makes a non-vanishing first-order contribution to the limiting distribution. Developing the large-sample theory therefore requires tracking this nonlinear propagation explicitly and characterizing the resulting two-step   adjustment, which is highly non-trivial.


\subsection{Estimation of the Long-Run Component}
\label{subsec:estimation-tau}

Under the normalization $\mathbb{E}(u_t^2)=1$, we have
$\mathbb{E}(r_t^2)=\tau_t=\tau(t/T)$.
The squared return can therefore be written as
\begin{equation}
r_t^2
=
\tau(t/T)+v_{t,T},
\qquad
v_{t,T}
=
\tau(t/T)\bigl(u_t^2-1\bigr),
\label{eq:nonparametric-regression}
\end{equation}
where $\mathbb{E}(v_{t,T})=0$. Equation
\eqref{eq:nonparametric-regression} defines a fixed-design
nonparametric regression with serially dependent and
heteroskedastic errors.

Following \citet{jiang2021adaptive}, we estimate $\tau(x)$ using
the simplified Nadaraya--Watson kernel estimator
\begin{equation}
\widehat{\tau}(x)
=
\frac{1}{T}
\sum_{s=1}^{T}
K_h\left(x-\frac{s}{T}\right)r_s^2,
\qquad x\in[0,1],
\label{eq:estimate-tau}
\end{equation}
where $h>0$ is the smoothing bandwidth and  $K_h(v)=h^{-1}K(v/h)$.
In the practical analysis, we use the
Epanechnikov kernel
$K(v)=\frac34(1-v^2)\mathbb{I}(|v|\leq1)$.

To derive the asymptotic properties of
$\widehat{\tau}(x)$, we impose the following assumptions.

\begin{assumption}\label{Asm2}
The function $\tau:[0,1]\rightarrow(0,\infty)$ is twice continuously differentiable, and there exist finite constants
$0<\underline{\tau}<\overline{\tau}<\infty$ such that
$\underline{\tau}\leq\tau(x)\leq\overline{\tau}$ for all
$x\in[0,1]$.
\end{assumption}

\begin{assumption}\label{Asm3}
The kernel $K$ is symmetric, bounded, and Lipschitz continuous,
has compact support $[-1,1]$, and satisfies
$\int_{-1}^{1}K(v)\,dv=1$. Let
$C_j=\int_{-1}^{1}v^jK(v)\,dv$ for $j=1,2$. As
$T\rightarrow\infty$, the bandwidth satisfies $h\rightarrow0$,
$Th\rightarrow\infty$, and $Th^5=O(1)$.
\end{assumption}

\begin{assumption}\label{Asm4}
For $z_t=u_t^2-1$, $\mathbb E(z_t^2)<\infty$.
\end{assumption}

To analyze the kernel estimator $\widehat\tau(x)$ and the subsequent second-step likelihood, we must control the serial dependence of the short-run process $\{u_t^2\}$ as well as the nonlinear observed-data volatility recursion and its parameter derivatives. Instead of  imposing high-level $\alpha$- or $\beta$-mixing conditions on the unobservable process, we follow \citet{smetanina2021asymptotic} and  use the physical dependence measure introduced by
\citet{wu2005nonlinear} to establish geometric moment contraction
(GMC) property in \cite{shao2007asymptotic} for these quantities.

\begin{definition}[Geometric moment contraction]
\label{def:gmc}
Let $X_t=F(\epsilon_t,\epsilon_{t-1},\ldots)$ be a causal process,
where $F$ is measurable and $\{\epsilon_t\}$ is i.i.d. Let
$\{\epsilon_t^\circ\}$ be an independent copy of the innovation
sequence, and define
\[
X_t^\circ
=F(\epsilon_t,\ldots,\epsilon_1,
   \epsilon_0^\circ,\epsilon_{-1}^\circ,\ldots),
\qquad t\ge1.
\]
For $p \geq 1$, the process is $L^p$ geometrically moment contracting
(GMC) if there exist $C<\infty$ and $\rho_p\in(0,1)$ such that
\[
\|X_t-X_t^\circ\|_p\le C\rho_p^t,
\qquad t\ge1,
\]
where $\|X\|_p=(E\|X\|^p)^{1/p}$.
\end{definition}
Moreover, let
$X_t^*
=F(\epsilon_t,\ldots,\epsilon_1,
\epsilon_0^\circ,\epsilon_{-1},\ldots)$
denote the single-innovation coupled version obtained
by replacing only $\epsilon_0$ with an independent copy.
By Minkowski's inequality and the i.i.d. structure of
the innovations,
\[
\begin{aligned}
\|X_t-X_t^*\|_p
&\le
\|X_t-X_t^\circ\|_p
+\|X_t^\circ-X_t^*\|_p\\
&=
\|X_t-X_t^\circ\|_p
+\|X_{t+1}-X_{t+1}^\circ\|_p\\
&\le C(1+\rho_p)\rho_p^t
\le 2C\rho_p^t.
\end{aligned}
\]
Hence, GMC also implies geometric decay of the
single-innovation physical dependence measure.

Furthermore, GMC can be extended to higher moment
orders through interpolation. Specifically, if
$\{X_t\}$ is $L^r$ GMC for some $r>1$ and
$E\|X_t\|^a<\infty$ for some $a>r$, then
H\"older's inequality implies $L^p$ GMC for every
$p\in(r,a)$, possibly with a different geometric
decay rate; see Remark~3.1 of
\citet{shao2007asymptotic}.
 
 Lemma~\ref{lem:contractive-master} in the Appendix collects this moment interpolation and the resulting recursive contraction and covariance bounds, which we apply to obtain the GMC properties for the short-run model process and score functions in Lemma~\ref{lem:augmented-gmc}. The following proposition  
establishes GMC for the short-run process.

\begin{proposition}[Moment contraction of the short-run process]
\label{prop:rt-gmc}
Under Assumption~\ref{Asm1}, suppose $\mathbb E|u_t|^{2p}<\infty$ for some
$p\ge2$. Then $b_t=\omega_0+\alpha_0u_t^2+\beta_0h_t$, $h_t$, and
$z_t=u_t^2-1$ are $L^p$ GMC.
\end{proposition}

Assumptions~\ref{Asm2} and~\ref{Asm3} impose the usual smoothness,
positivity, kernel, and bandwidth conditions. The restrictions
$h\to0$ and $Th\to\infty$ balance local bias and variance, while
$Th^5=O(1)$ controls the remainder in the pointwise bias expansion.
Assumption~\ref{Asm4} only requires a finite variance
for squared short-run returns; {by Proposition~\ref{prop:normalization-proof},
it is equivalent to $\mathbb E(u_t^4)<\infty$.} For the RT-GARCH
recursion, Proposition~\ref{prop:rt-gmc} shows that this moment
condition and Assumption~\ref{Asm1} imply that
$\{z_t\}$ is $L^2$ GMC; see, e.g.,
\citet{wu2004limit}. The argument uses the random-coefficient
representation of the true short-run state and the coupling approach
of \citet{wu2005nonlinear} and \citet{smetanina2021asymptotic}. It is worth noting that the established GMC property implies
absolute summability of the autocovariances and the functional central
limit theorem
\[
\frac{1}{\sqrt T}
\sum_{t=1}^{\lfloor Tr\rfloor}z_t
\Rightarrow
\Omega_z^{1/2}B(r),
\qquad 0\leq r\leq1,
\]
where
$\Omega_z=\sum_{j=-\infty}^{\infty}
\operatorname{Cov}(z_0,z_j)$ and $B(\cdot)$ is a standard Brownian motion.
In contrast to \citet{jiang2021adaptive}, who rely on $\beta$-mixing conditions for standard GARCH processes, our functional-dependence derivation requires no external mixing assumptions. 
Because $K$ is Lipschitz continuous and has compact support,
summation by parts then gives
\[
\frac{1}{\sqrt{Th}}
\sum_{t=1}^{T}
K\!\left(\frac{x-t/T}{h}\right)z_t
\to_d
N\!\left(
0,\,
\Omega_z\int_{-1}^{1}K^2(v)\,dv
\right),
\]
for every fixed $x\in(0,1)$. Hence neither covariance summability nor
a kernel-specific central limit theorem is needed to be imposed separately. 

The preceding result, together with the usual second-order bias expansion for the kernel estimator, leads directly to the pointwise asymptotic normality of $\widehat{\tau}(x)$ as below.

\begin{theorem}\label{thm:tau-asymptotic}
Suppose Assumptions~\ref{Asm1}--
\ref{Asm4} hold. Then, for each fixed
$x\in(0,1)$,
\begin{equation*}
\sqrt{Th}
\left\{
\widehat{\tau}(x)
-\tau(x)
-h^2B_\tau(x)
\right\}
\to_d
N\bigl(0,V_\tau(x)\bigr),
\end{equation*}
where
$B_\tau(x)=C_2\tau''(x)/2$ and
$V_\tau(x)=\tau^2(x)\{\int_{-1}^{1}K^2(v)\,dv\}\Omega_z$, with
$\Omega_z=\sum_{j=-\infty}^{\infty}\operatorname{Cov}(z_0,z_j)$.
\end{theorem}

The proof derives the kernel central limit theorem for $\{z_t\}$
under the GMC condition. This parallels Theorem~2.1 of
\citet{jiang2021adaptive} using mixing type conditions.


For $t=1,\ldots,T$, define
$\widehat{\tau}_t=\widehat{\tau}(t/T)$.
While the unadjusted estimator in \eqref{eq:estimate-tau} achieves an $O(h^2)$ smoothing bias at interior points $t/T\in[h,1-h]$, its kernel weights lose a nonvanishing fraction of their mass near the boundaries $x=0$ and $x=1$, causing an $O(1)$ boundary bias. To correct this loss, we adopt the reflection
method used by \citet{chen2016detecting} and
\citet{jiang2021adaptive}.

Specifically, let $m_T=\lfloor Th\rfloor$, define the interior index set $\mathcal I_T=\{m_T+1,\ldots,T-m_T\}$ and the boundary index set $\mathcal B_T=\{1,\ldots,T\}\setminus\mathcal I_T$, and reflect the squared returns across the endpoints via $r_{1-j}^2=r_{1+j}^2$ and $r_{T+j}^2=r_{T+1-j}^2$ for $j=1,\ldots,m_T$.
The boundary-adjusted estimator is then defined as
\begin{equation*}
\widehat{\tau}_t^{\,\mathrm{ref}}
=
\frac{1}{T}
\sum_{s=t-m_T}^{t+m_T}
K_h\left(\frac{t-s}{T}\right)r_s^2,
\qquad t=1,\ldots,T.
\end{equation*}
For notational simplicity, $\widehat\tau_t$ denotes
$\widehat{\tau}_t^{\,\mathrm{ref}}$ throughout the remainder of the
paper.
\begin{remark}
    Note that unless $\tau'(0)=\tau'(1)=0$, simple reflection eliminates the $O(1)$ loss of kernel mass but leaves a first-order $O(h)$ bias within the boundary band $\mathcal{B}_T$, while preserving the $O(h^2)$ bias on the interior block $\mathcal{I}_T$. Importantly, we do not trim the boundary observations when constructing the second-step likelihood: the volatility recursion and log-likelihood are evaluated over the full sample. In our asymptotic analysis, we partition the first-step estimation errors into $\mathcal{B}_T$ and $\mathcal{I}_T$ and show that because $|\mathcal{B}_T|=2m_T=O(Th)$, the cumulative impact of the $O(h)$ boundary errors—including their recursive propagation into subsequent interior scores—is of order $O_p(\sqrt{h\log T}+\sqrt{T}h^2)=o_p(1)$ under Assumption~\ref{Asm5}. Consequently, the $O(h)$ endpoint bias does not affect the first-order asymptotic distribution of the short-run parameter estimator.
\end{remark}

\subsection{Estimation of the Short-Run Parameters}
\label{subsec:estimation-theta}

Let $\theta=(\alpha,\beta,\varphi)^\top\in\Theta$ denote the
finite-dimensional parameter vector. Under Gaussian
innovations, the normalization condition in
\eqref{omega-normalization} becomes
$\omega(\theta)=1-\alpha-\beta-3\varphi+2\beta\varphi$.
Thus, $\omega$ is not estimated as an additional free parameter.

We first define an infeasible estimator based on the unobservable
short-run process $\{u_t\}$. Conditional on
$\mathcal{F}_{t-1}$, its nonlinear scale recursion is given by
\begin{equation*}
\begin{aligned}
b_{t-1}(\theta)
&=\omega(\theta)+\alpha u_{t-1}^2+\beta h_{t-1}(\theta),\\
h_t(\theta)
&=\frac{b_{t-1}(\theta)+
\sqrt{b_{t-1}^2(\theta)+4\varphi u_t^2}}{2}.
\end{aligned}
\end{equation*}

Under the standard Gaussian innovation, the conditional negative
log-likelihood contribution, up to a constant independent of
$\theta$, is
\begin{equation}
\ell_t(\theta)
=
\frac{1}{2}\frac{u_t^2}{h_t(\theta)}
-\frac{1}{2}\log h_t(\theta)
+
\log\left\{
h_t(\theta)
+
\varphi\frac{u_t^2}{h_t(\theta)}
\right\}.
\label{eq:log-likelihood}
\end{equation}
The corresponding infeasible objective function is
$L_T(\theta)=\sum_{t=1}^{T}\ell_t(\theta)$.

For finite-sample computation, let $\widetilde h_t(\theta)$
denote the initialized infeasible process based on
$u_1^2,\ldots,u_T^2$ with the initial values
$\widetilde h_0(\theta)=u_0^2=1$. Let $\widetilde\ell_t(\theta)$ denote
\eqref{eq:log-likelihood} evaluated at
$(u_t,\widetilde h_t(\theta))$. Moreover, because $\tau_t$ is unobservable, we replace $u_t$ by
$\widehat u_t=r_t/\sqrt{\widehat\tau_t}$ for
$t=1,\ldots,T$. Using the same initialization convention,   we then obtain $\widehat h_t(\theta)$ from the
preceding recursion by replacing the squared inputs $u_t^2$
with $\widehat u_t^2$ and the previous state
$h_{t-1}(\theta)$ with $\widehat h_{t-1}(\theta)$. Let
$\widehat\ell_t(\theta)$ be \eqref{eq:log-likelihood} evaluated at
$(\widehat u_t,\widehat h_t(\theta))$. The feasible Gaussian
conditional maximum likelihood estimator is
\begin{equation}
\widehat\theta_T
=
\arg\min_{\theta\in\Theta}
\widehat L_T(\theta),
\qquad
\widehat L_T(\theta)=\sum_{t=1}^{T}\widehat\ell_t(\theta).
\label{eq:feasible-log-likelihood}
\end{equation}

\begin{assumption}\label{Asm5}
The bandwidth satisfies
$h=c_hT^{-\lambda_h}$, where
$1/4<\lambda_h<1/2$ and $0<c_h<\infty$.
\end{assumption}
Assumption~\ref{Asm5} specifies an  undersmoothing bandwidth range for the two-step procedure. On the one hand, the lower bound $\lambda_h>1/4$ implies $\sqrt{T}h^2\to0$, which simultaneously eliminates the $O(h^2)$ interior smoothing bias and the aggregate $O(h)$ boundary bias over the $O(Th)$ observations in $\mathcal{B}_T$.  On the other hand, the upper bound $\lambda_h<1/2$ guarantees $\sqrt{T}h\to\infty$ and $\log T/(\sqrt{T}h)\to0$. Together with $\sqrt{T}h^2\to0$, this can be used to control the second-order Taylor remainder of the feasible score and to ensure that the cross-term in the interior kernel projection vanishes at rate $O_p((\sqrt{T}h)^{-1})=o_p(1)$.

\begin{assumption}\label{Asm6}
For some $q>4$, $E|z_t|^q<\infty$, where $z_t=u_t^2-1$.
\end{assumption}

Assumption~\ref{Asm6} provides the moment order needed to establish uniform consistency of $\widehat\tau_t$ over $1\le t\le T$ and to control the higher-order derivatives of the log-likelihood. By Proposition~\ref{prop:rt-gmc}, $\mathbb{E}|z_t|^q<\infty$ implies that $\{z_t\}$ is also $L^q$ GMC.

\begin{proposition}[Uniform first-step rate]
\label{prop:first-step-uniform}
Under Assumptions~\ref{Asm1}--\ref{Asm6}, the boundary-adjusted
first-step estimator satisfies that, with probability approaching one,
(i). $
\max_{1\leq t\leq T}
\left|
\widehat\tau_t-\tau_t
\right|
=
O_p(\kappa_T),$
where $
\kappa_T
=
\sqrt{{\log T}/{Th}}+h;$
(ii). for  some $c_\tau>0$, $
\inf_{1\leq t\leq T}\widehat\tau_t
\geq c_\tau;
$
and (iii). for the interior block, 
$
\max_{t\in\mathcal I_T}
\left|\widehat\tau_t-\tau_t\right|
=O_p(\kappa_{I,T}),
$ where $
\kappa_{I,T}
=\sqrt{{\log T}/{Th}}+h^2 .
$
\end{proposition}
The proof of Proposition~\ref{prop:first-step-uniform}  relies on a maximal inequality for $L^q$ GMC processes \citep[Theorem~2(ii)]{liu2013probability}. The difference between the full-sample rate $\kappa_T$ and the interior rate $\kappa_{I,T}$ reflects the $O(h)$ boundary bias on $\mathcal{B}_T$.
Equipped with the uniform first-step convergence rate in Proposition~\ref{prop:first-step-uniform}, we first establish the consistency of $\widehat\theta_T$.

\begin{theorem}\label{thm:consistency}
 Suppose the innovations are standard Gaussian and
Assumptions~\ref{Asm1}--\ref{Asm6} hold, and
$\alpha_0+\varphi_0>0$.  Then
$
\widehat\theta_T
\to_p
\theta_0
$ as $T\to\infty$.
\end{theorem}
When $\alpha_0+\varphi_0>0$, the short-run dynamic parameters are uniquely identified from the Gaussian conditional likelihood \citep[Theorem~5]{smetanina2021asymptotic}.Therefore, by the  uniform convergence of the feasible criterion $T^{-1}\widehat L_T(\theta)$ to the population likelihood $\mathbb{E}\{\ell_t(\theta)\}$, we thus obtain the consistency from the classical argmax continuous mapping theorem \citep[Theorem 2.1]{newey1994large}.

We now turn to the asymptotic distribution of $\widehat\theta_T$. Let $s_t(\theta)=\partial\ell_t(\theta)/\partial\theta$ denote the infeasible score and denote the population Hessian matrix as \begin{equation}\label{hessian}
    H=\mathbb{E}[\partial^2\ell_t(\theta_0)/(\partial\theta\,\partial\theta^\top)].
\end{equation}    Because the feasible score $\partial\widehat L_T(\theta_0)/\partial\theta$ evaluates the model at the standardized returns $\widehat u_i^2=u_i^2\tau_i/\widehat\tau_i$, a perturbation in $\widehat\tau_i$ at time $i$ affects not only the time-$i$ score $s_i(\theta_0)$ directly, but also every subsequent score $s_t(\theta_0)$ ($t>i$) through the recursive volatility process $h_t(\theta_0)$. The key idea of the proof is to show that the cumulative sensitivity of future scores to a relative scale perturbation at time $t$ is governed by the mean loading vector
\begin{equation}\label{eq:main-Ctau-def}
C_\tau
=
\mathbb{E}\left\{
\sum_{j=0}^{\infty}
u_{t-j}^2\frac{\partial s_t(\theta_0)}{\partial(u_{t-j}^2)}
\right\},
\end{equation}
where $\partial s_t(\theta_0)/\partial(u_{t-j}^2)$ measures the partial derivative of the time-$t$ score with respect to the squared observation $u_{t-j}^2$ holding all other observations fixed. Projecting the first-step kernel estimation error $\widehat\tau_i-\tau_i$ onto the score expansion yields the first-order score adjustment
\[
\psi_t=-C_\tau z_t,
\]
so that the effective influence vector at time $t$ is $$\xi_t=s_t(\theta_0)+\psi_t=s_t(\theta_0)-C_\tau z_t.$$ 

\begin{theorem}\label{thm:asymptotic-normality}
Suppose the innovations are standard Gaussian,
Assumptions~\ref{Asm1}--\ref{Asm6} hold, $\theta_0$ is interior to
$\Theta$, and $H$ is nonsingular.
Then
\[
\sqrt{T}
\left(
\widehat\theta_T-\theta_0
\right)
\to_d
N(0,\Sigma),
\qquad
\Sigma=H^{-1}VH^{-\top},
\]
where $H$ is defined in \eqref{hessian}, 
and $V=\sum_{j=-\infty}^{\infty}\operatorname{Cov}(\xi_t,\xi_{t-j})$ is the long-run covariance matrix of $\{\xi_t\}$.
\end{theorem}

In general two-step semiparametric time-series models, conducting inference based on $\Sigma=H^{-1}VH^{-1}$ is complicated by two practical difficulties. First, although the exact conditional score $\{s_t(\theta_0)\}$ is a martingale difference sequence with $\mathbb{E}\{s_t(\theta_0)s_t(\theta_0)^\top\}=H$, the first-step error sequence $\{z_t\}$ is serially correlated across all lags $j\neq 0$, which typically induces non-vanishing autocovariances $\operatorname{Cov}(z_t,z_{t-j})$ as well as cross-covariances $\operatorname{Cov}(s_t(\theta_0),z_{t+j})$ for $j>0$. Estimating $V=\sum_{j=-\infty}^\infty\operatorname{Cov}(\xi_t,\xi_{t-j})$ directly from sample residuals would therefore require a heteroskedasticity and autocorrelation consistent (HAC) kernel estimator \citep[e.g.,][]{newey1994large} with a user-selected lag-truncation bandwidth, which can suffer from substantial finite-sample distortion when volatility persistence $\alpha_0+\beta_0$ is high. Second, the definition of the loading vector $C_\tau$ in \eqref{eq:main-Ctau-def} involves an infinite series of recursive observation derivatives $\partial s_t(\theta_0)/\partial(u_{t-j}^2)$, which would ordinarily require an additional truncation lag to approximate, see such treatment in \cite{Jiang2023Testing} in the multivariate semiparametric models. 

Under the normalized Gaussian SRT-GARCH specification, however, two structural properties of the model completely eliminate the need for HAC estimation or lag truncation. This mirrors the adaptive inference phenomenon documented for the nested S-GARCH model in \citet{jiang2021adaptive}, where the two-step estimator is shown to be adaptive to the unknown functional form of $\tau(\cdot)$ and to share the exact same asymptotic efficiency as the parametric variance-targeting estimator \citep{francq2011merits} when the process is stationary. Specifically, the limiting covariance matrix $\Sigma$ depends only on $\theta_0$ and is invariant to $\tau(\cdot)$; moreover, because the kernel projection of $\widehat\tau_i-\tau_i$ aggregates to the same influence term $-C_\tau T^{-1/2}\sum_{s=1}^T z_s$ as the parametric sample-variance error $(\overline\tau_T-\tau_0)/\tau_0=T^{-1}\sum_{s=1}^T z_s$, estimating a time-varying scale $\tau(\cdot)$ nonparametrically incurs no first-order asymptotic efficiency loss relative to parametric variance targeting.  

Nevertheless, establishing this simplification for SRT-GARCH differs fundamentally from the S-GARCH analysis in \citet{jiang2021adaptive}. When $\varphi_0>0$, contemporaneous feedback makes $u_t^2=b_{t-1}\epsilon_t^2+\varphi_0\epsilon_t^4$ a quadratic polynomial in $\epsilon_t^2$, renders the score $s_t(\theta_0)$ a nonlinear function of the innovation, and couples the scale of $u_t^2$ with both the intercept $\omega(\theta)$ and the feedback parameter $\varphi$ inside the square-root recursion $h_t(\theta)$. Consequently, neither the score--error orthogonality nor the truncation-free representation of $C_\tau$ follows from standard GARCH models; instead, we establish them through the following two results:
\begin{enumerate}[1.]
\item \emph{Rank-one covariance decomposition and score--error orthogonality.} In Lemma~\ref{lem:covariance-simplification} of the Appendix, we show that $z_t=u_t^2-1$ admits the ARMA-type representation
\[
z_t=(\alpha_0+\beta_0)z_{t-1}+\zeta_t-\beta_0\zeta_{t-1}+\beta_0\varphi_0(\epsilon_{t-1}^2-1),
\]
where $\zeta_t=u_t^2-b_{t-1}-3\varphi_0$ and $\epsilon_t^2-1$ are martingale differences with respect to $\mathcal{F}_t$. By a Beveridge--Nelson telescoping decomposition, the partial sum $T^{-1/2}\sum_{t=1}^T z_t$ is $L^2$-equivalent to the partial sum of the martingale difference sequence
\[
\chi_t=\frac{(1-\beta_0)\zeta_t+\beta_0\varphi_0(\epsilon_t^2-1)}{1-\alpha_0-\beta_0}.
\]
Because both $s_t(\theta_0)$ and $\chi_t$ are martingale difference sequences with respect to $\mathcal{F}_t$, all of their serial autocovariances and cross-covariances at nonzero lags ($j\neq 0$) vanish identically. Even more importantly, at lag zero, the conditional score identities implied by the variance normalization $\omega(\theta)=1-\alpha-\beta-3\varphi+2\beta\varphi$ ensure exact contemporaneous orthogonality between the second-step score and the first-step martingale innovation:
\[
\mathbb{E}\bigl\{s_t(\theta_0)\chi_t\bigr\}=0.
\]
Consequently, the cross-terms between $s_t(\theta_0)$ and $\psi_t=-C_\tau z_t$ cancel completely in the long-run covariance matrix $V$, yielding the exact rank-one decomposition
\begin{equation}\label{eq:main-V-Sigma-decomp}
V=H+\Omega_z C_\tau C_\tau^\top,
\qquad
\Sigma=H^{-1}+\Omega_z(H^{-1}C_\tau)(H^{-1}C_\tau)^\top,
\end{equation}
where $\Omega_z=\sum_{j=-\infty}^\infty\operatorname{Cov}(z_0,z_j)=\mathbb{E}(\chi_t^2)$. Thus, nonparametric estimation of the unknown long-run scale $\tau(\cdot)$ inflates the oracle asymptotic covariance $H^{-1}$ by a positive semidefinite matrix of rank at most one.

\item \emph{Closed-form $\Omega_z$ and truncation-free representation of $C_\tau$.} Lemma~\ref{lem:covariance-simplification} further shows that under Gaussian innovations, $\Omega_z=\mathbb{E}(\chi_t^2)$ admits the explicit closed-form rational expression in terms of $(\alpha_0,\beta_0,\varphi_0)$:
\begin{equation}\label{eq:main-Omegaz-closed}
\Omega_z
=
\frac{2(1-\beta_0)^2\operatorname{Var}(b_t)+2\{1-\beta_0+(3-2\beta_0)\varphi_0\}^2+24(1-\beta_0)^2\varphi_0^2}{(1-\alpha_0-\beta_0)^2},
\end{equation}
with $\operatorname{Var}(b_t)=[2\{\alpha_0+(3\alpha_0+\beta_0)\varphi_0\}^2+24\alpha_0^2\varphi_0^2]/(1-3\alpha_0^2-2\alpha_0\beta_0-\beta_0^2)$. Moreover, Lemma~\ref{lem:Ctau-estimation} of the Appendix re-expresses the infinite series $C_\tau$ in \eqref{eq:main-Ctau-def} as a single expectation:
\begin{equation}\label{eq:main-Ctau-product}
C_\tau
=
-\mathbb{E}\left[
\frac{\partial\ell_t(\theta_0)}{\partial\theta}
\left\{
\varphi_0\frac{\partial\ell_t(\theta_0)}{\partial\varphi}
+(1-\alpha_0-\beta_0)\frac{\partial\ell_t(\theta_0)}{\partial\omega}
\right\}
\right],
\end{equation}
where $\partial\ell_t(\theta)/\partial\omega$ denotes the partial derivative of $\ell_t(\theta)$ with respect to an unnormalized intercept $\omega$ (holding $(\alpha,\beta,\varphi)$ and the observations fixed) and is computed via one scalar recursion; see \eqref{eq:intercept-process-derivative}--\eqref{eq:intercept-likelihood-derivative} in the Appendix.
\end{enumerate}

Together, \eqref{eq:main-V-Sigma-decomp}--\eqref{eq:main-Ctau-product} yield a plug-in covariance estimator requiring neither lag truncation nor additional tuning parameters
\[
\widehat\Sigma
=
\widehat H^{-1}
+\widehat\Omega_z(\widehat H^{-1}\widehat C_\tau)(\widehat H^{-1}\widehat C_\tau)^\top,
\qquad
\widehat H
=
\frac{1}{T}\sum_{t=1}^{T}\frac{\partial^2\widehat\ell_t(\widehat\theta_T)}{\partial\theta\,\partial\theta^\top},
\]
where $\widehat\Omega_z$ is obtained by substituting $\widehat\theta_T$ into \eqref{eq:main-Omegaz-closed} and $\widehat C_\tau$ is the sample average counterpart of \eqref{eq:main-Ctau-product} evaluated at $\widehat\theta_T$.

\section{Volatility Forecasting}
\label{sec:volatility-forecasting}
Having established the estimation and asymptotic properties of the
model, we now turn to volatility forecasting. The SRT-GARCH
specification naturally separates the forecast into two components:
a short-run conditional-moment recursion and a forecast of the smooth
long-run volatility component.

For forecasts beyond the observed sample, suppose that the long-run
component $\tau(\cdot)$ admits a positive deterministic continuation
over the required forecast horizon. Conditional on this continuation,
symmetry implies
\[
\mathbb{E}(r_t\mid\mathcal{F}_{t-1})=0,
\]
so forecasting the conditional variance is equivalent to forecasting
\[
\mathbb{E}(r_{T+k}^2\mid\mathcal{F}_T)
=
\tau_{T+k}
\mathbb{E}(u_{T+k}^2\mid\mathcal{F}_T).
\]
The short-run expectation is determined by the model recursion,
whereas forecasting $\tau_{T+k}$ requires a separate extrapolation of
the long-run component.

Under Gaussian innovations and the normalization
\eqref{omega-normalization}, the feasible one-step-ahead volatility
forecast is
\begin{equation*}
\widehat{\mathbb{E}}
\left(
r_{T+1}^2\mid\mathcal{F}_T
\right)
=
\widehat{\tau}_{T+1\mid T}
\left[
1-\widehat{\alpha}-\widehat{\beta}
+2\widehat{\beta}\widehat{\varphi}
+\widehat{\alpha}\widehat{u}_T^2
+\widehat{\beta}\widehat{h}_T
\right],
\end{equation*}
where $\widehat{\tau}_{T+1\mid T}$ denotes the forecast of the
long-run volatility component. The corresponding population
recursion and its multi-step extension are derived in
\ref{app:forecast-proof}.

In the empirical rolling-window forecasts, the long-run component is
extrapolated using
\begin{equation*}
\widehat{\tau}_{s+k\mid s}
=
\max\left\{
(k+1)\widehat{\tau}_s
-k\widehat{\tau}_{s-1},
c_\tau
\right\},
\end{equation*}
where $c_\tau>0$ is a small numerical floor that preserves positivity.
This two-point extrapolation is an empirical forecasting convention
rather than a consequence of the population moment recursion.

\section{Simulations}
\label{sec:simulations}
We use pilot-based cross-validation to select the smoothing bandwidth
and examine the finite-sample behavior of the two-step estimator under
 the three DGPs described in Section~\ref{sub:distributional}. They represent, respectively,
periodic variation, a temporary rise followed by a decline, and
a monotone nonlinear increase in the unconditional variance. For each DGP, we consider sample sizes
$T\in\{1000,2000\}$.
The innovations are independently generated from
$\mathcal N(0,1)$. The intercept $\omega_0$ is determined by the
Gaussian normalization
$\omega_0=1-\alpha_0-\beta_0-3\varphi_0+2\beta_0\varphi_0$. The number of Monte Carlo replications is
$M=1000$.
For each replication, we first generate a stationary short-run
RT-GARCH process after discarding the first 500 observations as
burn-in, and then multiply the short-run process by
$\sqrt{\tau(t/T)}$ to obtain the observed return series $\{r_t\}$.

\subsection{Bandwidth Selection}
\label{subsec:bandwidth-selection}

For both the simulations and empirical applications, we follow the
pilot-based cross-validation approach of
\citet{jiang2021adaptive}. The bandwidth for estimating
$\tau(\cdot)$ is selected as follows.

\begin{algorithm}[Cross-validation bandwidth selection]
\label{alg:bandwidth-selection}
~~
\begin{enumerate}
\item
Choose a pilot bandwidth $h_0=T^{-\lambda_0}$ with
$\lambda_0\in(1/4,1/2)$.
Using the kernel estimator in \eqref{eq:estimate-tau}, compute the
pilot estimates
$\widehat{\tau}_{t,0}=\widehat{\tau}_0(t/T)$ and
$\widehat u_{t,0}=r_t/\sqrt{\widehat{\tau}_{t,0}}$.

\item
Estimate the short-run SRT-GARCH model from
$\{\widehat u_{t,0}\}_{t=1}^T$ by Gaussian conditional maximum
likelihood, and obtain the pilot scale estimates
$\{\widehat h_{t,0}\}_{t=1}^T$.

\item
For each candidate bandwidth $h\in\mathcal H_T$, compute the
leave-one-out kernel estimator
\[
\widehat{\tau}_{-t,h}
=
\frac{1}{T}
\sum_{\substack{s=1\\s\neq t}}^{T}
K_h\left(\frac{t-s}{T}\right)r_s^2.
\]
Define the cross-validation criterion
\begin{equation}
CV(h)
=
\frac1T
\sum_{t=1}^{T}
\left(
\frac{r_t^2}
{\widehat{\tau}_{-t,h}\widehat h_{t,0}}
-1
\right)^2.
\label{eq:cv-bandwidth}
\end{equation}

\item
Select
$\widehat h_{\mathrm{CV}}=\arg\min_{h\in\mathcal H_T}CV(h)$, where
$\mathcal H_T=[c_{\min}T^{-\lambda_0},
c_{\max}T^{-\lambda_0}]$.
\end{enumerate}
\end{algorithm}


\subsection{Finite-Sample Performance of the Parameter Estimator}
\label{subsec:parameter-simulation}


For the $m$th replication, we select the bandwidth using
Algorithm~\ref{alg:bandwidth-selection}, estimate
$\widehat\tau_t^{(m)}$, construct
$\widehat u_t^{(m)}
=r_t^{(m)}/\sqrt{\widehat\tau_t^{(m)}}$,
and obtain the feasible estimator
$\widehat\theta_T^{(m)}$ by Gaussian conditional maximum
likelihood in \eqref{eq:feasible-log-likelihood}.
For each component of
$\theta=(\alpha,\beta,\varphi)^\top$, 
Table~\ref{table_simulation} reports the empirical bias and empirical standard deviation (ESD) of the parameter estimates across the $M$ Monte Carlo replications, together with the average estimated
asymptotic standard deviation (ASD) and the empirical coverage probability of the corresponding 95\% confidence intervals ($\mathrm{CP}_{0.95}$).


\begin{table}[htbp]
\centering
\caption{Finite-sample performance of the parameter estimates.}
\label{table_simulation}
\begin{tabular*}{0.75\textwidth}{
@{\extracolsep{\fill}}
ll
S[table-format=-1.4]
S[table-format=-1.4]
S[table-format=-1.4]
@{}
}
\toprule
& & {$\alpha$} & {$\beta$} & {$\varphi$} \\
\midrule

\multicolumn{5}{@{}l}{
\textit{DGP 1: }
$\tau(x)=1+0.5\sin(2\pi x)$,
$\theta_0=(0.20,0.60,0.05)^\top$
}\\

\midrule

$T=1000$ 
& Bias & -0.0102 & -0.0381 & -0.0013 \\*
& ESD & 0.0494 & 0.0916 & 0.0151 \\*
& ASD & 0.0510 & 0.0915 & 0.0151 \\*
& $\mathrm{CP}_{0.95}$ & 0.9400 & 0.9530 & 0.9470 \\[2pt]
$T=2000$ 
& Bias & -0.0071 & -0.0205 & 0.0002 \\*
& ESD & 0.0356 & 0.0603 & 0.0104 \\*
& ASD & 0.0361 & 0.0610 & 0.0105 \\*
& $\mathrm{CP}_{0.95}$ & 0.9340 & 0.9380 & 0.9510 \\[2pt]

\midrule
\multicolumn{5}{@{}l}{
\textit{DGP 2: }
$\tau(x)=1+4(x-x^2)$,
$\theta_0=(0.20,0.50,0.03)^\top$
}\\*
\midrule

$T=1000$ 
& Bias & -0.0089 & -0.0571 & -0.0038 \\*
& ESD & 0.0483 & 0.1172 & 0.0140 \\*
& ASD & 0.0506 & 0.1231 & 0.0147 \\*
& $\mathrm{CP}_{0.95}$ & 0.9460 & 0.9390 & 0.9530 \\[2pt]
$T=2000$ 
& Bias & -0.0084 & -0.0265 & -0.0026 \\*
& ESD & 0.0352 & 0.0831 & 0.0102 \\*
& ASD & 0.0356 & 0.0823 & 0.0102 \\*
& $\mathrm{CP}_{0.95}$ & 0.9330 & 0.9460 & 0.9330 \\[2pt]

\midrule
\multicolumn{5}{@{}l}{
\textit{DGP 3: }
$\tau(x)=1+4x^2$,
$\theta_0=(0.15,0.65,0.04)^\top$
}\\*
\midrule

$T=1000$ 
& Bias & -0.0054 & -0.0708 & -0.0040 \\*
& ESD & 0.0442 & 0.1129 & 0.0157 \\*
& ASD & 0.0460 & 0.1112 & 0.0154 \\*
& $\mathrm{CP}_{0.95}$ & 0.9480 & 0.9460 & 0.9130 \\[2pt]
$T=2000$ & Bias & -0.0009 & -0.0431 & -0.0030 \\*
& ESD & 0.0327 & 0.0708 & 0.0107 \\*
& ASD & 0.0324 & 0.0709 & 0.0106 \\*
& $\mathrm{CP}_{0.95}$ & 0.9470 & 0.9340 & 0.9370 \\[2pt]

\bottomrule
\end{tabular*}
\end{table}

Table~\ref{table_simulation} shows a common pattern across the three designs. The absolute bias and empirical dispersion generally decline as the sample size $T$ increases, broadly consistent with root-$T$ convergence. The empirical coverage probabilities are generally close to the nominal 95\% level, 
supporting the finite-sample accuracy of the proposed asymptotic inference procedure.

\begin{figure}[htb]
	\centering
	\includegraphics[width=.8\linewidth]{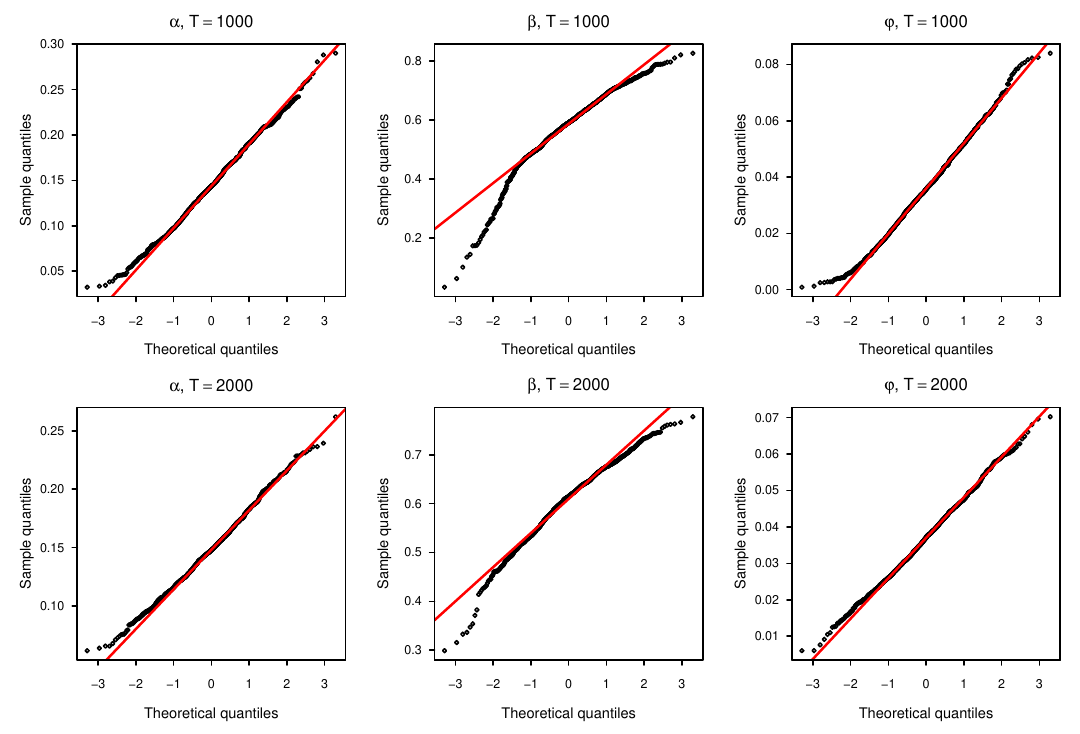}
	\caption{QQ plots of parameter estimates under DGP~3.}
	\label{fig:simulation_qq}
\end{figure}

Figure~\ref{fig:simulation_qq} provides a complementary distributional check for the parameter estimates under DGP~3. The QQ plots move closer to the reference line as the sample size increases, with the closest agreement observed at $T=2000$. Similar patterns are observed for DGPs~1 and~2, and are therefore omitted for brevity. Together with Table~\ref{table_simulation}, these results confirm that the asymptotic normal approximation provides
reliable finite-sample inference.

\section{Applications}
\label{sec:applications}

The empirical analysis considers two financial return series: the FTSE~100 index and U.S. natural gas futures.  For each dataset, we first test covariance stationarity and examine the fitted long-run and short-run volatility components. We then compare residual diagnostics, out-of-sample volatility forecasts, and Value-at-Risk (VaR) forecasts against the nested RT-GARCH and S-GARCH models.

For reference, we also report randomly weighted bootstrap standard deviation (BSD) alongside the asymptotic standard errors implied by Theorem~\ref{thm:asymptotic-normality} as a robustness check. The implementation details are provided in \ref{app:weighted-bootstrap}.

To assess the forecasting performance of the proposed model, we consider both volatility and tail-risk forecasts. While the multi-step volatility forecasts follow directly from the forecasting formulas developed above, the monotone observation equation provides a convenient way to construct one-step-ahead VaR forecasts. Let $q_p$ denote the $p$th quantile of the innovation distribution. Under the Gaussian specification, $q_p=\Phi^{-1}(p)$. Since $r_{t+1}$ is a strictly increasing function of $\epsilon_{t+1}$ conditional on $\mathcal{F}_t$, its conditional $p$th quantile is obtained by evaluating the observation equation at $\epsilon_{t+1}=q_p$. Under the positive-loss convention, the resulting one-step-ahead SRT-GARCH VaR is
\begin{equation*}
\operatorname{VaR}_{t+1\mid t}(p)
=
-
\sqrt{\tau_{t+1}}\,
q_p
\sqrt{
\omega+\alpha u_t^2+\beta h_t+\varphi q_p^2
}.
\end{equation*}
In implementation, $\tau_{t+1}$ is replaced by its forecast $\widehat{\tau}_{t+1\mid t}$, and the unknown parameters and state variables are replaced by their estimated counterparts. The same procedure is applied to the benchmark models. In particular, the RT-GARCH model imposes a constant long-run component, whereas the S-GARCH model is obtained by setting $\varphi=0$. The empirical analysis below therefore compares the three models in terms of both volatility forecasting accuracy and one-step-ahead VaR performance.

\subsection{FTSE 100 Index}

The first application uses $T=2700$ daily FTSE~100 returns from January~2, 2004, to September~10, 2014. The covariance-stationarity test of \citet{xiao2007testing}, implemented with the Bartlett kernel, produces $C_T=2.8374$ and rejects stationarity at the 1\% level. 
This provides evidence consistent with a time-varying long-run volatility component and motivates the semiparametric specification before comparing the alternative short-run dynamics.

The parameter estimates in Table~\ref{tab:ftse-estimates} are based on the first $2050$ observations, covering January 2, 2004, through February 13, 2012. 
Algorithm~\ref{alg:bandwidth-selection} selects the same bandwidth, $\widehat h_{\mathrm{CV}}=0.0981$, for both SRT-GARCH and S-GARCH.
Table~\ref{tab:ftse-estimates} reports the SRT-GARCH parameter estimates together with their asymptotic standard errors. 
Randomly weighted bootstrap standard deviations based on 500 replications are also reported as a finite-sample robustness check.

The large estimate of $\beta$ indicates persistent short-run volatility dynamics. 
The positive estimate $\widehat\varphi=0.0262$ is consistent with contemporaneous volatility feedback. Relative to its estimated uncertainty, the conventional $t-$statistic based on the asymptotic and bootstrap standard errors are $2.7579$ and $2.9438$, respectively, providing clear evidence of real-time return feedback.  Although these statistics do not account for the boundary restriction $\varphi\ge0$, they still indicate that the estimated contemporaneous feedback effect is substantial.

\begin{table}[htbp]
\centering
\caption{SRT-GARCH(1,1) parameter estimates for FTSE 100 returns.}
\label{tab:ftse-estimates}
\begin{tabular}{lccc}
\toprule
& $\alpha$ & $\beta$ & $\varphi$\\
\midrule
Estimate
& 0.0700 & 0.8791 & 0.0262\\
ASD
&0.0191 & 0.0265 &0.0095 \\
BSD
& 0.0186 & 0.0256 & 0.0089\\
\bottomrule
\end{tabular}
\end{table}

\begin{figure}[htb]
	\centering
	\includegraphics[width=0.8\linewidth]{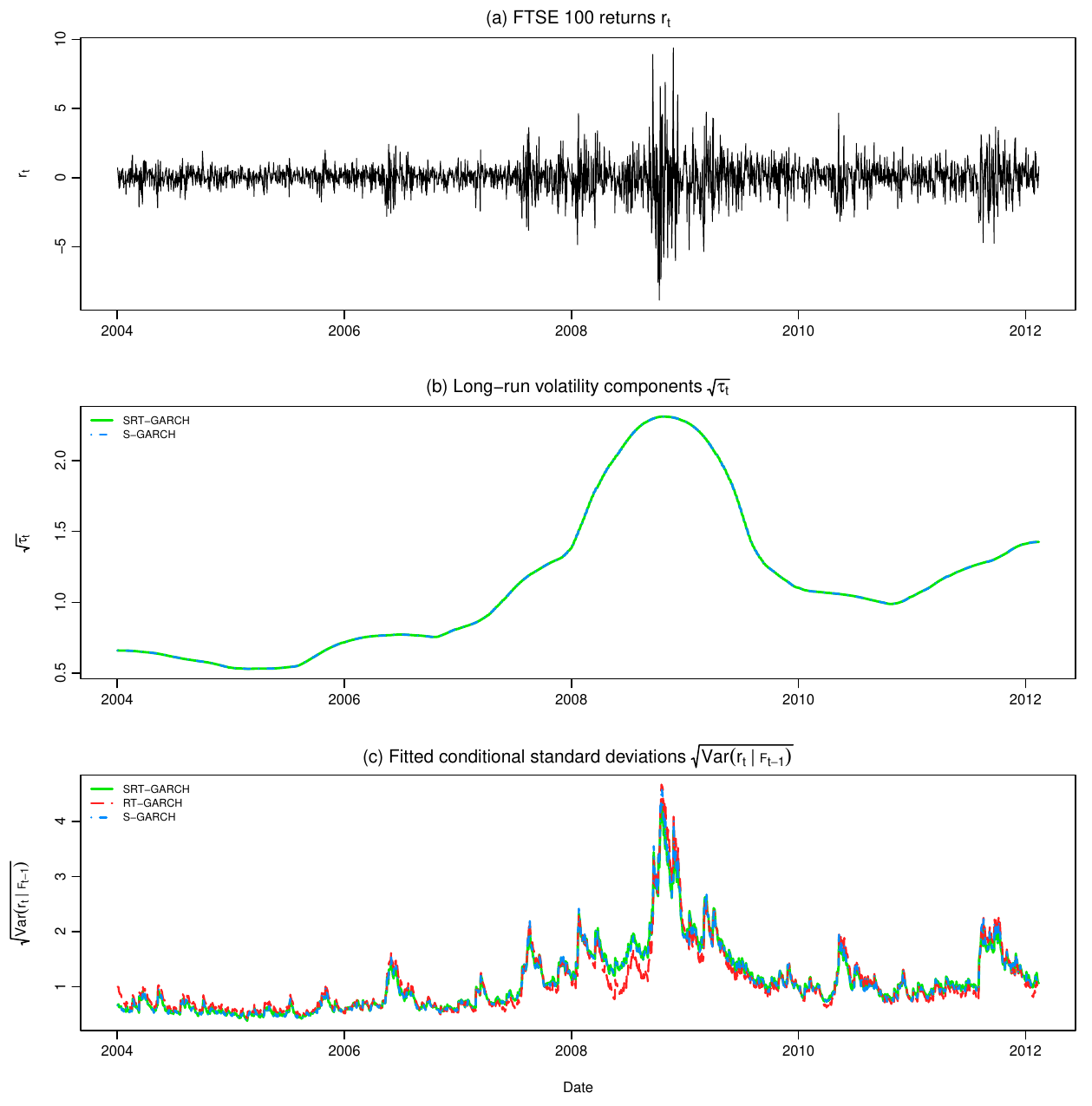}
\caption{FTSE 100 returns and estimated volatility components over the period from January 2, 2004, to February 13, 2012.}
	\label{fig:the FTSE 100 Index return timeplot}
\end{figure}

Figure~\ref{fig:the FTSE 100 Index return timeplot} illustrates the resulting volatility decomposition. 
Panel~(b) shows gradual changes in the estimated long-run scale. 
The SRT-GARCH and S-GARCH long-run estimates coincide because their cross-validation procedures select the same bandwidth and both models consequently use the same first-step kernel estimate of $\tau_t$.
Panel~(c) compares the fitted conditional standard deviations $[\widehat{\mathbb E}(r_t^2\mid\mathcal F_{t-1})]^{1/2}$ across the three models.

\begin{figure}[htb]
\centering
\includegraphics[width=0.68\linewidth]{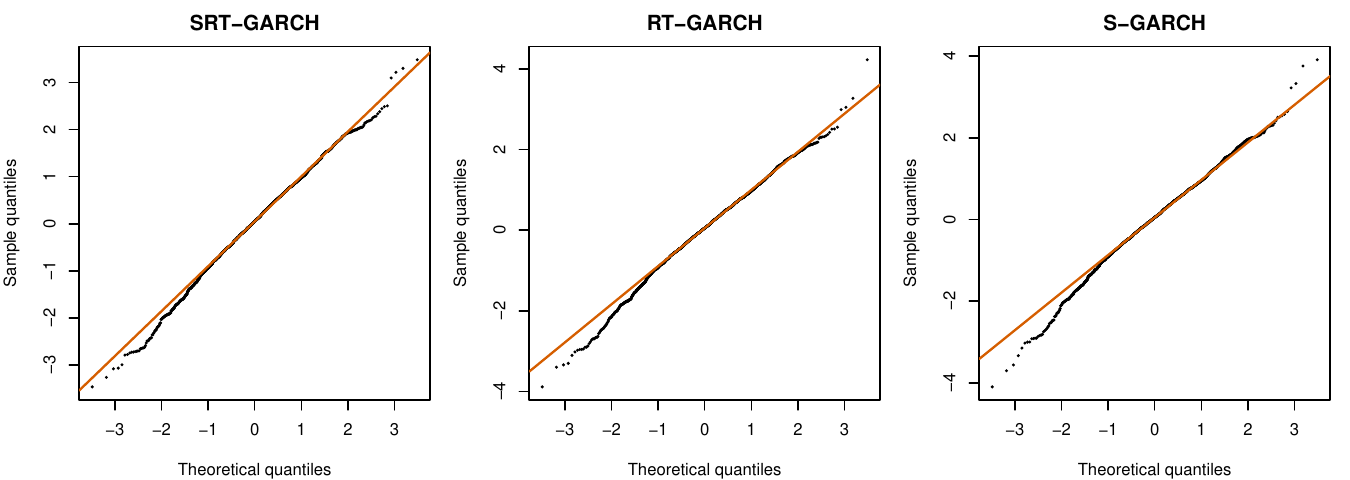}
\caption{QQ plots of FTSE 100 standardized
residuals.}
\label{fig:the FTSE 100 Index return resi QQ}
\end{figure}

The residual diagnostics provide further evidence on the adequacy of the alternative short-run specifications.
Figure~\ref{fig:the FTSE 100 Index return resi QQ} places the SRT-GARCH residuals closer to the normal reference line. 
The corresponding Kolmogorov--Smirnov normality tests yield $p$-values of $0.0550$, $0.0340$, and $0.0143$ for SRT-GARCH, RT-GARCH, and S-GARCH, respectively.
At the 5\% significance level, the standard-normal innovation specification is not rejected for SRT-GARCH, whereas it is rejected
for both RT-GARCH and S-GARCH.

We next evaluate out-of-sample volatility forecasting performance over the final $N=650$ trading days of the sample using a rolling window of fixed length $W=T-N$. 
In particular, 
at each forecast date $s$, all three models are re-estimated using the most recent $W$ observations $\{r_{s-W+1},\ldots,r_s\}$, and $k$-step-ahead conditional variance forecasts $\widehat v_{s+k\mid s}=\widehat{\mathbb E}(r_{s+k}^2\mid\mathcal F_s)$ are generated for horizons $k\in\{1,5,10,15\}$.

Following \citet{patton2011volatility}, forecast accuracy is evaluated using the mean squared error (MSE) and QLIKE loss. 
For a forecast horizon $k$, the average losses are defined as
\begin{align*}
\operatorname{MSE}(k)
&=
\frac{1}{N-k+1}
\sum_{s=T-N}^{T-k}
\left[
r_{s+k}^2-\widehat v_{s+k\mid s}
\right]^2,
\\
\operatorname{QLIKE}(k)
&=
\frac{1}{N-k+1}
\sum_{s=T-N}^{T-k}
\left[
\frac{r_{s+k}^2}{\widehat v_{s+k\mid s}}
+
\log\left(\widehat v_{s+k\mid s}\right)
\right].
\end{align*}
Lower values of MSE and QLIKE indicate more accurate volatility forecasts.

To further assess whether differences in average losses are statistically distinguishable, we use the Diebold--Mariano test with the Harvey--Leybourne--Newbold finite-sample correction (DM--HLN) \citep{diebold1995comparing,harvey1997testing}. The null hypothesis is equal unconditional predictive accuracy. All reported tests are two-sided and
use Bartlett--Newey--West estimates of the long-run variance.

Table~\ref{table:forcas_ftse}  reports the out-of-sample volatility forecasting performance. SRT-GARCH yields the smallest MSE at all four forecast horizons. The SRT-GARCH model achieves the lowest MSE across all four forecast horizons ($0.9492$, $0.9689$, $0.9855$, and $0.9866$ at $k=1,5,10,15$) and the lowest QLIKE at the 5-, 10-, and 15-step horizons ($0.3601$, $0.3717$, and $0.3696$). Relative to the stationary RT-GARCH model, SRT-GARCH delivers uniformly lower losses at every horizon, and the gains grow increasingly pronounced when the forecast horizon is large. More specifically,  the DM--HLN test rejects equal predictive accuracy in favor of SRT-GARCH at the 5\% level for 10- and 15-step MSE and for 5-, 10-, and 15-step QLIKE (and at the 10\% level for 1-step QLIKE). This pattern demonstrates the substantial improvement of accounting for smooth changes in the unconditional variance when forming multi-step volatility forecasts. Relative to the S-GARCH model, which shares the nonparametric long-run component, SRT-GARCH still yields slightly lower average losses in most cases,  although the pairwise differences between the two semiparametric specifications are not statistically significant.

\setlength{\tabcolsep}{4pt}

\begin{table}[htbp]
\centering
\caption{Out-of-sample volatility forecasting performance for FTSE 100 returns.}
\label{table:forcas_ftse}
\begin{tabular}{lcccccccc}
\toprule
& \multicolumn{2}{c}{1-step}
& \multicolumn{2}{c}{5-step}
& \multicolumn{2}{c}{10-step}
& \multicolumn{2}{c}{15-step}\\
\cmidrule(lr){2-3}\cmidrule(lr){4-5}
\cmidrule(lr){6-7}\cmidrule(lr){8-9}
Model & MSE & QLIKE & MSE & QLIKE
      & MSE & QLIKE & MSE & QLIKE\\
\midrule

SRT-GARCH
& \textbf{0.9492} & 0.3342
& \textbf{0.9689} & \textbf{0.3601}
& \textbf{0.9855} & \textbf{0.3717}
& \textbf{0.9866} & \textbf{0.3696} \\
\addlinespace[2pt]

RT-GARCH
& 0.9616 & 0.3544
& 0.9981 & 0.4015
& 1.0419 & 0.4348
& 1.0609 & 0.4448 \\
\multicolumn{1}{r}{$p_{\mathrm{DM}}$}
& {\footnotesize $(0.2376)$}
& {\footnotesize $(0.0778^{*})$}
& {\footnotesize $(0.1215)$}
& {\footnotesize $(0.0177^{*})$}
& {\footnotesize $(0.0409^{*})$}
& {\footnotesize $(0.0041^{*})$}
& {\footnotesize $(0.0298^{*})$}
& {\footnotesize $(0.0042^{*})$} \\
\addlinespace[2pt]

S-GARCH
& 0.9500 & \textbf{0.3335}
& 0.9700 & 0.3607
& 0.9900 & 0.3731
& 0.9936 & 0.3703 \\
\multicolumn{1}{r}{$p_{\mathrm{DM}}$}
& {\footnotesize $(0.7531)$}
& {\footnotesize $(0.7284)$}
& {\footnotesize $(0.8032)$}
& {\footnotesize $(0.8023)$}
& {\footnotesize $(0.5437)$}
& {\footnotesize $(0.6859)$}
& {\footnotesize $(0.3930)$}
& {\footnotesize $(0.8599)$} \\

\bottomrule
\end{tabular}

\vspace{2pt}
\parbox{0.97\linewidth}{\footnotesize
\textit{Notes:}
Lower values indicate better forecasting performance, and bold entries denote the smallest loss at each horizon. Values in parentheses are DM--HLN $p$-values, denoted by $p_{\mathrm{DM}}$, for testing equal unconditional predictive accuracy between SRT-GARCH and the model in the corresponding row. An asterisk indicates $p_{\mathrm{DM}}<0.10$.
}
\end{table}

We next evaluate one-step-ahead Value-at-Risk forecasts at the
left-tail probability level $p$ by the conditional coverage test of
\citet{christoffersen1998evaluating}. Let
$I_t(p)=\mathbb{I}\{r_t<
-\widehat{\operatorname{VaR}}_{t\mid t-1}(p)\}$
denote the VaR violation indicator, where VaR is reported as a
positive loss. Under the null hypothesis,
the sequence $\{I_t(p)\}$ is independently and identically
distributed as $\operatorname{Bernoulli}(p)$. Thus, the test jointly
examines whether the empirical violation frequency is correctly
calibrated and whether VaR violations are serially independent.

We also compare the VaR forecasts using the quantile loss function
considered by \citet{francq2020virtual}. For an out-of-sample
evaluation period containing $N$ observations, the average loss is
defined as
\begin{equation*}
\operatorname{QLoss}(p)
=
\frac{1}{N}
\sum_{t=T-N+1}^T
\left[
r_t+
\widehat{\operatorname{VaR}}_{t\mid t-1}(p)
\right]
\left[
p-
\mathbb{I}
\left\{
r_t<
-\widehat{\operatorname{VaR}}_{t\mid t-1}(p)
\right\}
\right].
\end{equation*}
Lower values of $\operatorname{QLoss}(p)$ indicate more accurate
VaR forecasts.

\setlength{\tabcolsep}{14pt}

\begin{table}[htbp]
\centering
\caption{Evaluation of one-step-ahead VaR forecasts for FTSE 100 returns.}
\label{table:FTSE100-VaR}
\begin{tabular}{clcc}
\toprule
Tail probability & Model & $p_{\mathrm{CC}}$ & QLoss \\
\midrule

\multirow{3}{*}{$p=1\%$}
& SRT-GARCH & 0.9091 & \textbf{0.0224} \\
& RT-GARCH  & 0.9091 & 0.0225 \\
& S-GARCH   & 0.2248 & 0.0230 \\
\midrule

\multirow{3}{*}{$p=5\%$}
& SRT-GARCH & 0.8719 & \textbf{0.0834} \\
& RT-GARCH  & 0.8719 & 0.0848 \\
& S-GARCH   & 0.8358 & 0.0836 \\
\midrule

\multirow{3}{*}{$p=10\%$}
& SRT-GARCH & 0.6140 & \textbf{0.1376} \\
& RT-GARCH  & 0.1528 & 0.1396 \\
& S-GARCH   & 0.3617 & 0.1378 \\
\bottomrule
\end{tabular}
\end{table}

Table~\ref{table:FTSE100-VaR} reports the 1\%, 5\%, and 10\% tail probabilities together with the corresponding quantile losses. 
 For all three models and at all three tail probabilities, the conditional coverage null is not rejected at conventional significance levels.
SRT-GARCH achieves the smallest quantile loss at each tail probability. The differences in quantile loss are modest, but the results show that the SRT-GARCH VaR forecasts are at least as well calibrated as those of the benchmark models while delivering slightly lower tail-loss values across all three probability levels.

\subsection{U.S. Natural Gas Futures}
 
Natural gas futures provide a second application in a market exposed to changes in weather, inventories, and supply--demand fundamentals. We use $T=3000$ daily returns from May~14, 2013, to October~4, 2024. The \citet{xiao2007testing} statistic is $C_T=4.7216$, again rejecting covariance stationarity at the 1\% level and motivating a time-varying long-run volatility component and the semiparametric specification.

Before turning to the forecasting exercise, we estimate the models using the first $2350$ observations from May 14, 2013, to April 21, 2022. 
Algorithm~\ref{alg:bandwidth-selection} selects bandwidths of \(0.0843\) and \(0.0734\) for SRT-GARCH and S-GARCH, respectively,
and Table~\ref{tab:gas-estimates} reports the estimation results.
The estimate $\widehat\varphi=0.0345$ is positive and consistent with contemporaneous volatility feedback. 
The conventional $t-$statistics based on the asymptotic and bootstrap standard errors are  $4.0588$ and $4.3125$, respectively.

 \begin{table}[htbp]
\centering
\caption{SRT-GARCH(1,1) parameter estimates for U.S. Natural Gas Futures. }
\label{tab:gas-estimates}
\begin{tabular}{lccc}
\toprule
& $\alpha$ & $\beta$ & $\varphi$\\
\midrule
Estimate
& 0.0502 & 0.9090 & 0.0345\\
ASD
& 0.0141 & 0.0194 & 0.0085 \\
BSD
& 0.0119 & 0.0105 & 0.0080 \\
\bottomrule
\end{tabular}
\end{table}

\begin{figure}[H]
 \centering
 \includegraphics[width=.72\linewidth]{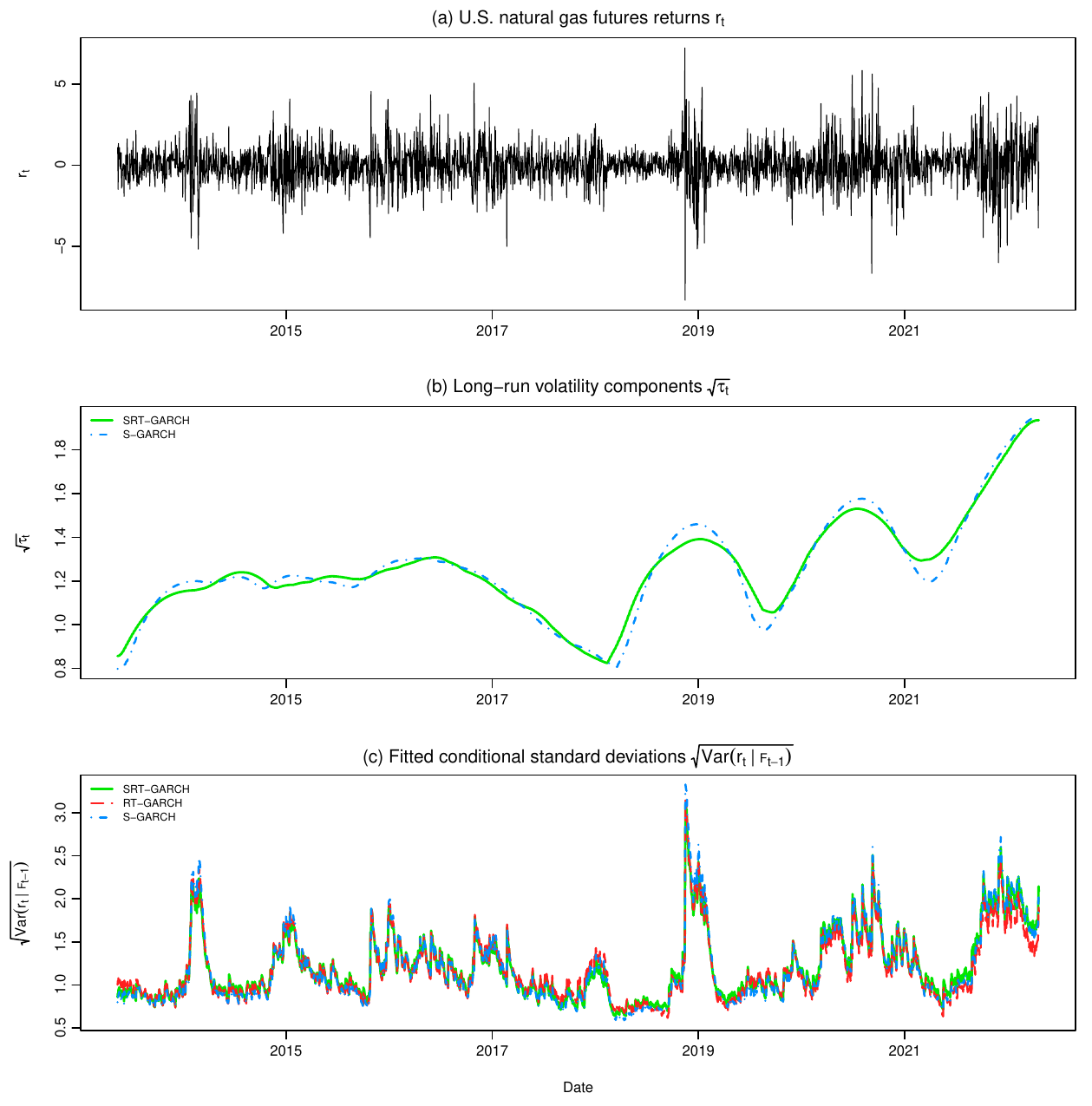}
 \caption{U.S. natural gas futures returns and estimated volatility components over the period from May 14, 2013, to April 21, 2022.}
 \label{fig:US-natural-gas-timeplot}
 \end{figure}

Figure~\ref{fig:US-natural-gas-timeplot} plots the return series and the estimated volatility components. In Panel~(b), the nonparametric scale $\sqrt{\widehat{\tau}_t}$ captures the sharp structural rise in  energy market volatility around 2021--2022. In Panel~(c), the fitted conditional standard deviations diverge markedly during extreme return episodes: the SRT-GARCH specification responds immediately to large contemporaneous shocks while adapting smoothly to the elevated long-run volatility regime.

\begin{figure}[H]
 \centering
 \includegraphics[width=0.68\linewidth]{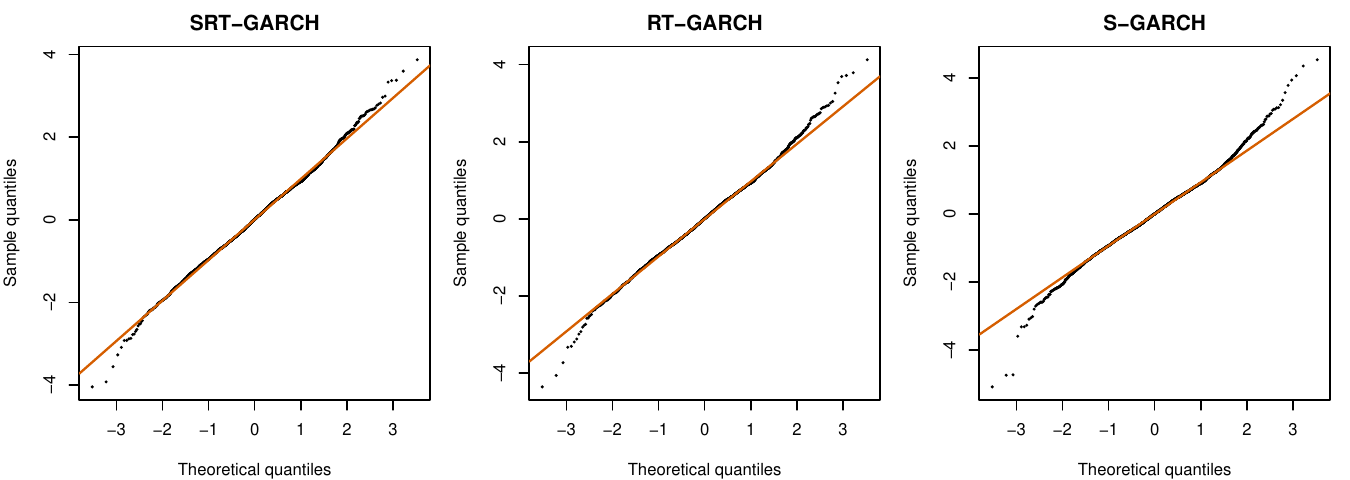}
 \caption{Normal QQ plots of U.S. natural gas futures
 standardized residuals.}
 \label{fig:the US natural gas futures return resi QQ}
 \end{figure}

The residual diagnostics provide further evidence on the adequacy of the alternative short-run specifications.
Figure~\ref{fig:the US natural gas futures return resi QQ} shows that the SRT-GARCH standardized residuals exhibit substantially closer agreement with the normal reference line than those of the benchmark models.

Using the same rolling-window design as before 
with a fixed estimation-window length of $W=2350$ and an out-of-sample evaluation period of $N=650$,
we next compare multi-step-ahead volatility forecasts across horizons $k\in\{1,5,10,15\}$  using MSE and QLIKE loss.

Table~\ref{table:forcas_gas} shows that SRT-GARCH yields the smallest sample-average MSE and QLIKE loss at every forecast horizon. The pairwise DM--HLN tests confirm that SRT-GARCH significantly improves upon RT-GARCH under QLIKE at the 5\% level for $k=1$ and $k=10$ ($p_{\mathrm{DM}}=0.0296$ and $0.0415$) and at the 10\% level for $k=5$ and $k=15$ ($p_{\mathrm{DM}}=0.0575$ and $0.0641$). At the same time, SRT-GARCH improves over S-GARCH under MSE at the 5\% level for $k=1$ ($p_{\mathrm{DM}}=0.0234$) and at the 10\% level for $k=5$ and $k=15$ ($p_{\mathrm{DM}}=0.0595$ and $0.0735$). Thus, in this example, both the nonparametric long-run scale and the real-time short-run feedback contribute meaningfully to out-of-sample forecast accuracy.

\setlength{\tabcolsep}{4pt}

\begin{table}[htbp]
\centering
\caption{Out-of-sample volatility forecast losses for U.S. natural gas futures.}
\label{table:forcas_gas}
\begin{tabular}{lcccccccc}
\toprule
& \multicolumn{2}{c}{1-step}
& \multicolumn{2}{c}{5-step}
& \multicolumn{2}{c}{10-step}
& \multicolumn{2}{c}{15-step}\\
\cmidrule(lr){2-3}\cmidrule(lr){4-5}
\cmidrule(lr){6-7}\cmidrule(lr){8-9}
Model & MSE & QLIKE & MSE & QLIKE
      & MSE & QLIKE & MSE & QLIKE\\
\midrule

SRT-GARCH
& \textbf{85.6325} & \textbf{2.4721}
& \textbf{85.9291} & \textbf{2.4623}
& \textbf{86.6901} & \textbf{2.4551}
& \textbf{85.9769} & \textbf{2.4526} \\
\addlinespace[2pt]

RT-GARCH
& 86.3373 & 2.5480
& 86.8783 & 2.5446
& 87.8039 & 2.5482
& 87.2140 & 2.5647 \\
\multicolumn{1}{r}{\footnotesize $p_{\mathrm{DM}}$}
& {\footnotesize $(0.1258)$}
& {\footnotesize $(0.0296^{*})$}
& {\footnotesize $(0.1219)$}
& {\footnotesize $(0.0575^{*})$}
& {\footnotesize $(0.1605)$}
& {\footnotesize $(0.0415^{*})$}
& {\footnotesize $(0.2198)$}
& {\footnotesize $(0.0641^{*})$} \\
\addlinespace[2pt]

S-GARCH
& 86.2642 & 2.4906
& 86.6905 & 2.4832
& 87.4232 & 2.4701
& 86.8122 & 2.4790 \\
\multicolumn{1}{r}{\footnotesize $p_{\mathrm{DM}}$}
& {\footnotesize $(0.0234^{*})$}
& {\footnotesize $(0.1368)$}
& {\footnotesize $(0.0595^{*})$}
& {\footnotesize $(0.1680)$}
& {\footnotesize $(0.1624)$}
& {\footnotesize $(0.2865)$}
& {\footnotesize $(0.0735^{*})$}
& {\footnotesize $(0.2264)$} \\

\bottomrule
\end{tabular}

\vspace{2pt}
\parbox{0.97\linewidth}{\footnotesize
\textit{Notes:}
Lower values indicate better forecasting performance, and bold entries denote the smallest loss at each horizon. Values in parentheses are DM--HLN $p$-values, denoted by $p_{\mathrm{DM}}$, for testing equal unconditional predictive accuracy between SRT-GARCH and the model in the corresponding row. An asterisk indicates $p_{\mathrm{DM}}<0.10$.
}
\end{table}

Finally, we evaluate one-step-ahead VaR forecasts at the $p=1\%$, $5\%$, and $10\%$ left-tail probability levels. Table~\ref{table:gas-VaR} reports the $p$-values from the Christoffersen conditional coverage test together with the corresponding quantile losses.

As reported in Table~\ref{table:gas-VaR}, none of the three specifications is rejected by the Christoffersen conditional coverage test at any of the three tail probability levels, with SRT-GARCH exhibiting the highest $p$-value ($0.9091$) at the extreme $1\%$ tail. In terms of quantile loss, all three models perform nearly identically at $p=1\%$ ($0.0666$--$0.0670$), while SRT-GARCH achieves the lowest quantile losses at both the $5\%$ and $10\%$ levels ($0.2138$ and $0.3446$).

 \begin{table}[H]
 \centering
 \setlength{\tabcolsep}{14pt}
 \renewcommand{\arraystretch}{1.08}
 \caption{Evaluation of one-step-ahead VaR forecasts for U.S. natural gas futures.}
 \label{table:gas-VaR}
 \begin{tabular}{clcc}
 \toprule
 Tail probability & Model & $p_{\mathrm{CC}}$ & QLoss\\
 \midrule

 \multirow{3}{*}{$p=1\%$}
 & SRT-GARCH & 0.9091 & 0.0670\\
 & RT-GARCH & 0.3776 & 0.0667\\
 & S-GARCH & 0.7690 & \textbf{0.0666}\\
 \midrule

 \multirow{3}{*}{$p=5\%$}
 & SRT-GARCH & 0.2329 & \textbf{0.2138}\\
 & RT-GARCH & 0.3357 & 0.2144\\
 & S-GARCH & 0.5577 & 0.2150\\
 \midrule

 \multirow{3}{*}{$p=10\%$}
 & SRT-GARCH & 0.1824 & \textbf{0.3446}\\
 & RT-GARCH & 0.8364 & 0.3469\\
 & S-GARCH & 0.1824 & 0.3481\\
 \bottomrule
 \end{tabular}

 \vspace{1mm}
 \begin{minipage}{0.94\textwidth}
 \footnotesize
 \textit{Notes:}
 $p_{\mathrm{CC}}$ denotes the $p$-value of the Christoffersen conditional coverage test. Bold entries indicate the smallest quantile loss at each tail probability.
 \end{minipage}
 \end{table}


\section{Conclusion}
\label{sec:conclusion}
This paper develops a semiparametric real-time GARCH model that combines a smooth deterministic long-run volatility component with stationary short-run dynamics that respond to the contemporaneous innovation. The model nests both the semiparametric GARCH and real-time GARCH specifications as special cases.

We propose a two-step estimation procedure in which the long-run component is estimated nonparametrically by kernel smoothing and the short-run parameters are subsequently estimated by conditional Gaussian maximum likelihood. Under the stated conditions, we establish consistency and asymptotic normality of the resulting estimator. A key feature of the asymptotic theory is that estimation of the long-run component has a non-vanishing first-order effect on the short-run parameter estimates. By combining the kernel projection with a martingale approximation, we show that this contribution enters the asymptotic covariance matrix through a positive semidefinite rank-one adjustment to the oracle covariance. Under Gaussian innovations, the resulting covariance matrix admits a tuning-free plug-in estimator that requires neither HAC estimation nor lag-truncation parameters. The dependence conditions needed for these results are derived from model-specific geometric moment contraction rather than imposed through generic mixing assumptions.

Monte Carlo experiments show finite-sample behavior broadly consistent with the asymptotic theory. The empirical applications to FTSE~100 returns and U.S. natural gas futures indicate that allowing for a time-varying long-run component improves substantially on the RT-GARCH benchmark, while the additional real-time feedback mechanism provides more pronounced forecasting gains in the natural gas market than in the FTSE application. The VaR results are also broadly supportive of the proposed specification.

Two extensions are particularly worth pursuing. First, it would be useful to extend the boundary inference developed for the parametric GARCH/RT-GARCH model to the present two-step semiparametric setting, where nonparametric estimation of the long-run component contributes to the first-order distribution of the short-run parameter estimator. Second, extending the exact-likelihood analysis beyond Gaussian innovations would broaden the applicability of the model and allow more flexible treatment of heavy-tailed and asymmetric returns.

\section*{Data and Code Availability}

The data and code required to reproduce the simulation and empirical results will be made available to the editors for replicability review. Subject to any licensing restrictions imposed by the data providers, the replication files will be deposited in a public repository upon acceptance.

\bibliography{refer_JFEC_refs}

\newpage
\appendix
\section{
{ Proofs}}\label{app:proofs}
\renewcommand{\theequation}{A.\arabic{equation}}
\renewcommand{\theHequation}{A.\arabic{equation}}
\setcounter{equation}{0}

Throughout the proofs, let
$\bar\beta=\sup_{\theta\in\Theta}\beta<1$ and
$\underline\omega=\inf_{\theta\in\Theta}\omega(\theta)>0$,
which follow from Assumption~\ref{Asm1}. The constant $C$ below may change from line to line and does not depend on
$t$, $T$, or $\theta\in\Theta$.

Write
\[
b_{t-1}(\theta)=\omega(\theta)+\alpha u_{t-1}^2+\beta h_{t-1}(\theta),
\]
and define, for $b>0$, {$v\ge0$}, and $\varphi\ge0$,
\begin{equation}\label{eq:g-and-f-def}
   g(b,{v};\varphi)
   :=\frac{b+\sqrt{b^2+4\varphi {v}}}{2},
\qquad
{f(v,h,\varphi):=\frac{v}{2h}+\log(h^2+\varphi v)-\frac32\log h.} 
\end{equation}
Thus,
$h_t(\theta)=g\bigl(b_{t-1}(\theta),u_t^2;\varphi\bigr)$ {and $\ell_t(\theta)=f\bigl(u_t^2,h_t(\theta),\varphi\bigr)$.}
Throughout the proof, we will use the fact that 
\begin{equation}\label{eq:H-derivative-b}
\frac{\partial g(b,{v};\varphi)}{\partial b}
 =\frac12\left(1+\frac{b}{\sqrt{b^2+4\varphi {v}}}\right)\in[1/2,1],
\end{equation}
and, whenever $b_1,b_2\ge \underline\omega>0$ {and $v_1,v_2\ge0$},
\begin{equation}\label{eq:H-lipschitz}
\left|g(b_1,{v_1};\varphi)-g(b_2,{v_2};\varphi)\right|
 \le |b_1-b_2|+\frac{\varphi}{\underline\omega}|{v_1-v_2}|.
\end{equation}

We first outline some useful results on the derivatives. More generally, direct differentiation on  $\{(b,v,\varphi):b\ge\underline\omega>0,\,v\ge0,\,\theta\in\Theta\}$ yields that 
\begin{align}
\left|\frac{\partial g}{\partial v}\right|
&=\frac{\varphi}{\sqrt{b^2+4\varphi v}}\le\frac{\varphi}{\underline\omega}\le C,\qquad
\left|\frac{\partial g}{\partial\varphi}\right|
=\frac{v}{\sqrt{b^2+4\varphi v}}\le\frac{v}{\underline\omega}\le Cv,
\label{eq:g-first-bounds}\\
\left|\frac{\partial^2g}{\partial b^2}\right|
&=\frac{2\varphi v}{(b^2+4\varphi v)^{3/2}}\le\frac{1}{2\underline\omega}\le C,\qquad
\left|\frac{\partial^2g}{\partial b\,\partial v}\right|
=\frac{\varphi b}{(b^2+4\varphi v)^{3/2}}\le\frac{\varphi}{\underline\omega^2}\le C,
\notag\\
\left|\frac{\partial^2g}{\partial v^2}\right|
&=\frac{2\varphi^2}{(b^2+4\varphi v)^{3/2}}\le\frac{2\varphi^2}{\underline\omega^3}\le C,\qquad
\left|\frac{\partial^2g}{\partial v\,\partial\varphi}\right|
=\frac{b^2+2\varphi v}{(b^2+4\varphi v)^{3/2}}\le\frac{1}{\underline\omega}\le C,
\label{eq:g-second-bounds}\\
\left|\frac{\partial^2g}{\partial b\,\partial\varphi}\right|
&=\frac{bv}{(b^2+4\varphi v)^{3/2}}\le\frac{v}{\underline\omega^2}\le Cv,\qquad
\left|\frac{\partial^2g}{\partial\varphi^2}\right|
=\frac{2v^2}{(b^2+4\varphi v)^{3/2}}\le\frac{2v^2}{\underline\omega^3}\le Cv^2,
\notag
\end{align}
where we use $2\varphi v\le (b^2+4\varphi v)/2$ and $\sqrt{b^2+4\varphi v}\ge\underline\omega$. Similarly,  the third-order partial derivatives of $g$ satisfy
\begin{equation}\label{eq:g-third-bounds}
\sum_{i+j=3}\left|\frac{\partial^3g}{\partial b^i\,\partial v^j}\right|
+\left|\frac{\partial^3g}{\partial b\,\partial v\,\partial\varphi}\right|
+\left|\frac{\partial^3g}{\partial v^2\,\partial\varphi}\right|\le C,\qquad
\left|\frac{\partial^3g}{\partial b^2\,\partial\varphi}\right|\le C(1+v),\qquad
\left|\frac{\partial^3g}{\partial b\,\partial\varphi^2}\right|\le Cv^2.
\end{equation}
Consequently, by the mean value theorem, for any $b_1,b_2\ge\underline\omega$, $v_1,v_2\ge0$, and $\theta\in\Theta$, the partial derivatives of $g$ satisfy the explicit Lipschitz bounds
\begin{equation}\label{eq:g-deriv-lipschitz}
\begin{aligned}
&\left|\frac{\partial g^{(1)}}{\partial b}-\frac{\partial g^{(2)}}{\partial b}\right|
+\left|\frac{\partial g^{(1)}}{\partial v}-\frac{\partial g^{(2)}}{\partial v}\right|
+\left|\frac{\partial^2g^{(1)}}{\partial b^2}-\frac{\partial^2g^{(2)}}{\partial b^2}\right|
\le C\bigl(|b_1-b_2|+|v_1-v_2|\bigr),\\
&\left|\frac{\partial g^{(1)}}{\partial\varphi}-\frac{\partial g^{(2)}}{\partial\varphi}\right|
+\left|\frac{\partial^2g^{(1)}}{\partial b\,\partial\varphi}-\frac{\partial^2g^{(2)}}{\partial b\,\partial\varphi}\right|
\le C(1+v_1+v_2)\bigl(|b_1-b_2|+|v_1-v_2|\bigr),\\
&\left|\frac{\partial^2g^{(1)}}{\partial\varphi^2}-\frac{\partial^2g^{(2)}}{\partial\varphi^2}\right|
\le C(1+v_1+v_2)^2\bigl(|b_1-b_2|+|v_1-v_2|\bigr),
\end{aligned}
\end{equation}
where $g^{(m)}:=g(b_m,v_m;\varphi)$ ($m=1,2$).

Next, for $f(v,h,\varphi)$ defined in \eqref{eq:g-and-f-def}, direct differentiation on $\{(v,h,\varphi):v\ge0,\,h\ge\underline\omega>0,\,\theta\in\Theta\}$ gives
\begin{equation}\label{eq:f-explicit-partials}
\begin{aligned}
\frac{\partial f}{\partial h}&=-\frac{v}{2h^2}+\frac{2h}{h^2+\varphi v}-\frac{3}{2h},\qquad
\frac{\partial f}{\partial\varphi}=\frac{v}{h^2+\varphi v},\qquad
\frac{\partial f}{\partial v}=\frac{1}{2h}+\frac{\varphi}{h^2+\varphi v},\\
\frac{\partial^2 f}{\partial h^2}&=\frac{v}{h^3}+\frac{2(\varphi v-h^2)}{(h^2+\varphi v)^2}+\frac{3}{2h^2},\qquad
\frac{\partial^2 f}{\partial h\,\partial\varphi}=-\frac{2hv}{(h^2+\varphi v)^2},\qquad
\frac{\partial^2 f}{\partial\varphi^2}=-\frac{v^2}{(h^2+\varphi v)^2},\\
\frac{\partial^2 f}{\partial h\,\partial v}&=-\frac{1}{2h^2}-\frac{2h\varphi}{(h^2+\varphi v)^2},\qquad
\frac{\partial^2 f}{\partial\varphi\,\partial v}=\frac{h^2}{(h^2+\varphi v)^2}.
\end{aligned}
\end{equation}
Since $h^2+\varphi v\ge h^2\ge\underline\omega^2>0$, the partial derivatives of $f$ up to order three satisfy that uniformly on $\Theta$
\begin{equation}\label{eq:f-global-bounds}
\begin{aligned}
\sum_{j=1}^{3}\left|\frac{\partial^j f}{\partial h^j}\right|
&\le C(1+v),\qquad
\left|\frac{\partial f}{\partial\varphi}\right|
+\left|\frac{\partial^2 f}{\partial h\,\partial\varphi}\right|
+\left|\frac{\partial^3 f}{\partial h^2\,\partial\varphi}\right|
\le Cv,\qquad
\left|\frac{\partial^2 f}{\partial\varphi^2}\right|
+\left|\frac{\partial^3 f}{\partial h\,\partial\varphi^2}\right|
\le Cv^2,\\
\left|\frac{\partial f}{\partial v}\right|
+\left|\frac{\partial^2 f}{\partial h\,\partial v}\right|
&+\left|\frac{\partial^2 f}{\partial\varphi\,\partial v}\right|
+\left|\frac{\partial^3 f}{\partial h^2\,\partial v}\right|
+\left|\frac{\partial^3 f}{\partial h\,\partial\varphi\,\partial v}\right|
\le C,\qquad
\left|\frac{\partial^2 f}{\partial\varphi^2\,\partial v}\right|\le Cv.
\end{aligned}
\end{equation}
By the mean value theorem and \eqref{eq:f-global-bounds}, for any $v_1,v_2\ge0$, $h_1,h_2\ge\underline\omega$, and $\theta\in\Theta$, we have the explicit Lipschitz inequalities
\begin{equation}\label{eq:f-deriv-lipschitz}
\begin{aligned}
&\bigl|f^{(1)}-f^{(2)}\bigr|
+\left|\frac{\partial f^{(1)}}{\partial h}-\frac{\partial f^{(2)}}{\partial h}\right|
+\left|\frac{\partial f^{(1)}}{\partial\varphi}-\frac{\partial f^{(2)}}{\partial\varphi}\right|
\le C|v_1-v_2|+C(1+v_1+v_2)|h_1-h_2|,\\
&\left|\frac{\partial^2 f^{(1)}}{\partial h^2}-\frac{\partial^2 f^{(2)}}{\partial h^2}\right|
+\left|\frac{\partial^2 f^{(1)}}{\partial h\,\partial\varphi}-\frac{\partial^2 f^{(2)}}{\partial h\,\partial\varphi}\right|
\le C|v_1-v_2|+C(1+v_1+v_2)|h_1-h_2|,\\
&\left|\frac{\partial^2 f^{(1)}}{\partial\varphi^2}-\frac{\partial^2 f^{(2)}}{\partial\varphi^2}\right|
\le C(1+v_1+v_2)|v_1-v_2|+C(1+v_1+v_2)^2|h_1-h_2|,
\end{aligned}
\end{equation}
where $f^{(m)}:=f(v_m,h_m,\varphi)$ ($m=1,2$).

Furthermore, on any compact subset $\mathcal N$ of the interior of $\Theta$, we have $\underline\varphi:=\inf_{\theta\in\mathcal N}\varphi>0$. At any state $h=g(b,v;\varphi)$ with $b\ge\underline\omega$, the quadratic relation $h^2=bh+\varphi v\ge\max(\underline\omega^2,\underline\varphi v)$ implies $v/h^2\le 1/\underline\varphi$ and $h^2+\varphi v\ge\max(h^2,\underline\varphi v)$. Consequently, at any  state on $\mathcal N$, all partial derivatives of $f$ up to order three satisfy 
\begin{equation}\label{eq:f-local-bounds}
\sum_{1\le j+k\le 3}\left|\frac{\partial^{j+k}f(v,h,\varphi)}{\partial h^j\,\partial\varphi^k}\right|
+\sum_{i=1}^{2}\sum_{0\le j+k\le 3-i}v^{i-1}\left|\frac{\partial^{i+j+k}f(v,h,\varphi)}{\partial v^i\,\partial h^j\,\partial\varphi^k}\right|
\le C.
\end{equation}


\begin{proof}[Proof of Proposition~\ref{prop:stationary-short-run}]
Since
$h_t=b_{t-1}+\varphi\epsilon_t^2$ and
$u_t^2=b_{t-1}\epsilon_t^2+\varphi\epsilon_t^4$, we have that 
\[
b_t=C_t+A_tb_{t-1},
\]
where $
A_t=\alpha\epsilon_t^2+\beta, $ and $
C_t=\omega+\varphi
\left(\alpha\epsilon_t^4+\beta\epsilon_t^2\right)$, with $(A_t,C_t)$ being independent of $b_{t-1}$.

Note that $\{(A_t,C_t)\}$ are i.i.d.,
$E(A_t)=\alpha+\beta<1$, and $E(C_t)<\infty$ by that 
$E(\epsilon_t^4)<\infty$.By iteration, we obtain that 
\[
b_t=C_t+\sum_{k=1}^{\infty}
\left(\prod_{j=0}^{k-1}A_{t-j}\right)C_{t-k},
\]
which converges in $L^1$ and almost surely because $\sum_{k\ge0}E(C_t)\{E(A_t)\}^k<\infty$.  The rest of the proof follows directly from Theorem 1 in \cite{smetanina2021asymptotic}. 
\end{proof}

\begin{proof}[Proof of Proposition~\ref{prop:normalization-proof}]
Let $m_u=E(u_t^2)$ and $m_h=E(h_t)$.
Since $b_{t-1}$ is $\mathcal F_{t-1}$-measurable and
$\epsilon_t$ is independent of $\mathcal F_{t-1}$, taking expectations
in $h_t=b_{t-1}+\varphi\epsilon_t^2$ and
$u_t^2=b_{t-1}\epsilon_t^2+\varphi\epsilon_t^4$ gives
\begin{align}
m_h&=\omega+\alpha m_u+\beta m_h+\varphi,\label{eq:Eh-step}\\
m_u&=\omega+\alpha m_u+\beta m_h+\varphi\mu_4.\label{eq:Eu2-step}
\end{align}
Hence, $m_u-m_h=\varphi(\mu_4-1)$.
Substituting this  and \eqref{omega-normalization} into
\eqref{eq:Eh-step} gives $(1-\alpha-\beta)(m_u-1)=0$.
Thus $m_u=1$ and $m_h=1-\varphi(\mu_4-1)$.

For fixed $b_{t-1}>0$, the mapping 
$e\mapsto e\sqrt{b_{t-1}+\varphi e^2}$ is odd. Using symmetry and integrability, we then obtain that 
$E(u_t\mid\mathcal F_{t-1})=0$.
For $j\ge1$, $u_{t-j}$ is $\mathcal F_{t-1}$-measurable and
$E|u_tu_{t-j}|<\infty$ by Cauchy-Schwarz inequality.  Hence
$E(u_tu_{t-j})=E\{u_{t-j}E(u_t\mid\mathcal F_{t-1})\}=0$, the result follows.
\end{proof}

\begin{proof}[Proof of Proposition~\ref{pro_moment}]
Conditional on $\mathcal F_{t-1}$, $b_{t-1}$ is fixed and
\[
r_t^2
=\tau_t\left(b_{t-1}\epsilon_t^2
+\varphi_0\epsilon_t^4\right).
\]
Using the independence of $\epsilon_t$ from $\mathcal F_{t-1}$, we have 
\[
E(r_t^2\mid\mathcal F_{t-1})
=\tau_t(b_{t-1}+\varphi_0\mu_4).
\]
Similarly,
\[
r_t^4
=\tau_t^2\left(
b_{t-1}^2\epsilon_t^4
+2\varphi_0b_{t-1}\epsilon_t^6
+\varphi_0^2\epsilon_t^8
\right).
\]
 Dividing the
latter expression by the square of the conditional second moment
proves \eqref{standardized-fourth-moment}.
\end{proof}


\begin{proof}[Proof of Proposition~\ref{prop:rt-gmc}]
Recall in the proof of Proposition \ref{prop:stationary-short-run}, we have $
b_t=C_t+A_tb_{t-1}, $ $
A_t=\alpha\epsilon_t^2+\beta, $ $
C_t=\omega+\varphi\{\alpha\epsilon_t^4+\beta\epsilon_t^2\},
$
where $\{(A_t,C_t)\}$ is i.i.d.\  and  independent of
$b_{t-1}$. Moreover,
\[
u_t^2=b_{t-1}\epsilon_t^2+\varphi\epsilon_t^4.
\]
Since $b_{t-1}\ge\omega>0$ is independent of $\epsilon_t$, the
moment $E|u_t|^{2p}<\infty$ implies $E(b_t^p)<\infty$ and
$E|\epsilon_t|^{2p}<\infty$. Also, $b_t\ge C_t\ge\omega$
gives $0<E(C_t^p)<\infty$. By the fact that
$(a+c)^p\ge a^p+c^p$ for $a,c\ge0$, we then obtain that 
\[
E(b_t^p)\ge E(C_t^p)+E(A_t^p)E(b_{t-1}^p).
\]
Consequently, $E(A_t^p)<1$ using the stationarity of $b_t$ and that $0<E(C_t^p)<\infty$. Without loss of generality, we  choose $\rho_p\in((E(A_t^p))^{1/p},1)$.

Let $b_t^\circ$ be the coupled stationary solution by Definition~\ref{def:gmc}, then iterating the recursion
gives
\[
\|b_t-b_t^\circ\|_p
=
\{E(A_t^p)\}^{t/p}\|b_0-b_0^\circ\|_p.
\]
For $t\ge1$, we have  $h_t-h_t^\circ=b_{t-1}-b_{t-1}^\circ$ and
$z_t-z_t^\circ=\epsilon_t^2(b_{t-1}-b_{t-1}^\circ)$.  
 
 Therefore, replacing only $\epsilon_0$ changes the
initial difference in $b_0$ and $b_0^\circ$ but leaves the same product recursion. This proves the result. 
\end{proof}

\begin{proof}[Proof of Theorem~\ref{thm:tau-asymptotic}]
Applying Proposition~\ref{prop:rt-gmc} with $p=2$ and Lemma~\ref{lem:contractive-master}(ii) to the stationary sequence $\{z_t\}$ yields that $|\operatorname{Cov}(z_0,z_j)|\le C\rho^{|j|}$ for all $j\in\mathbb Z$. Hence, the absolute convergence of $\Omega_z=\sum_{j\in\mathbb Z}\operatorname{Cov}(z_0,z_j)$, and the functional central limit theorem \citet{wu2005nonlinear} is satisfied, and therefore  
\begin{equation}\label{eq:FCLT}
    \frac{1}{\sqrt n}
\sum_{t=1}^{\lfloor nr\rfloor}z_t
\Rightarrow
\Omega_z^{1/2}B(r),
\qquad 0\leq r\leq1.
\end{equation}

For fixed $x\in(0,1)$, the kernel uses only a consecutive block
of about $2Th$ observations. Let $M=Th$ and
$l_T=\lfloor Tx-M\rfloor$. By stationarity, the partial sums
of $z_{l_T+1},z_{l_T+2},\ldots$ have the same distribution as
those starting at $t=1$. \eqref{eq:FCLT}
therefore gives
$$M^{-1/2}\sum_{j=1}^{\lfloor Mr\rfloor}z_{l_T+j}
\Rightarrow\Omega_z^{1/2}B(r), \quad 0\le r\le2.$$

 Since $K$ is Lipschitz and has bounded variation,
summation by parts gives the limit
$\Omega_z^{1/2}\int_0^2K(1-r)\,dB(r)$.
Integer endpoints change the weighted sum by $o_p(1)$.
Consequently,
\[
\frac{1}{\sqrt{Th}}\sum_{s=1}^T
K\!\left(\frac{x-s/T}{h}\right)z_s
\to_d
N\!\left(0,\,
\Omega_z\int_{-1}^1K^2(v)\,dv\right).
\]

For a fixed interior point $x$, write
\[
\widehat\tau(x)-E\{\widehat\tau(x)\}
=
\frac{1}{Th}\sum_{s=1}^T
K\!\left(\frac{x-s/T}{h}\right)\tau(s/T)z_s .
\]
A second-order Taylor expansion of $\tau(s/T)$ around $x$, symmetry
of $K$, and a Riemann-sum approximation give
\[
E\{\widehat\tau(x)\}-\tau(x)
=
\frac{h^2C_2}{2}\tau''(x)+o(h^2)+O\!\left((Th)^{-1}\right).
\]
Because $Th^5=O(1)$ and $Th\to\infty$, this implies that 
$$
\sqrt{Th}\left(E\{\widehat\tau(x)\}-\tau(x)\right)
=
\sqrt{Th}\frac{h^2C_2}{2}\tau''(x)+o(1).
$$
Furthermore, using the boundedness of $\tau$, and Assumption \ref{Asm3}, we can see that 
\[
\left\|\frac1{\sqrt{Th}}\sum_{s=1}^T
K\!\left(\frac{x-s/T}{h}\right)
\{\tau(s/T)-\tau(x)\}z_s\right\|_2^2
\le C h^2\sum_{j\in\mathbb Z}|\operatorname{Cov}(z_0,z_j)|
=O(h^2).
\]
Here the weights vanish outside $O(Th)$ indices and
$|\tau(s/T)-\tau(x)|\le Ch$ on that block.
The kernel central limit theorem therefore yields
\[
\frac{1}{\sqrt{Th}}\sum_{s=1}^T
K\!\left(\frac{x-s/T}{h}\right)\tau(s/T)z_s
\ \to_d\
N\!\left(
0,\,
\tau^2(x)\int_{-1}^1K^2(v)\,dv\,\Omega_z
\right).
\]
The result thus follows.
\end{proof}



\begin{proof}[Proof of Proposition~\ref{pro:conditional-density}]
Write $b=b_{t-1}>0$ and
$r=\sqrt{\tau_t}\,e\sqrt{b+\varphi_0e^2}$.
This transformation is strictly increasing, with derivative
$\sqrt{\tau_t}(b+2\varphi_0e^2)/\sqrt{b+\varphi_0e^2}$.
Its inverse satisfies $e(r)=r/\sqrt{\tau_t h_t(r)}$, where
$h_t(r)=\{b+\sqrt{b^2+4\varphi_0r^2/\tau_t}\}/2$.
The change-of-variables  therefore gives
\begin{equation}\label{eq:density-positive-phi}
f_t(r\mid\mathcal F_{t-1})
=\frac{\sqrt{h_t(r)}}
{\sqrt{\tau_t}\sqrt{b^2+4\varphi_0r^2/\tau_t}}
f_\epsilon\!\left(\frac{r}{\sqrt{\tau_t h_t(r)}}\right).
\end{equation}
At $\varphi_0=0$ this reduces to
\begin{equation}\label{eq:density-zero-phi}
f_t(r\mid\mathcal F_{t-1})
=\frac1{\sqrt{\tau_t b}}
f_\epsilon\!\left(\frac{r}{\sqrt{\tau_t b}}\right),
\end{equation}
the result follows.
\end{proof}

For $\varphi_0>0$, the inverse transformation above satisfies
\[
e(r)^2
=\frac{-b+\sqrt{b^2+4\varphi_0r^2/\tau_t}}{2\varphi_0}
=\frac{|r|}{\sqrt{\tau_t\varphi_0}}+O(1).
\]
The logarithm of the Jacobian factor in
\eqref{eq:density-positive-phi} is $O(\log|r|)$. Therefore,
under Gaussian innovations,
\[
\lim_{|r|\to\infty}
\frac{\log f_t(r\mid\mathcal F_{t-1})}{|r|}
=-\frac{1}{2\sqrt{\tau_t\varphi_0}},
\qquad \varphi_0>0.
\]
When $\varphi_0=0$, \eqref{eq:density-zero-phi} instead gives
\[
\lim_{|r|\to\infty}
\frac{\log f_t(r\mid\mathcal F_{t-1})}{r^2}
=-\frac{1}{2\tau_tb_{t-1}}.
\]
Thus the conditional tails are exponential-type for
$\varphi_0>0$ and Gaussian-type for $\varphi_0=0$.

\subsection{{Auxiliary Lemmas}}\label{app:auxiliary-lemmas}
We first establish three key lemmas: Lemma~\ref{lem:contractive-master} handles contractive triangular recursions and GMC moment interpolation; Lemma~\ref{lem:process-moment-bounds} establishes the stationary moment envelopes; and Lemma~\ref{lem:master-perturbation} provides explicit perturbation inequalities for the volatility process and the composite log-likelihood with initial values. 
\begin{lemma}
\label{lem:contractive-master}
\begin{enumerate}[(i)]
\item Let $\bar\beta\in(0,1)$ and let $\{x_{0,t},x_{1,t},x_{2,t}\}_{t\ge0}$ be nonnegative random sequences satisfying, for $t\ge1$,
\begin{equation}\label{eq:master-triangular-system}
\begin{aligned}
x_{0,t}&\le\bar\beta x_{0,t-1}+\xi_{0,t},\\
x_{1,t}&\le\bar\beta x_{1,t-1}+B_t(x_{0,t-1}+\xi_{0,t})+\xi_{1,t},\\
x_{2,t}&\le\bar\beta x_{2,t-1}+B_t x_{1,t-1}+C_t(x_{0,t-1}+\xi_{0,t})+\xi_{2,t},
\end{aligned}
\end{equation}
where $B_t,C_t,\xi_{r,t}\ge0$ are random sequences.  Fix any $\rho\in(\bar\beta,1)$.
\begin{enumerate}[(a)]
\item  Suppose $\sup_t\|B_t\|_q<\infty$, $\sup_t\|C_t\|_{q/2}<\infty$, $\|x_{r,0}\|_{q/(r+1)}<\infty$, and $\|\xi_{r,t}\|_{q/(r+1)}\le C_0\rho^t$ for $r=0,1,2$ and $t\ge1$. Then there exists $C<\infty$ depending only on $(\bar\beta,\rho,C_0)$ and the moment bounds of $(B_t,C_t,x_{r,0})$ such that \begin{equation}\label{eq:master-decay-conclusion}
\|x_{0,t}\|_q\le C\rho^t,\qquad
\|x_{1,t}\|_{q/2}\le C\rho^t,\qquad
\|x_{2,t}\|_{q/3}\le C\rho^t,
\qquad t\ge1.
\end{equation}
\item For any stationary sequence $u_t^2\in L^q$ ($q>4$), define the stationary ergodic envelopes
\begin{equation}\label{eq:master-envelopes-UG}
\begin{aligned}
U_t&:=C\sum_{j=0}^{\infty}\rho^j(1+u_{t-j}^2)\in L^q,\qquad
G_{1,t}:=C\sum_{j=0}^{\infty}\rho^j U_{t-j}^2\in L^{q/2},\\
G_{2,t}&:=C\sum_{j=0}^{\infty}\rho^j\{U_{t-j}^3+U_{t-j}G_{1,t-j-1}\}\in L^{q/3}.
\end{aligned}
\end{equation}
If $x_{r,0}=0$, $B_t\le CU_t$, $C_t\le C(U_t^2+G_{1,t-1})$, and for some nonnegative random variable $\Delta_T$, $\xi_{0,t}\le C\Delta_T(1+u_t^2+u_{t-1}^2)$, $\xi_{1,t}\le C\Delta_T U_t^2$, and $\xi_{2,t}\le C\Delta_T(U_t^3+U_tG_{1,t-1})$, then for $t\geq 1$, \begin{equation}\label{eq:master-envelope-conclusion}
x_{0,t}\le C\Delta_T U_t,\qquad
x_{1,t}\le C\Delta_T G_{1,t},\qquad
x_{2,t}\le C\Delta_T G_{2,t}.
\end{equation}
\end{enumerate}
\item Let $X_t=F(\epsilon_t,\epsilon_{t-1},\ldots)$ be a stationary causal process with coupled version $X_t^\circ$ (from Definition~\ref{def:gmc}) and single-innovation coupled version $X_t^*=F(\epsilon_t,\ldots,\epsilon_1,\epsilon_0^\circ,\epsilon_{-1},\ldots)$.
\begin{enumerate}[(a)]
\item  If $\|X_t\|_a<\infty$ and $\|X_t-X_t^\circ\|_r\le C\rho^t$ for some $1<r<a$ and $\rho\in(0,1)$, then for every $p\in[r,a)$, \begin{equation}\label{eq:gmc-interpolation-bound}
\|X_t-X_t^\circ\|_p\le C_p\rho_p^t,\qquad
\|X_t-X_t^*\|_p\le 2\|X_t-X_t^\circ\|_p\le 2C_p\rho_p^t,
\end{equation}
where $\rho_p:=\rho^{\frac{r(a-p)}{p(a-r)}}\in(0,1).$
\item If $E(X_t)=0$ and $\{X_t\}$ is $L^2$ GMC with decay rate $\rho\in(0,1)$, then $\|\operatorname{Cov}(X_0,X_j)\|\le\|X_0\|_2\|X_j-X_j^\circ\|_2\le C\rho^{|j|}$ for all $j\in\mathbb Z$, the series $\Omega_X:=\sum_{j\in\mathbb Z}\operatorname{Cov}(X_0,X_j)$ converges absolutely, $\operatorname{Var}(T^{-1/2}\sum_{t=1}^T X_t)\to\Omega_X$, and $n^{-1/2}\sum_{t=1}^{\lfloor nr\rfloor}X_t\Rightarrow\Omega_X^{1/2}B(r)$ on $[0,1]$.
\end{enumerate}
\end{enumerate}
\end{lemma}

\begin{proof}
For (i), iterating $x_{0,t}\le\bar\beta x_{0,t-1}+\xi_{0,t}$ gives $x_{0,t}\le\bar\beta^t x_{0,0}+\sum_{j=0}^{t-1}\bar\beta^j\xi_{0,t-j}$. 

\noindent Under (a), Minkowski's inequality and   $\bar\beta<\rho$ give $$\|x_{0,t}\|_q\le\bar\beta^t\|x_{0,0}\|_q+C_0\sum_{j=0}^{t-1}\bar\beta^j\rho^{t-j}\le C\rho^t.$$ By H\"older's inequality, $$\|B_t(x_{0,t-1}+\xi_{0,t})+\xi_{1,t}\|_{q/2}\le\|B_t\|_q(\|x_{0,t-1}\|_q+\|\xi_{0,t}\|_q)+\|\xi_{1,t}\|_{q/2}\le C'\rho^t,$$ iterating  $x_{1,t}$ and using the fact that  $\sum_{j=0}^{t-1}(\bar\beta/\rho)^j\le(1-\bar\beta/\rho)^{-1}$ gives $\|x_{1,t}\|_{q/2}\le C\rho^t$. The bound $\|x_{2,t}\|_{q/3}\le C\rho^t$ follows identically using $\|B_tx_{1,t-1}\|_{q/3}\le\|B_t\|_q\|x_{1,t-1}\|_{q/2}$ and $\|C_t(x_{0,t-1}+\xi_{0,t})\|_{q/3}\le\|C_t\|_{q/2}\|x_{0,t-1}+\xi_{0,t}\|_q$.

\noindent Under (b), Minkowski's and H\"older's inequalities give $U_t\in L^q$, $G_{1,t}\in L^{q/2}$, and $G_{2,t}\in L^{q/3}$, and direct substitution of $\bar\beta\le\rho$ and $U_{t-1}\le\rho^{-1}U_t$ into the iterated sums yields $x_{0,t}\le C\Delta_T U_t$, $x_{1,t}\le C\Delta_T G_{1,t}$, and $x_{2,t}\le C\Delta_T G_{2,t}$.

For (ii)(a), let $1/p=\theta/r+(1-\theta)/a$ with $\theta=r(a-p)/\{p(a-r)\}\in(0,1]$. Then, H\"older's inequality gives $\|X_t-X_t^\circ\|_p\le\|X_t-X_t^\circ\|_r^\theta(\|X_t\|_a+\|X_t^\circ\|_a)^{1-\theta}\le C_p\rho^{\theta t}$; see Remark~3.1 of \citet{shao2007asymptotic}. Since $X_t^*$ and $X_t$ share $\epsilon_s$ for $s\ne0$, then $X_t-X_t^*=(X_t-X_t^{\circ})+(X_t^\circ-X_t^*)$ with $\|X_t-X_t^*\|_p\le\|X_t-X_t^{\circ}\|_p+\|X_t^\circ-X_t^*\|_p=\|X_t-X_t^{\circ}\|_p+\|X_{t+1}-X_{t+1}^\circ\|_p\le C_p\rho_p^t+C_p\rho_p^{t+1}$. 





For (ii)(b), independence of $X_j^\circ$ from $X_0$ for $j\ge1$ gives $\|\operatorname{Cov}(X_0,X_j)\|=\|E\{X_0(X_j-X_j^\circ)^\top\}\|\le\|X_0\|_2\|X_j-X_j^\circ\|_2\le C\rho^j$, and stationarity covers $j<0$. Moreover, the physical dependence coefficient of \citet{wu2005nonlinear} satisfies $\delta_2(j)=\|X_j-X_j^*\|_2\le 2C\rho^j$, so Theorems~1(i) and~3(i) of \citet{wu2005nonlinear} yield the FCLT.
\end{proof}

\begin{lemma}
\label{lem:process-moment-bounds}
Under Assumptions~\ref{Asm1} and~\ref{Asm6}, there exist
$C<\infty$ and $\rho\in(\bar\beta,1)$ such that
\begin{equation}
1+u_t^2+\sup_{\theta\in\Theta}
\left\{h_t(\theta)+
\left\|\frac{\partial h_t(\theta)}{\partial\theta}\right\|\right\}
\le U_t:=C\sum_{j=0}^{\infty}\rho^j(1+u_{t-j}^2)\in L^q,
\label{eq:process-level-first-bound}
\end{equation}
and
\begin{equation}
\sup_{\theta\in\Theta}
\left\|\frac{\partial^2h_t(\theta)}{\partial\theta\,\partial\theta^\top}\right\|
\le G_{1,t}:=C\sum_{j=0}^{\infty}\rho^jU_{t-j}^2\in L^{q/2}.
\label{eq:derived-moment-bounds}
\end{equation}
The score function satisfies that 
$\sup_{\theta\in\Theta}\|s_t(\theta)\|\le CU_t^2\in L^{q/2}$.
\end{lemma}

\begin{proof}
Let $e_\alpha=(1,0,0)^\top$, $e_\beta=(0,1,0)^\top$, and
$e_\varphi=(0,0,1)^\top$ be the standard basis vectors.
All derivatives with respect to $\theta$ hold the observations
$\{u_t^2\}$ fixed. Since
$b_{t-1}=\omega+\alpha u_{t-1}^2+\beta h_{t-1}$, we have
\begin{align}
\frac{\partial b_{t-1}}{\partial\theta}
&=\frac{\partial\omega}{\partial\theta}
+e_\alpha u_{t-1}^2+e_\beta h_{t-1}
+\beta\frac{\partial h_{t-1}}{\partial\theta},
\label{eq:b-first-deriv}\\
\frac{\partial^2 b_{t-1}}{\partial\theta\,\partial\theta^\top}
&=\frac{\partial^2\omega}{\partial\theta\,\partial\theta^\top}
+e_\beta\frac{\partial h_{t-1}}{\partial\theta^\top}
+\frac{\partial h_{t-1}}{\partial\theta}e_\beta^\top
+\beta\frac{\partial^2 h_{t-1}}{\partial\theta\,\partial\theta^\top}.
\label{eq:b-second-deriv}
\end{align}
Applying the chain rule to $h_t=g(b_{t-1},u_t^2;\varphi)$ gives
\begin{align}
\frac{\partial h_t}{\partial\theta}
&=\frac{\partial g}{\partial b}
  \frac{\partial b_{t-1}}{\partial\theta}
+\frac{\partial g}{\partial\varphi}e_\varphi,
\label{eq:h-first-deriv-recursion}\\
\frac{\partial^2 h_t}{\partial\theta\,\partial\theta^\top}
&=\frac{\partial g}{\partial b}
  \frac{\partial^2 b_{t-1}}{\partial\theta\,\partial\theta^\top}
+\frac{\partial^2 g}{\partial b^2}
  \frac{\partial b_{t-1}}{\partial\theta}
  \frac{\partial b_{t-1}}{\partial\theta^\top}\notag\\
&\quad+\frac{\partial^2 g}{\partial b\,\partial\varphi}
\left\{\frac{\partial b_{t-1}}{\partial\theta}e_\varphi^\top
+e_\varphi\frac{\partial b_{t-1}}{\partial\theta^\top}\right\}
+\frac{\partial^2 g}{\partial\varphi^2}e_\varphi e_\varphi^\top.
\label{eq:process-derivative-recursions}
\end{align}
Substituting \eqref{eq:b-first-deriv} into \eqref{eq:h-first-deriv-recursion} and using $\beta(\partial g/\partial b)\le\bar\beta$ from \eqref{eq:H-derivative-b} together with \eqref{eq:H-lipschitz} and \eqref{eq:g-first-bounds}, we obtain
\begin{align*}
\sup_{\theta\in\Theta}h_t(\theta)
&\le\bar\beta\sup_{\theta\in\Theta}h_{t-1}(\theta)+C(1+u_t^2+u_{t-1}^2),\\
\sup_{\theta\in\Theta}\left\|\frac{\partial h_t(\theta)}{\partial\theta}\right\|
&\le\bar\beta\sup_{\theta\in\Theta}\left\|\frac{\partial h_{t-1}(\theta)}{\partial\theta}\right\|
+C\left\{1+u_t^2+u_{t-1}^2+\sup_{\theta\in\Theta}h_{t-1}(\theta)\right\}.
\end{align*}
Therefore, applying Lemma~\ref{lem:contractive-master}(i)(b) with $\Delta_T=1$ proves \eqref{eq:process-level-first-bound}, which in turn implies that $$\sup_{\theta\in\Theta}\|\partial b_{t-1}(\theta)/\partial\theta\|\le C(1+u_{t-1}^2+\sup_{\theta\in\Theta}\{h_{t-1}(\theta)+\|\partial_\theta h_{t-1}(\theta)\|\})\le CU_t.$$ Substituting \eqref{eq:b-second-deriv} into \eqref{eq:process-derivative-recursions} and bounding the second-order derivatives of $g$ via \eqref{eq:g-second-bounds} yields
\[
\sup_{\theta\in\Theta}\left\|\frac{\partial^2 h_t(\theta)}{\partial\theta\,\partial\theta^\top}\right\|
\le\bar\beta\sup_{\theta\in\Theta}\left\|\frac{\partial^2 h_{t-1}(\theta)}{\partial\theta\,\partial\theta^\top}\right\|+CU_t^2,
\]
and Lemma~\ref{lem:contractive-master}(i)(b) gives \eqref{eq:derived-moment-bounds}.

Finally, differentiating $\ell_t(\theta)=f(u_t^2,h_t(\theta),\varphi)$ gives the score and Hessian matrix as \begin{align}
s_t(\theta)=\frac{\partial\ell_t(\theta)}{\partial\theta}
&=\frac{\partial f}{\partial h}\frac{\partial h_t}{\partial\theta}+\frac{\partial f}{\partial\varphi}e_\varphi,
\label{eq:score-explicit-form}\\
\frac{\partial^2\ell_t(\theta)}{\partial\theta\,\partial\theta^\top}
&=\frac{\partial f}{\partial h}
\frac{\partial^2h_t}{\partial\theta\,\partial\theta^\top}
+\frac{\partial^2f}{\partial h^2}
\frac{\partial h_t}{\partial\theta}
\frac{\partial h_t}{\partial\theta^\top}
+\frac{\partial^2f}{\partial h\,\partial\varphi}
\left\{\frac{\partial h_t}{\partial\theta}e_\varphi^\top
+e_\varphi\frac{\partial h_t}{\partial\theta^\top}\right\}
+\frac{\partial^2f}{\partial\varphi^2}e_\varphi e_\varphi^\top.
\label{eq:likelihood-hessian-form}
\end{align}
Combining \eqref{eq:score-explicit-form} with \eqref{eq:f-global-bounds} and \eqref{eq:process-level-first-bound}, we obtain $$\sup_{\theta\in\Theta}\|s_t(\theta)\|\le C(1+u_t^2)U_t\le CU_t^2\in L^{q/2}.$$
 \end{proof}
\begin{lemma}
\label{lem:master-perturbation}
For any nonnegative input sequence $\mathbf v=\{v_t\}_{t\ge0}$ and initial state $h_0(\theta)>0$, let $h_t(\theta;\mathbf v,h_0)$ be generated by $b_{t-1}(\theta)=\omega(\theta)+\alpha v_{t-1}+\beta h_{t-1}(\theta)$ and $h_t(\theta)=g(b_{t-1}(\theta),v_t;\varphi)$, and let $\ell_t(\theta;\mathbf v,h_0):=f(v_t,h_t(\theta;\mathbf v,h_0),\varphi)$. Given two input sequences $\mathbf v^{(1)},\mathbf v^{(2)}$ satisfying $0\le v_t^{(m)}\le 2(u_t^{(m)})^2$ (where $u_t^{(m)}$ has envelope $U_t^{(m)}$ and $G_{1,t}^{(m)}$ as in \eqref{eq:master-envelopes-UG}) and initial states $h_0^{(m)}$ for $m=1,2$. Write $h_t^{(m)}(\theta):=h_t(\theta;\mathbf v^{(m)},h_0^{(m)})$ and $\ell_t^{(m)}(\theta):=\ell_t(\theta;\mathbf v^{(m)},h_0^{(m)})$ for $m=1,2$, and define\begin{equation}\label{eq:Delta-notation}
\begin{aligned}
\Delta v_t&:=|v_t^{(1)}-v_t^{(2)}|,\qquad
\Delta_0 h_t:=\sup_{\theta\in\Theta}\bigl|h_t^{(1)}(\theta)-h_t^{(2)}(\theta)\bigr|,\\
\Delta_1 h_t&:=\sup_{\theta\in\Theta}\left\|\frac{\partial h_t^{(1)}(\theta)}{\partial\theta}-\frac{\partial h_t^{(2)}(\theta)}{\partial\theta}\right\|,\qquad
\Delta_2 h_t:=\sup_{\theta\in\Theta}\left\|\frac{\partial^2 h_t^{(1)}(\theta)}{\partial\theta\,\partial\theta^\top}-\frac{\partial^2 h_t^{(2)}(\theta)}{\partial\theta\,\partial\theta^\top}\right\|.
\end{aligned}
\end{equation} Let  $B_t:=C(U_t^{(1)}+U_t^{(2)})$ and $C_t:=C\{B_t^2+G_{1,t-1}^{(1)}+G_{1,t-1}^{(2)}\}$. Then for all $t\ge1$:
\begin{enumerate}[(i)]
\item The process differences $\{\Delta_0 h_t,\Delta_1 h_t,\Delta_2 h_t\}$ satisfy the contractive triangular recursions
\begin{equation}\label{eq:master-h-recursions}
\begin{aligned}
\Delta_0 h_t&\le\bar\beta\Delta_0 h_{t-1}+C(\Delta v_t+\Delta v_{t-1}),\\
\Delta_1 h_t&\le\bar\beta\Delta_1 h_{t-1}+CB_t(\Delta_0 h_{t-1}+\Delta v_t+\Delta v_{t-1}),\\
\Delta_2 h_t&\le\bar\beta\Delta_2 h_{t-1}+CB_t\Delta_1 h_{t-1}+CC_t(\Delta_0 h_{t-1}+\Delta v_t+\Delta v_{t-1}).
\end{aligned}
\end{equation}
\item The log-likelihood, score, and Hessian differences satisfy
\begin{equation}\label{eq:master-likelihood-lipschitz}
\begin{aligned}
\sup_{\theta\in\Theta}\bigl|\ell_t^{(1)}(\theta)-\ell_t^{(2)}(\theta)\bigr|
&\le C\Delta v_t+CB_t\Delta_0 h_t,\\
\sup_{\theta\in\Theta}\left\|\frac{\partial\ell_t^{(1)}(\theta)}{\partial\theta}-\frac{\partial\ell_t^{(2)}(\theta)}{\partial\theta}\right\|
&\le CB_t^2(\Delta v_t+\Delta_0 h_t)+CB_t\Delta_1 h_t,\\
\sup_{\theta\in\Theta}\left\|\frac{\partial^2\ell_t^{(1)}(\theta)}{\partial\theta\,\partial\theta^\top}-\frac{\partial^2\ell_t^{(2)}(\theta)}{\partial\theta\,\partial\theta^\top}\right\|
&\le C(B_tC_t+B_t^3)(\Delta v_t+\Delta_0 h_t)+CB_t^2\Delta_1 h_t+CB_t\Delta_2 h_t.
\end{aligned}
\end{equation}
\end{enumerate}
\end{lemma}

\begin{proof}
Write $g^{(m)}$ and $f^{(m)}$ for $g$ and $f$ evaluated at $(b_{t-1}^{(m)}(\theta),v_t^{(m)};\varphi)$ and $(v_t^{(m)},h_t^{(m)}(\theta),\varphi)$, respectively. By \eqref{eq:b-first-deriv}--\eqref{eq:b-second-deriv}, the differences of $b_{t-1}^{(1)}(\theta)$ and $b_{t-1}^{(2)}(\theta)$ satisfy
\begin{equation}\label{eq:Delta-b-bounds}
\begin{aligned}
\sup_{\theta\in\Theta}\bigl|b_{t-1}^{(1)}(\theta)-b_{t-1}^{(2)}(\theta)\bigr|
&\le\alpha\Delta v_{t-1}+\bar\beta\Delta_0 h_{t-1},\\
\sup_{\theta\in\Theta}\left\|\frac{\partial b_{t-1}^{(1)}(\theta)}{\partial\theta}-\frac{\partial b_{t-1}^{(2)}(\theta)}{\partial\theta}\right\|
&\le\Delta v_{t-1}+\Delta_0 h_{t-1}+\bar\beta\Delta_1 h_{t-1},\\
\sup_{\theta\in\Theta}\left\|\frac{\partial^2 b_{t-1}^{(1)}(\theta)}{\partial\theta\,\partial\theta^\top}-\frac{\partial^2 b_{t-1}^{(2)}(\theta)}{\partial\theta\,\partial\theta^\top}\right\|
&\le 2\Delta_1 h_{t-1}+\bar\beta\Delta_2 h_{t-1}.
\end{aligned}
\end{equation}
Applying \eqref{eq:H-derivative-b}--\eqref{eq:H-lipschitz} to $h_t^{(1)}(\theta)-h_t^{(2)}(\theta)=g(b_{t-1}^{(1)},v_t^{(1)};\varphi)-g(b_{t-1}^{(2)},v_t^{(2)};\varphi)$ gives the first inequality in \eqref{eq:master-h-recursions}. Next, by \eqref{eq:b-first-deriv} and \eqref{eq:h-first-deriv-recursion}, subtracting the first-derivative recursions gives
\[
\begin{aligned}
\frac{\partial h_t^{(1)}}{\partial\theta}-\frac{\partial h_t^{(2)}}{\partial\theta}
={}&\beta\frac{\partial g^{(1)}}{\partial b}\left(\frac{\partial h_{t-1}^{(1)}}{\partial\theta}-\frac{\partial h_{t-1}^{(2)}}{\partial\theta}\right)
+\frac{\partial g^{(1)}}{\partial b}\bigl\{e_\alpha(v_{t-1}^{(1)}-v_{t-1}^{(2)})+e_\beta(h_{t-1}^{(1)}-h_{t-1}^{(2)})\bigr\}\\
&+\left(\frac{\partial g^{(1)}}{\partial b}-\frac{\partial g^{(2)}}{\partial b}\right)\frac{\partial b_{t-1}^{(2)}}{\partial\theta}
+\left(\frac{\partial g^{(1)}}{\partial\varphi}-\frac{\partial g^{(2)}}{\partial\varphi}\right)e_\varphi.
\end{aligned}
\]

Bounding the partial derivatives of $g$ and their differences via \eqref{eq:H-derivative-b} and \eqref{eq:g-deriv-lipschitz}, and using $\sup_{\theta\in\Theta}\|\partial_\theta b_{t-1}^{(2)}(\theta)\|\le CB_t$ and $1+v_t^{(1)}+v_t^{(2)}\le CB_t$, yields the second inequality in \eqref{eq:master-h-recursions}. Similarly, subtracting \eqref{eq:process-derivative-recursions} for $m=1,2$ gives
\[
\begin{aligned}
\frac{\partial^2 h_t^{(1)}}{\partial\theta\,\partial\theta^\top}-\frac{\partial^2 h_t^{(2)}}{\partial\theta\,\partial\theta^\top}
={}&\frac{\partial g^{(1)}}{\partial b}\left(\frac{\partial^2 b_{t-1}^{(1)}}{\partial\theta\,\partial\theta^\top}-\frac{\partial^2 b_{t-1}^{(2)}}{\partial\theta\,\partial\theta^\top}\right)
+\left(\frac{\partial g^{(1)}}{\partial b}-\frac{\partial g^{(2)}}{\partial b}\right)\frac{\partial^2 b_{t-1}^{(2)}}{\partial\theta\,\partial\theta^\top}\\
&+\frac{\partial^2 g^{(1)}}{\partial b^2}\left(\frac{\partial b_{t-1}^{(1)}}{\partial\theta}\frac{\partial b_{t-1}^{(1)}}{\partial\theta^\top}-\frac{\partial b_{t-1}^{(2)}}{\partial\theta}\frac{\partial b_{t-1}^{(2)}}{\partial\theta^\top}\right)
+\left(\frac{\partial^2 g^{(1)}}{\partial b^2}-\frac{\partial^2 g^{(2)}}{\partial b^2}\right)\frac{\partial b_{t-1}^{(2)}}{\partial\theta}\frac{\partial b_{t-1}^{(2)}}{\partial\theta^\top}\\
&+\frac{\partial^2 g^{(1)}}{\partial b\,\partial\varphi}\left\{\left(\frac{\partial b_{t-1}^{(1)}}{\partial\theta}-\frac{\partial b_{t-1}^{(2)}}{\partial\theta}\right)e_\varphi^\top+e_\varphi\left(\frac{\partial b_{t-1}^{(1)}}{\partial\theta^\top}-\frac{\partial b_{t-1}^{(2)}}{\partial\theta^\top}\right)\right\}\\
&+\left(\frac{\partial^2 g^{(1)}}{\partial b\,\partial\varphi}-\frac{\partial^2 g^{(2)}}{\partial b\,\partial\varphi}\right)\left(\frac{\partial b_{t-1}^{(2)}}{\partial\theta}e_\varphi^\top+e_\varphi\frac{\partial b_{t-1}^{(2)}}{\partial\theta^\top}\right)
+\left(\frac{\partial^2 g^{(1)}}{\partial\varphi^2}-\frac{\partial^2 g^{(2)}}{\partial\varphi^2}\right)e_\varphi e_\varphi^\top.
\end{aligned}
\]
Applying \eqref{eq:g-second-bounds}, \eqref{eq:g-deriv-lipschitz}, and \eqref{eq:Delta-b-bounds} together with $\sup_{\theta\in\Theta}\|\partial^2_{\theta\theta^\top}b_{t-1}^{(2)}(\theta)\|\le CC_t$ gives the third inequality in \eqref{eq:master-h-recursions}. This proves (i).

To prove (ii), the first inequality in \eqref{eq:master-likelihood-lipschitz} follows directly from the first line of \eqref{eq:f-deriv-lipschitz}. For the score difference, subtracting \eqref{eq:score-explicit-form} for $m=1,2$ gives
\begin{equation}\label{eq:score-diff-decomposition}
\frac{\partial\ell_t^{(1)}}{\partial\theta}-\frac{\partial\ell_t^{(2)}}{\partial\theta}
=\frac{\partial f^{(1)}}{\partial h}\left(\frac{\partial h_t^{(1)}}{\partial\theta}-\frac{\partial h_t^{(2)}}{\partial\theta}\right)
+\left(\frac{\partial f^{(1)}}{\partial h}-\frac{\partial f^{(2)}}{\partial h}\right)\frac{\partial h_t^{(2)}}{\partial\theta}
+\left(\frac{\partial f^{(1)}}{\partial\varphi}-\frac{\partial f^{(2)}}{\partial\varphi}\right)e_\varphi,
\end{equation}
and applying \eqref{eq:f-global-bounds} and \eqref{eq:f-deriv-lipschitz} with $\sup_{\theta\in\Theta}\|\partial_\theta h_t^{(2)}(\theta)\|\le CB_t$ yields the second inequality in \eqref{eq:master-likelihood-lipschitz}. Finally, subtracting the Hessian  \eqref{eq:likelihood-hessian-form} for $m=1,2$ gives
\[
\begin{aligned}
\frac{\partial^2\ell_t^{(1)}}{\partial\theta\,\partial\theta^\top}-\frac{\partial^2\ell_t^{(2)}}{\partial\theta\,\partial\theta^\top}
={}&\frac{\partial f^{(1)}}{\partial h}\left(\frac{\partial^2 h_t^{(1)}}{\partial\theta\,\partial\theta^\top}-\frac{\partial^2 h_t^{(2)}}{\partial\theta\,\partial\theta^\top}\right)
+\left(\frac{\partial f^{(1)}}{\partial h}-\frac{\partial f^{(2)}}{\partial h}\right)\frac{\partial^2 h_t^{(2)}}{\partial\theta\,\partial\theta^\top}\\
&+\frac{\partial^2 f^{(1)}}{\partial h^2}\left(\frac{\partial h_t^{(1)}}{\partial\theta}\frac{\partial h_t^{(1)}}{\partial\theta^\top}-\frac{\partial h_t^{(2)}}{\partial\theta}\frac{\partial h_t^{(2)}}{\partial\theta^\top}\right)
+\left(\frac{\partial^2 f^{(1)}}{\partial h^2}-\frac{\partial^2 f^{(2)}}{\partial h^2}\right)\frac{\partial h_t^{(2)}}{\partial\theta}\frac{\partial h_t^{(2)}}{\partial\theta^\top}\\
&+\frac{\partial^2 f^{(1)}}{\partial h\,\partial\varphi}\left\{\left(\frac{\partial h_t^{(1)}}{\partial\theta}-\frac{\partial h_t^{(2)}}{\partial\theta}\right)e_\varphi^\top+e_\varphi\left(\frac{\partial h_t^{(1)}}{\partial\theta^\top}-\frac{\partial h_t^{(2)}}{\partial\theta^\top}\right)\right\}\\
&+\left(\frac{\partial^2 f^{(1)}}{\partial h\,\partial\varphi}-\frac{\partial^2 f^{(2)}}{\partial h\,\partial\varphi}\right)\left(\frac{\partial h_t^{(2)}}{\partial\theta}e_\varphi^\top+e_\varphi\frac{\partial h_t^{(2)}}{\partial\theta^\top}\right)
+\left(\frac{\partial^2 f^{(1)}}{\partial\varphi^2}-\frac{\partial^2 f^{(2)}}{\partial\varphi^2}\right)e_\varphi e_\varphi^\top.
\end{aligned}
\]
Bounding each term by \eqref{eq:f-global-bounds}, \eqref{eq:f-deriv-lipschitz}, $\sup_{\theta\in\Theta}\|\partial_\theta h_t^{(m)}(\theta)\|\le CB_t$, and $\sup_{\theta\in\Theta}\|\partial^2_{\theta\theta^\top}h_t^{(2)}(\theta)\|\le CC_t$ gives the third inequality in \eqref{eq:master-likelihood-lipschitz}.
\end{proof}

\begin{lemma}\label{lem:augmented-gmc}
Under Assumptions~\ref{Asm1} and~\ref{Asm6}, we have that uniformly for $\theta\in\Theta$,
\begin{enumerate}[(i)]
    \item $z_t$ and $h_t(\theta)$ are $L^q$ GMC;
    \item $\partial h_t(\theta)/\partial\theta$ is $L^p$ GMC for $1\le p<q$;
    \item $\partial^2h_t(\theta)/(\partial\theta\partial\theta^\top)$ and $s_t(\theta_0)$ are $L^p$ GMC for $1\le p<q/2$.
\end{enumerate}
\end{lemma}

\begin{proof}
Apply Lemma~\ref{lem:master-perturbation} with
$\mathbf v^{(1)}=\{u_t^2\}$ and
$\mathbf v^{(2)}=\{(u_t^\circ)^2\}$.
We retain its differences $\Delta_r h_t$ and
$\Delta v_t=|u_t^2-(u_t^\circ)^2|$.
Proposition~\ref{prop:rt-gmc} with $p=q$ gives
$\|\Delta v_t+\Delta v_{t-1}\|_q\le C\rho^t$, after increasing
$\rho<1$ if necessary.
The initial process differences have the moments required by
Lemma~\ref{lem:contractive-master}(i)(a), since each version
satisfies Lemma~\ref{lem:process-moment-bounds}.
Hence, Lemma~\ref{lem:master-perturbation}(i) yields
\[
\|\Delta_0h_t\|_q\le C\rho^t,\qquad
\|\Delta_1h_t\|_{q/2}\le C\rho^t,\qquad
\|\Delta_2h_t\|_{q/3}\le C\rho^t.
\]
Moreover,  the second inequality in \eqref{eq:master-likelihood-lipschitz} of Lemma~\ref{lem:master-perturbation}(ii) gives 
\begin{equation}\label{ineq:diffs}
\sup_{\theta\in\Theta}\|s_t(\theta)-s_t^\circ(\theta)\|
=\Delta_1\ell_t
\le C\{B_t^2(\Delta_0h_t+\Delta v_t)+B_t\Delta_1h_t\},
\end{equation}
where $\|B_t\|_q\le C$.
H\"older's inequality therefore yields
\[
\left\|\sup_{\theta\in\Theta}\|s_t(\theta)-s_t^\circ(\theta)\|\right\|_{q/3}
\le C\|B_t\|_q^2(\|\Delta_0h_t\|_q+\|\Delta v_t\|_q)
+C\|B_t\|_q\|\Delta_1h_t\|_{q/2}
\le C\rho^t.
\]
The first bound proves $L^q$ GMC for $h_t$, and
Proposition~\ref{prop:rt-gmc} already gives it for $z_t$.
For the first parameter derivative, combine its $L^{q/2}$
coupling bound with its finite $q$th moment.
For the second parameter derivative and score, combine their
$L^{q/3}$ coupling bounds with their finite $(q/2)$th moments.
The interpolation in Lemma~\ref{lem:contractive-master}(ii)(a)
then gives  (ii) and (iii).
\end{proof}

\begin{proof}[Proof of Proposition~\ref{prop:first-step-uniform}]
The proof adapts the bias and stochastic error decomposition in
Lemma~A.1 of \citet{jiang2021adaptive}, and the key idea is to use a GMC maximal
inequality for the stochastic term. 
In particular, we can write
\[
\widehat\tau_t=\frac1{Th}\sum_{s=1}^T w_{t,s}r_s^2,
\]
where $w_{t,s}$ is the sum of the kernel weights at 
$r_s^2$ and its reflected copies. For $t\in\mathcal I_T$,
no reflection is involved and $w_{t,s}=K((t-s)/(Th))$.
Since $r_s^2=\tau_s(1+z_s)$ and $Ez_s=0$, we have
\[
\begin{aligned}
\widehat\tau_t-\tau_t
&=\bigl(E\widehat\tau_t-\tau_t\bigr)
  +\bigl(\widehat\tau_t-E\widehat\tau_t\bigr)\\
&=\left\{\frac1{Th}\sum_{s=1}^T w_{t,s}\tau_s-\tau_t\right\}
  +\frac1{Th}\sum_{s=1}^T w_{t,s}\tau_s z_s:=D_{t,T}+ S_{t,T}.
\end{aligned}
\]
The first term is the deterministic bias, while the second
is the centered stochastic error. We bound these two terms
separately.

Since the weights are uniformly bounded, and are nonzero for at most $CTh$ observations, then by standard calculation, 
\[
\begin{aligned}
\max_{1\leq t\leq T}|D_{t,T}|
&=O\!\left(h+\frac1{Th}\right),\\
\max_{t\in\mathcal I_T}|D_{t,T}|
&=O\!\left(h^2+\frac1{Th}\right).
\end{aligned}
\]
The first bound uses Lipschitz continuity of $\tau$ at the
reflected boundaries. In the interior, kernel symmetry removes
the first-order Taylor term.

Let $m=\lceil Th\rceil$ and partition $\{1,\ldots,T\}$ into
$\lceil T/m\rceil$ consecutive blocks, each containing $m$ or $m-1$ observations. For the $j$th block,
write $a=(j-1)m+1$ and $b=\min(jm,T)$ for its endpoints. Then, we have 
\[
\begin{aligned}
\sum_{s=a}^{b}w_{t,s}\tau_s z_s
&=w_{t,b}\tau_b\sum_{v=a}^{b}z_v +
\sum_{s=a}^{b-1}
\bigl(w_{t,s}\tau_s-w_{t,s+1}\tau_{s+1}\bigr)
\sum_{v=a}^{s}z_v.
\end{aligned}
\]
By Assumption \ref{Asm3}, the kernel is bounded and has bounded variation. In addition, $\tau$
and its first derivative are bounded.  Consequently,
uniformly over $t$ and the blocks,
\[
|w_{t,b}\tau_b|
+\sum_{s=a}^{b-1}
|w_{t,s}\tau_s-w_{t,s+1}\tau_{s+1}|
\le C.
\]
It follows that
\[
\left|\sum_{s=a}^{b}w_{t,s}\tau_s z_s\right|
\le C\max_{a\le k\le b}
\left|\sum_{s=a}^{k}z_s\right|.
\]

For each $t$, the nonzero weights arise from a fixed number
of original or reflected kernel windows, each of length
$O(m)$. Hence only a bounded number of blocks contribute to
the weighted sum. Since $Th\asymp m$, we obtain
\[
\max_{1\le t\le T}
|\widehat\tau_t-E\widehat\tau_t|
\le \frac{C}{m}
\max_{1\le j\le\lceil T/m\rceil}
\max_{(j-1)m<k\le\min(jm,T)}
\left|\sum_{s=(j-1)m+1}^{k}z_s\right|.
\]

By Proposition~\ref{prop:rt-gmc}, we have that $z_t$ is $L^q$ GMC. Therefore, by 
Theorem~2(ii) of \citet{liu2013probability}, we have that 
\[
P\left\{
\max_{1\le k\le m}\left|\sum_{s=1}^{k}z_s\right|>x
\right\}
\le \frac{Cm}{x^q}
+C\exp\left(-\frac{cx^2}{m}\right)
\]
whenever $x\ge C_0\sqrt m$, for a sufficiently large fixed
constant $C_0$.

By stationarity, this bound applies to every block. Hence, a  union bound over the
$\lceil T/m\rceil$ blocks yields
\[
\begin{aligned}
&P\left\{
\max_{1\le t\le T}
|\widehat\tau_t-E\widehat\tau_t|>\frac{Cx}{m}
\right\}\le
\left\lceil\frac{T}{m}\right\rceil
P\left\{
\max_{1\le k\le m}
\left|\sum_{s=1}^{k}z_s\right|>x
\right\}\\
&\quad\le
\frac{CT}{x^q}
+\frac{CT}{m}\exp\left(-\frac{cx^2}{m}\right).
\end{aligned}
\]
Taking $x=C_1\sqrt{m\log T}$ for a sufficiently large $C_1$, we thus obtain that 
\[
\begin{aligned}
&P\left\{
\max_{1\le t\le T}
|\widehat\tau_t-E\widehat\tau_t|
>C_1\sqrt{\frac{\log T}{Th}}
\right\}\\
&\quad\le
CTm^{-q/2}(\log T)^{-q/2}
+C(T/m)T^{-cC_1^2}
=o(1).
\end{aligned}
\]
By Assumptions~\ref{Asm5} and~\ref{Asm6},
$m\asymp T^{1-\lambda_h}$, $q>4$, and
$\lambda_h<1/2$. These imply that
$1-(1-\lambda_h)q/2<0$, so the first term tends to zero. The second term tends to zero for a sufficiently large $C_1.$

In other words, we have that 
\[
\max_{1\le t\le T}
|\widehat\tau_t-E\widehat\tau_t|
=O_p\left(\sqrt{\frac{\log T}{Th}}\right).
\]
Combining this bound with the two bias bounds gives
the convergence rate. Moreover, $\kappa_T\to0$ and
$\inf_t\tau_t\ge\underline\tau>0$ imply
$\inf_t\widehat\tau_t\ge\underline\tau/2$ with probability
approaching one.
\end{proof}

In the following, we distinguish three  processes. The stationary infeasible one  is
$h_t(\theta)$. The  infeasible one with initial values based on the true
$\{u_t\}$ is denoted by $\widetilde h_t(\theta)$, while the feasible
one based on
$\widehat u_t=r_t/\sqrt{\widehat\tau_t}$ is
$\widehat h_t(\theta)$. Both initialized processes use fixed positive
initial values at the beginning of the sample.
The corresponding negative log-likelihood functions  are defined as
$\ell_t(\theta)$, $\widetilde\ell_t(\theta)$, and
$\widehat\ell_t(\theta)$, respectively.

\begin{lemma}\label{lem:first-step-expansion}
Under Assumptions~\ref{Asm1}--\ref{Asm6}, let $\kappa_T$ be defined in
Proposition~\ref{prop:first-step-uniform}. Then
\begin{equation}\label{eq:uhat-expansion}
\widehat u_t^2-u_t^2
=-u_t^2\frac{\widehat\tau_t-\tau_t}{\tau_t}+u_t^2R_{tT},
\qquad
\max_{1\le t\le T}|R_{tT}|=O_p(\kappa_T^2).
\end{equation}
In addition,
$\max_t|\widehat u_t^2-u_t^2|/(1+u_t^2)=O_p(\kappa_T)$.
For every $p\ge1$ such that $E|u_t|^{2p}<\infty$,
\[
\frac1T\sum_{t=1}^T|\widehat u_t^2-u_t^2|^p
=O_p(\kappa_T^p).
\]
\end{lemma}

\begin{proof}
Let $\delta_{t,T}=(\widehat\tau_t-\tau_t)/\tau_t$.
Proposition~\ref{prop:first-step-uniform} gives
$\max_t|\delta_{t,T}|=O_p(\kappa_T)$.
Since $\kappa_T\to0$, the event
$\max_t|\delta_{t,T}|\le1/2$ has probability approaching one.
The positivity truncation is also inactive with probability
approaching one. On these events, $1+\delta_{t,T}\ge1/2$ and
$\widehat u_t^2=u_t^2/(1+\delta_{t,T})$. Hence,
\[
\widehat u_t^2-u_t^2
=-u_t^2\delta_{t,T}
+u_t^2\frac{\delta_{t,T}^2}{1+\delta_{t,T}},
\]
which gives \eqref{eq:uhat-expansion} with
$R_{tT}=\delta_{t,T}^2/(1+\delta_{t,T})$, because
\[
\max_t|R_{tT}|
\le2\bigl(\max_t|\delta_{t,T}|\bigr)^2
=O_p(\kappa_T^2).
\]

Moreover, $|\widehat u_t^2-u_t^2|
\le2u_t^2|\delta_{t,T}|$. Dividing by $1+u_t^2$ and
taking the maximum over $t$ gives the stated ratio bound.
For every $p\ge1$ with $E|u_t|^{2p}<\infty$, the same
inequality yields
\[
\frac1T\sum_{t=1}^T|\widehat u_t^2-u_t^2|^p
\le
2^p\bigl(\max_t|\delta_{t,T}|\bigr)^p
\frac1T\sum_{t=1}^T|u_t|^{2p}
=O_p(\kappa_T^p),
\]
where the last equality follows by the ergodic theorem
  $T^{-1}\sum_{t=1}^T|u_t|^{2p}
\to E|u_t|^{2p}<\infty$ a.s.
\end{proof}

By combining Lemma~\ref{lem:contractive-master} and Lemma~\ref{lem:master-perturbation}, we now establish the initialization error bounds and the first-step estimation error bounds as follows.

\begin{lemma}\label{lem:initial-values}
Suppose Assumptions~\ref{Asm1}--\ref{Asm6} hold with some fixed positive initial values. Then, for every $\rho\in(\bar\beta,1)$,
\begin{align}
\left\|\sup_{\theta\in\Theta}|\widetilde h_t(\theta)-h_t(\theta)|\right\|_q&\le C\rho^t,\label{eq:init-h}\\
\left\|\sup_{\theta\in\Theta}\left\|\frac{\partial\widetilde h_t(\theta)}{\partial\theta}-\frac{\partial h_t(\theta)}{\partial\theta}\right\|\right\|_{q/2}&\le C\rho^t,\label{eq:init-dh}\\
\left\|\sup_{\theta\in\Theta}\left\|\frac{\partial^2\widetilde h_t(\theta)}{\partial\theta\,\partial\theta^\top}-\frac{\partial^2 h_t(\theta)}{\partial\theta\,\partial\theta^\top}\right\|\right\|_{q/3}&\le C\rho^t.\label{eq:init-d2h}
\end{align}
Furthermore, the log-likelihood, score, and Hessian initialization errors satisfy
\begin{equation}\label{eq:init-likelihood-moments}
E\left[\sup_{\theta\in\Theta}|\widetilde\ell_t(\theta)-\ell_t(\theta)|\right]
+E\left[\sup_{\theta\in\Theta}\left\|\frac{\partial\widetilde\ell_t(\theta)}{\partial\theta}-\frac{\partial\ell_t(\theta)}{\partial\theta}\right\|\right]
+E\left[\sup_{\theta\in\Theta}\left\|\frac{\partial^2\widetilde\ell_t(\theta)}{\partial\theta\,\partial\theta^\top}-\frac{\partial^2\ell_t(\theta)}{\partial\theta\,\partial\theta^\top}\right\|\right]
\le C\rho^t,
\end{equation}
and consequently,
\begin{align}
\sup_{\theta\in\Theta}\left|\frac1T\sum_{t=1}^T\{\widetilde\ell_t(\theta)-\ell_t(\theta)\}\right|&=O_p(T^{-1})=o_p(1),\label{eq:criterion-init}\\
\sup_{\theta\in\Theta}\left\|\frac1T\sum_{t=1}^T\left\{\frac{\partial\widetilde\ell_t(\theta)}{\partial\theta}-\frac{\partial\ell_t(\theta)}{\partial\theta}\right\}\right\|&=O_p(T^{-1})=o_p(1),\label{eq:score-init}\\
\sup_{\theta\in\Theta}\left\|\frac1T\sum_{t=1}^T\left\{\frac{\partial^2\widetilde\ell_t(\theta)}{\partial\theta\,\partial\theta^\top}-\frac{\partial^2\ell_t(\theta)}{\partial\theta\,\partial\theta^\top}\right\}\right\|&=O_p(T^{-1})=o_p(1).\label{eq:hessian-init}
\end{align}
Moreover, the same bounds hold uniformly when $\{u_t^2\}$ is replaced by the interpolated inputs $u_t^2(a)=u_t^2+a(\widehat u_t^2-u_t^2)$, $a\in[0,1]$, on the event $\max_{1\le s\le T}|\tau_s/\widehat\tau_s-1|\le1/2$.
\end{lemma}

\begin{proof}
    Apply Lemma~\ref{lem:master-perturbation} with identical inputs $v_t^{(1)}=v_t^{(2)}=u_t^2$ (or $u_t^2(a)\le 2u_t^2$) for $t\ge1$, so that $\Delta v_t=|v_t^{(1)}-v_t^{(2)}|=0$ for $t\ge1$, while $h_0^{(1)}$ is the fixed initial value and $h_0^{(2)}=h_0(\theta)$ is the stationary initial state. Then \eqref{eq:master-h-recursions} in Lemma~\ref{lem:master-perturbation}(i) and \eqref{eq:master-decay-conclusion} in Lemma~\ref{lem:contractive-master}(i)(a) immediately yield \eqref{eq:init-h}--\eqref{eq:init-d2h}. Next, setting $\Delta v_t=0$ in \eqref{eq:master-likelihood-lipschitz} of Lemma~\ref{lem:master-perturbation}(ii) gives
\begin{align*}
\sup_{\theta\in\Theta}|\widetilde\ell_t(\theta)-\ell_t(\theta)|&\le CB_t\Delta_0 h_t,\\
\sup_{\theta\in\Theta}\left\|\frac{\partial\widetilde\ell_t(\theta)}{\partial\theta}-\frac{\partial\ell_t(\theta)}{\partial\theta}\right\|&\le CB_t^2\Delta_0 h_t+CB_t\Delta_1 h_t,\\
\sup_{\theta\in\Theta}\left\|\frac{\partial^2\widetilde\ell_t(\theta)}{\partial\theta\,\partial\theta^\top}-\frac{\partial^2\ell_t(\theta)}{\partial\theta\,\partial\theta^\top}\right\|&\le C(B_tC_t+B_t^3)\Delta_0 h_t+CB_t^2\Delta_1 h_t+CB_t\Delta_2 h_t.
\end{align*}
Since $\sup_t\|B_t\|_q<\infty$, $\sup_t\|C_t\|_{q/2}<\infty$, and $q>4$ (so $q/4>1$), H\"older's inequality implies that the $L^{q/4}$ norms (and hence the expectations) of the right-hand sides are bounded by $C\rho^t$, proving \eqref{eq:init-likelihood-moments}. Summing \eqref{eq:init-likelihood-moments} over $t=1,\ldots,T$ and applying Markov's inequality proves \eqref{eq:criterion-init}--\eqref{eq:hessian-init}.
\end{proof}

\begin{lemma}\label{lem:feasible-infeasible}\label{lem:likelihood-comparison}\label{lem:criterion-differences}
Define $u_t^2(a)=u_t^2+a(\widehat u_t^2-u_t^2)$, $a\in[0,1]$. 
Suppose Assumptions~\ref{Asm1}--\ref{Asm6} hold. Then with probability approaching one, for all $1\le t\le T$ and any $a,a'\in[0,1]$, the process $h_t(a,\theta)$ driven by $\{u_s^2(a)\}$ with fixed initial values satisfies
\begin{equation}\label{eq:path-h-lipschitz}
\begin{aligned}
\sup_{\theta\in\Theta}|h_t(a',\theta)-h_t(a,\theta)|
&\le C|a'-a|U_t\max_{1\le s\le T}|\tau_s/\widehat\tau_s-1|,\\
\sup_{\theta\in\Theta}\left\|\frac{\partial h_t(a',\theta)}{\partial\theta}-\frac{\partial h_t(a,\theta)}{\partial\theta}\right\|
&\le C|a'-a|G_{1,t}\max_{1\le s\le T}|\tau_s/\widehat\tau_s-1|,\\
\sup_{\theta\in\Theta}\left\|\frac{\partial^2 h_t(a',\theta)}{\partial\theta\,\partial\theta^\top}-\frac{\partial^2 h_t(a,\theta)}{\partial\theta\,\partial\theta^\top}\right\|
&\le C|a'-a|G_{2,t}\max_{1\le s\le T}|\tau_s/\widehat\tau_s-1|.
\end{aligned}
\end{equation}
In particular, at $(a',a)=(1,0)$ (so $h_t(1,\theta)=\widehat h_t(\theta)$ and $h_t(0,\theta)=\widetilde h_t(\theta)$). 

The pointwise likelihood, score, and Hessian differences satisfy that uniformly for $1\le t\le T$ with probability approaching one:
\begin{equation}\label{eq:feasible-hessian-envelope}
\begin{aligned}
\sup_{\theta\in\Theta}|\widehat\ell_t(\theta)-\widetilde\ell_t(\theta)|
&\le CU_t^2\max_{1\le s\le T}|\tau_s/\widehat\tau_s-1|,\\
\sup_{\theta\in\Theta}\left\|\frac{\partial\widehat\ell_t(\theta)}{\partial\theta}-\frac{\partial\widetilde\ell_t(\theta)}{\partial\theta}\right\|
&\le C\{U_tG_{1,t}+U_t^3\}\max_{1\le s\le T}|\tau_s/\widehat\tau_s-1|,\\
\sup_{\theta\in\Theta}\left\|\frac{\partial^2\widehat\ell_t(\theta)}{\partial\theta\,\partial\theta^\top}-\frac{\partial^2\widetilde\ell_t(\theta)}{\partial\theta\,\partial\theta^\top}\right\|
&\le CJ_t\max_{1\le s\le T}|\tau_s/\widehat\tau_s-1|,
\end{aligned}
\end{equation}
where $J_t:=C\{U_tG_{2,t}+U_t^2G_{1,t}+U_t^4\}\in L^{q/4}\subset L^1,$
and the sample averages satisfy
\begin{align}
\frac1T\sum_{t=1}^T\sup_{\theta\in\Theta}\left|\widehat h_t(\theta)-\widetilde h_t(\theta)\right|&=O_p(\kappa_T),\label{eq:feas-h}\\
\frac1T\sum_{t=1}^T\sup_{\theta\in\Theta}\left\|\frac{\partial\widehat h_t(\theta)}{\partial\theta}-\frac{\partial\widetilde h_t(\theta)}{\partial\theta}\right\|&=O_p(\kappa_T),\label{eq:feas-dh}\\
\frac1T\sum_{t=1}^T\sup_{\theta\in\Theta}\left\|\frac{\partial^2\widehat h_t(\theta)}{\partial\theta\,\partial\theta^\top}-\frac{\partial^2\widetilde h_t(\theta)}{\partial\theta\,\partial\theta^\top}\right\|&=O_p(\kappa_T),\label{eq:feas-d2h}\\
\frac1T\sum_{t=1}^T\sup_{\theta\in\Theta}\left|\widehat\ell_t(\theta)-\widetilde\ell_t(\theta)\right|&=O_p(\kappa_T)=o_p(1),\label{eq:criterion-feasible}\\
\frac1T\sum_{t=1}^T\sup_{\theta\in\Theta}\left\|\frac{\partial\widehat\ell_t(\theta)}{\partial\theta}-\frac{\partial\widetilde\ell_t(\theta)}{\partial\theta}\right\|&=O_p(\kappa_T)=o_p(1),\label{eq:score-feasible}\\
\frac1T\sum_{t=1}^T\sup_{\theta\in\Theta}\left\|\frac{\partial^2\widehat\ell_t(\theta)}{\partial\theta\,\partial\theta^\top}-\frac{\partial^2\widetilde\ell_t(\theta)}{\partial\theta\,\partial\theta^\top}\right\|&=O_p(\kappa_T)=o_p(1).\label{eq:hessian-feasible}
\end{align}
\end{lemma}

\begin{proof}
The proof is similar to that of 
Lemma~A.3 in \citet{jiang2021adaptive}.
By Lemma~\ref{lem:first-step-expansion}, the event $\delta_T:=\max_{1\le s\le T}|\tau_s/\widehat\tau_s-1|\le1/2$ has probability approaching one, and on this event $u_t^2(a)\le 2u_t^2$ and $$\Delta v_t=|u_t^2(a')-u_t^2(a)|=|a'-a|\,|\widehat u_t^2-u_t^2|\le|a'-a|\delta_T u_t^2.$$ Applying Lemma~\ref{lem:master-perturbation}(i) with identical initial values $h_0^{(1)}=h_0^{(2)}$ and \eqref{eq:master-envelope-conclusion} of Lemma~\ref{lem:contractive-master}(i)(b) with $\Delta_T=|a'-a|\delta_T$ immediately yields \eqref{eq:path-h-lipschitz}. Setting $(a',a)=(1,0)$ and substituting $\Delta v_t\le\delta_T U_t$, $\Delta_0 h_t\le C\delta_T U_t$, $\Delta_1 h_t\le C\delta_T G_{1,t}$, and $\Delta_2 h_t\le C\delta_T G_{2,t}$ (with $B_t\le CU_t$ and $C_t\le C(U_t^2+G_{1,t-1})\le C(U_t^2+G_{1,t})$) into \eqref{eq:master-likelihood-lipschitz} of Lemma~\ref{lem:master-perturbation}(ii) gives \eqref{eq:feasible-hessian-envelope}. Since $U_t\in L^q$, $G_{1,t}\in L^{q/2}$, $G_{2,t}\in L^{q/3}$, and $J_t\in L^{q/4}\subset L^1$ are stationary and ergodic, their sample averages are $O_p(1)$ by the ergodic theorem, which together with $\delta_T=O_p(\kappa_T)$ proves \eqref{eq:feas-h}--\eqref{eq:hessian-feasible}.
\end{proof}

\begin{proof}[Proof of
Theorem~\ref{thm:consistency}]
Let
$
Q_T(\theta)=T^{-1}\sum_{t=1}^T\ell_t(\theta),$ $
\widetilde Q_T(\theta)
=T^{-1}\sum_{t=1}^T\widetilde\ell_t(\theta),$ $
\widehat Q_T(\theta)
=T^{-1}\sum_{t=1}^T\widehat\ell_t(\theta),
$
and define the population criterion
$Q(\theta)=E\{\ell_t(\theta)\}$.

First, $\theta\mapsto\ell_t(\theta)$ is almost surely continuous. Indeed,
$\omega(\theta)$ is continuous, $h_t(\theta)$ is obtained from a continuous
stable recursion, and $h_t(\theta)\ge\underline\omega>0$ uniformly on
$\Theta$. Under the moment conditions in
Assumption~\ref{Asm6},
\[
E\left[\sup_{\theta\in\Theta}|\ell_t(\theta)|\right]<\infty.
\]
Since the stationary augmented process is ergodic, a uniform ergodic theorem
applies. More precisely, compactness of $\Theta$, almost-sure
continuity, the displayed integrable envelope, and the pointwise
ergodic theorem verify the conditions of Lemma~2.4 of
\citet{newey1994large}, which gives
\begin{equation}\label{eq:ulln-infeasible}
\sup_{\theta\in\Theta}|Q_T(\theta)-Q(\theta)|=o_p(1).
\end{equation}
Second, combining \eqref{eq:criterion-init} in Lemma~\ref{lem:initial-values} and \eqref{eq:criterion-feasible} in Lemma~\ref{lem:criterion-differences} implies
\begin{equation}\label{eq:criterion-all-close}
\sup_{\theta\in\Theta}
\left|\widehat Q_T(\theta)-Q_T(\theta)\right|
=o_p(1).
\end{equation}
Combining \eqref{eq:ulln-infeasible} and
\eqref{eq:criterion-all-close} gives
\[
\sup_{\theta\in\Theta}
\left|\widehat Q_T(\theta)-Q(\theta)\right|=o_p(1).
\]
Note that under Gaussian innovations, the identification
argument for condition (iv) in the proof of Theorem~5 of
\citet{smetanina2021asymptotic} implies that there is a unique minimizer
$\theta_0$ on $\Theta$. 
Theorem~2.1 of
\citet{newey1994large} then yields
\[
\widehat\theta_T\to_p\theta_0.
\]
\end{proof}


The next lemma gives the expansion needed when the true squared
returns are replaced by their estimates.
\begin{lemma} 
\label{lem:score-regularity}
Suppose Assumptions~\ref{Asm1}--\ref{Asm6} hold and
$\theta_0$ is an interior point of $\Theta$. Let $\mathcal N$ be a compact
neighborhood of $\theta_0$ contained in the interior of $\Theta$,
and let $p=q/2>2$.  Then, for some $\rho\in(0,1)$,
\begin{equation}
\|u_{t-j}^2\frac{\partial s_{t}(\theta_0)}{\partial (u_{t-j}^2)}\|_p\le C\rho^j,\qquad j\ge0.
\label{eq:path-derivative-decay}
\end{equation}
For $1\le i\le T$, consider the squared observations
$u_i^2(a)=u_i^2+a(\widehat u_i^2-u_i^2)$, where $0\le a\le1$.
Let $s_t(a,\theta)$ denote the resulting score function. Then, 
with probability approaching one,
\begin{equation}
\sup_{\theta\in\mathcal N}\sup_{0\le a\le1}
\left\|\frac{\partial^2s_t(a,\theta)}{\partial a^2}\right\|
\le CJ_t
\left(\max_{1\le i\le T}|\tau_i/\widehat\tau_i-1|\right)^2,
\quad 1\le t\le T,
\label{eq:score-directional-bound}
\end{equation}
where $J_t$ is defined in 
\eqref{eq:feasible-hessian-envelope}.
Moreover, the Hessian matrix for oracle process is almost surely continuous
on $\mathcal N$ and satisfies
\begin{equation}
E\sup_{\theta\in\mathcal N}
\|\frac{\partial^2 \ell_t(\theta)}{\partial \theta\,\partial \theta^\top}\|<\infty.
\label{eq:local-hessian-envelope}
\end{equation}
\end{lemma}
\begin{proof}
We use the partial-derivative bounds of $g$ and $f$ in \eqref{eq:g-first-bounds}--\eqref{eq:g-third-bounds} and \eqref{eq:f-local-bounds}, together with Lemma~\ref{lem:contractive-master}. 
First, taking first order derivative with respect to the squared return $u_i^2$ gives
\begin{equation}\label{eq:dh-du2-recursion}
\frac{\partial h_{t}}{\partial (u_{i}^2)}=\frac{\partial g}{\partial v}\mathbf1\{i=t\}
+\frac{\partial g}{\partial b}\left\{\alpha\mathbf1\{i=t-1\}+\beta \frac{\partial h_{t-1}}{\partial (u_{i}^2)}\right\}.
\end{equation}
A change in $u_i^2$ affects $h_i$ directly and $h_{i+1}$
through the terms $\alpha u_i^2$ and $\beta h_i$.
For $t\ge i+2$, its effect is multiplied at each step by
$\beta(\partial g/\partial b)\leq \overline{\beta}$.  By \eqref{eq:H-derivative-b} and \eqref{eq:g-first-bounds}, $|\partial g/\partial v|\le C$ and $0\le\beta(\partial g/\partial b)\le\bar\beta$, so iterating \eqref{eq:dh-du2-recursion} gives \begin{equation}\label{eq:dh-du2-deterministic}
\sup_{\theta\in\Theta}\left|\frac{\partial h_t(\theta)}{\partial(u_i^2)}\right|\le C\bar\beta^{t-i},\qquad t\ge i,
\end{equation}
and hence $\|\sup_{\theta\in\mathcal N}|u_i^2\{\partial h_t(\theta)/\partial(u_i^2)\}|\|_q\le C\|u_i^2\|_q\bar\beta^{t-i}\le C\rho^{t-i}$.  For $t\ge i+2$,
\begin{equation}\label{eq:d2h-dtheta-du2-recursion}
\frac{\partial^2 h_{t}}{\partial \theta\,\partial (u_{i}^2)}
=\beta \frac{\partial g}{\partial b}\frac{\partial^2 h_{t-1}}{\partial \theta\,\partial (u_{i}^2)}
+\frac{\partial}{\partial\theta}
\left(\beta\frac{\partial g}{\partial b}\right)\frac{\partial h_{t-1}}{\partial (u_{i}^2)},
\end{equation}
where by the chain rule and \eqref{eq:g-second-bounds},
\begin{equation}\label{eq:dtheta-beta-dgb}
\frac{\partial}{\partial\theta}\left(\beta\frac{\partial g}{\partial b}\right)
=\frac{\partial g}{\partial b}e_\beta+\beta\left(\frac{\partial^2 g}{\partial b^2}\frac{\partial b_{t-1}}{\partial\theta}+\frac{\partial^2 g}{\partial b\,\partial\varphi}e_\varphi\right),\qquad
\sup_{\theta\in\Theta}\left\|\frac{\partial}{\partial\theta}\left(\beta\frac{\partial g}{\partial b}\right)\right\|\le CU_t.
\end{equation}
Combining \eqref{eq:dh-du2-deterministic}--\eqref{eq:dtheta-beta-dgb} shows that for $t\ge i+2$,
\[
\sup_{\theta\in\mathcal N}\left\|u_i^2\frac{\partial^2 h_t(\theta)}{\partial\theta\,\partial(u_i^2)}\right\|
\le\bar\beta\sup_{\theta\in\mathcal N}\left\|u_i^2\frac{\partial^2 h_{t-1}(\theta)}{\partial\theta\,\partial(u_i^2)}\right\|
+CU_t u_i^2\bar\beta^{t-i},
\]
and applying Lemma~\ref{lem:contractive-master}(i)(a) (with H\"older's inequality $\|U_t u_i^2\|_{q/2}\le\|U_t\|_q\|u_i^2\|_q\le C$) yields
\begin{equation}\label{eq:d2h-du2-moment}
\left\|\sup_{\theta\in\mathcal N}
\left\|u_i^2\frac{\partial^2h_t(\theta)}{\partial\theta\,\partial(u_i^2)}\right\|
\right\|_{q/2}\le C\rho^{t-i},\qquad t\ge i.
\end{equation}
Differentiating the score  \eqref{eq:score-explicit-form} with respect to $u_i^2$ gives
\begin{equation}\label{eq:ds-du2-identity}
\frac{\partial s_t(\theta)}{\partial(u_i^2)}
=\frac{\partial f}{\partial h}\frac{\partial^2 h_t}{\partial\theta\,\partial(u_i^2)}
+\left(\frac{\partial^2 f}{\partial h^2}\frac{\partial h_t}{\partial\theta}+\frac{\partial^2 f}{\partial\varphi\,\partial h}e_\varphi\right)\frac{\partial h_t}{\partial(u_i^2)}
+\left(\frac{\partial^2 f}{\partial h\,\partial v}\frac{\partial h_t}{\partial\theta}+\frac{\partial^2 f}{\partial\varphi\,\partial v}e_\varphi\right)\mathbf1\{i=t\}.
\end{equation}
Applying the  bounds for the partial derivatives of $f$ on $\mathcal N$ in \eqref{eq:f-local-bounds} to \eqref{eq:ds-du2-identity} gives
\[
\left\|u_i^2\frac{\partial s_t(\theta_0)}{\partial(u_i^2)}\right\|
\le C\left[\left\|u_i^2
\frac{\partial^2h_t(\theta_0)}{\partial\theta\,\partial(u_i^2)}\right\|
+\left(1+\left\|\frac{\partial h_t(\theta_0)}{\partial\theta}\right\|\right)
\left\{\left|u_i^2\frac{\partial h_t(\theta_0)}{\partial(u_i^2)}\right|
+u_i^2\mathbf1\{i=t\}\right\}\right].
\]
Taking the $L^{q/2}$ norm and using \eqref{eq:dh-du2-deterministic}, \eqref{eq:d2h-du2-moment}, Lemma~\ref{lem:process-moment-bounds}, and H\"older's inequality proves \eqref{eq:path-derivative-decay}.

Next consider the interpolated observations $u_i^2(a)=u_i^2+a(\widehat u_i^2-u_i^2)$ on the event $\delta_T:=\max_{1\le i\le T}|\tau_i/\widehat\tau_i-1|\le1/2$, and let $h_t(a,\theta)$ and $b_{t-1}(a,\theta)$ denote the resulting processes evaluated at $(a,\theta)$. Dividing the Lipschitz bounds \eqref{eq:path-h-lipschitz} in Lemma~\ref{lem:feasible-infeasible} by $|a'-a|$ and letting $a'\to a$ gives the uniform first-order path bounds
\begin{equation}\label{eq:first-order-a-bounds}
\sup_{\theta\in\Theta}\left|\frac{\partial h_t(a,\theta)}{\partial a}\right|\le C\delta_T U_t,\qquad
\sup_{\theta\in\Theta}\left\|\frac{\partial^2 h_t(a,\theta)}{\partial a\,\partial\theta}\right\|\le C\delta_T G_{1,t},
\end{equation}
which in turn imply
\begin{equation}\label{eq:first-order-b-a-bounds}
\sup_{\theta\in\Theta}\left|\frac{\partial b_{t-1}(a,\theta)}{\partial a}\right|+\left|\frac{\partial u_t^2(a)}{\partial a}\right|\le C\delta_T U_t,\qquad
\sup_{\theta\in\Theta}\left\|\frac{\partial^2 b_{t-1}(a,\theta)}{\partial a\,\partial\theta}\right\|\le C\delta_T(U_t+G_{1,t-1}).
\end{equation}
Since $u_t^2(a)$ is linear in $a$, we can obtain $\partial^2 u_t^2(a)/\partial a^2=0$, $\partial^2 b_{t-1}/\partial a^2=\beta(\partial^2 h_{t-1}/\partial a^2)$ 
and $\partial^3 b_{t-1}/(\partial a^2\,\partial\theta)
=e_\beta(\partial^2 h_{t-1}/\partial a^2)+\beta\{\partial^3 h_{t-1}/(\partial a^2\,\partial\theta)\}$. 

Differentiating $h_t(a,\theta)=g(b_{t-1}(a,\theta),u_t^2(a);\varphi)$ twice with respect to $a$ gives
\[
\frac{\partial^2 h_t}{\partial a^2}
=\beta\frac{\partial g}{\partial b}\frac{\partial^2 h_{t-1}}{\partial a^2}
+\frac{\partial^2 g}{\partial b^2}\left(\frac{\partial b_{t-1}}{\partial a}\right)^2
+2\frac{\partial^2 g}{\partial b\,\partial v}\frac{\partial b_{t-1}}{\partial a}\frac{\partial u_t^2(a)}{\partial a}
+\frac{\partial^2 g}{\partial v^2}\left(\frac{\partial u_t^2(a)}{\partial a}\right)^2.
\]
Bounding the derivatives of $g$ via \eqref{eq:H-derivative-b} and \eqref{eq:g-second-bounds} and applying \eqref{eq:first-order-b-a-bounds} yields
\[
\sup_{\theta\in\mathcal N}\left|\frac{\partial^2 h_t(a,\theta)}{\partial a^2}\right|
\le\bar\beta\sup_{\theta\in\mathcal N}\left|\frac{\partial^2 h_{t-1}(a,\theta)}{\partial a^2}\right|+C\delta_T^2 U_t^2,
\]
so Lemma~\ref{lem:contractive-master}(i)(b) gives
\begin{equation}\label{eq:second-order-h-a-bound}
\sup_{\theta\in\mathcal N}\left|\frac{\partial^2 h_t(a,\theta)}{\partial a^2}\right|\le C\delta_T^2 G_{1,t},\qquad
\sup_{\theta\in\mathcal N}\left|\frac{\partial^2 b_{t-1}(a,\theta)}{\partial a^2}\right|\le C\delta_T^2 G_{1,t-1}.
\end{equation}
Similarly, differentiating $\partial h_t/\partial\theta=(\partial g/\partial b)(\partial b_{t-1}/\partial\theta)+(\partial g/\partial\varphi)e_\varphi$ twice with respect to $a$ gives
\[
\begin{aligned}
\frac{\partial^3 h_t}{\partial a^2\,\partial\theta}
={}&\beta\frac{\partial g}{\partial b}\frac{\partial^3 h_{t-1}}{\partial a^2\,\partial\theta}
+\frac{\partial g}{\partial b}\frac{\partial^2 h_{t-1}}{\partial a^2}e_\beta
+2\left(\frac{\partial^2 g}{\partial b^2}\frac{\partial b_{t-1}}{\partial a}+\frac{\partial^2 g}{\partial b\,\partial v}\frac{\partial u_t^2(a)}{\partial a}\right)\frac{\partial^2 b_{t-1}}{\partial a\,\partial\theta}\\
&+\biggl\{\frac{\partial^3 g}{\partial b^3}\left(\frac{\partial b_{t-1}}{\partial a}\right)^2
+2\frac{\partial^3 g}{\partial b^2\,\partial v}\frac{\partial b_{t-1}}{\partial a}\frac{\partial u_t^2(a)}{\partial a}\\
&\qquad+\frac{\partial^3 g}{\partial b\,\partial v^2}\left(\frac{\partial u_t^2(a)}{\partial a}\right)^2
+\frac{\partial^2 g}{\partial b^2}\frac{\partial^2 b_{t-1}}{\partial a^2}\biggr\}\frac{\partial b_{t-1}}{\partial\theta}\\
&+\biggl\{\frac{\partial^3 g}{\partial b^2\,\partial\varphi}\left(\frac{\partial b_{t-1}}{\partial a}\right)^2
+2\frac{\partial^3 g}{\partial b\,\partial v\,\partial\varphi}\frac{\partial b_{t-1}}{\partial a}\frac{\partial u_t^2(a)}{\partial a}\\
&\qquad+\frac{\partial^3 g}{\partial v^2\,\partial\varphi}\left(\frac{\partial u_t^2(a)}{\partial a}\right)^2
+\frac{\partial^2 g}{\partial b\,\partial\varphi}\frac{\partial^2 b_{t-1}}{\partial a^2}\biggr\}e_\varphi.
\end{aligned}
\]
Applying \eqref{eq:g-second-bounds}--\eqref{eq:g-third-bounds}, $\sup_{\theta\in\mathcal N}\|\partial_\theta b_{t-1}(a,\theta)\|\le CU_t$, \eqref{eq:first-order-b-a-bounds}, and \eqref{eq:second-order-h-a-bound} yields
\[
\sup_{\theta\in\mathcal N}\left\|\frac{\partial^3 h_t(a,\theta)}{\partial a^2\,\partial\theta}\right\|
\le\bar\beta\sup_{\theta\in\mathcal N}\left\|\frac{\partial^3 h_{t-1}(a,\theta)}{\partial a^2\,\partial\theta}\right\|
+C\delta_T^2\{U_t^3+U_tG_{1,t-1}\},
\]
and Lemma~\ref{lem:contractive-master}(i)(b) gives
\begin{equation}\label{eq:third-order-h-a-bound}
\sup_{\theta\in\mathcal N}\left\|\frac{\partial^3 h_t(a,\theta)}{\partial a^2\,\partial\theta}\right\|\le C\delta_T^2 G_{2,t}.
\end{equation}
Twice differentiating the score  $s_t(a,\theta)=(\partial f/\partial h)(\partial h_t/\partial\theta)+(\partial f/\partial\varphi)e_\varphi$ with respect to $a$ and bounding the partial derivatives of $f$ at $(u_t^2(a),h_t(a,\theta),\varphi)$ by the bound in \eqref{eq:f-local-bounds} gives
\[
\begin{aligned}
\left\|\frac{\partial^2s_t(a,\theta)}{\partial a^2}\right\|
\le C\bigg[&
\left\|\frac{\partial^3h_t(a,\theta)}{\partial a^2\,\partial\theta}\right\|
+\left(\left|\frac{\partial h_t(a,\theta)}{\partial a}\right|+|\widehat u_t^2-u_t^2|\right)
\left\|\frac{\partial^2h_t(a,\theta)}{\partial a\,\partial\theta}\right\|\\
&+\left\{\left|\frac{\partial^2h_t(a,\theta)}{\partial a^2}\right|
+\left(\left|\frac{\partial h_t(a,\theta)}{\partial a}\right|+|\widehat u_t^2-u_t^2|\right)^2\right\}
\left(1+\left\|\frac{\partial h_t(a,\theta)}{\partial\theta}\right\|\right)
\bigg].
\end{aligned}
\]
Substituting \eqref{eq:first-order-a-bounds}, \eqref{eq:second-order-h-a-bound}, and \eqref{eq:third-order-h-a-bound} into the right-hand side yields
\[
\sup_{\theta\in\mathcal N}\sup_{0\le a\le1}\left\|\frac{\partial^2 s_t(a,\theta)}{\partial a^2}\right\|
\le C\delta_T^2\{G_{2,t}+U_tG_{1,t}+U_t^3\}\le C\delta_T^2 J_t,
\]
which proves \eqref{eq:score-directional-bound}.

Finally, to prove \eqref{eq:local-hessian-envelope} and the almost-sure continuity of the stationary Hessian on $\mathcal N$, we apply the local constant bounds for the partial derivatives of $f$ in \eqref{eq:f-local-bounds} to the explicit Hessian  \eqref{eq:likelihood-hessian-form}: for all $\theta\in\mathcal N$,
\[
\left\|\frac{\partial^2\ell_t(\theta)}{\partial\theta\,\partial\theta^\top}\right\|
\le C\left(\left\|\frac{\partial^2 h_t(\theta)}{\partial\theta\,\partial\theta^\top}\right\|
+\left\|\frac{\partial h_t(\theta)}{\partial\theta}\right\|^2
+\left\|\frac{\partial h_t(\theta)}{\partial\theta}\right\|+1\right)
\le C\{G_{1,t}+U_t^2\}.
\]
Since $G_{1,t},U_t^2\in L^{q/2}\subset L^1$ ($q>4$), taking the supremum over $\theta\in\mathcal N$ and taking expectations immediately yields \eqref{eq:local-hessian-envelope}. Moreover, for each integer $m\ge1$, let $h_{t,m}(\theta)$ denote the truncated volatility process at time $t$ obtained by starting the recursion at time $t-m$ from the fixed constant $h_{t-m,0}=\underline\omega>0$. Because $h_{t,m}(\theta)$ is a finite $m$-fold composition of the smooth function $g$, the maps $\theta\mapsto h_{t,m}(\theta)$, $\theta\mapsto\partial_\theta h_{t,m}(\theta)$, and $\theta\mapsto\partial^2_{\theta\theta^\top}h_{t,m}(\theta)$ are continuous on $\Theta$. By stationarity and \eqref{eq:init-h}--\eqref{eq:init-d2h} in Lemma~\ref{lem:initial-values},
\[
\begin{aligned}
\sum_{m=1}^{\infty}E\Bigg[\sup_{\theta\in\Theta}\Bigg\{&\bigl|h_{t,m}(\theta)-h_t(\theta)\bigr|
+\left\|\frac{\partial h_{t,m}(\theta)}{\partial\theta}-\frac{\partial h_t(\theta)}{\partial\theta}\right\|\\
&+\left\|\frac{\partial^2 h_{t,m}(\theta)}{\partial\theta\,\partial\theta^\top}-\frac{\partial^2 h_t(\theta)}{\partial\theta\,\partial\theta^\top}\right\|\Bigg\}\Bigg]
\le C\sum_{m=1}^{\infty}\rho^m<\infty.
\end{aligned}
\]
By the monotone convergence theorem, the nonnegative series inside the expectation converges almost surely, which implies that $h_{t,m}(\theta)$ and its first and second parameter derivatives converge uniformly on $\Theta$ almost surely as $m\to\infty$ to $h_t(\theta)$ and its parameter derivatives. Being the uniform limits of continuous functions, $\theta\mapsto h_t(\theta)$, $\theta\mapsto\partial_\theta h_t(\theta)$, and $\theta\mapsto\partial^2_{\theta\theta^\top}h_t(\theta)$ are almost surely continuous on $\Theta$. Substituting these continuous processes into \eqref{eq:likelihood-hessian-form} establishes that $\theta\mapsto\partial^2\ell_t(\theta)/(\partial\theta\,\partial\theta^\top)$ is almost surely continuous on $\mathcal N$.
\end{proof}

Recall that $\partial s_{t}(\theta_0)/\partial (u_{i}^2)$ from Lemma~\ref{lem:score-regularity}
measures how changing $u_i^2$ changes the score at time $t$.
The effect decreases geometrically as $t-i$ increases. Define the mean coefficient of the first-step error by
\begin{equation}
C_\tau=E\left\{\sum_{j=0}^\infty
\frac{\partial s_{t}(\theta_0)}{\partial (u_{t-j}^2)}u_{t-j}^2\right\}.
\label{eq:Ctau}
\end{equation}
 The series converges in $L^p$, with $p=q/2>2$,
by Lemma~\ref{lem:score-regularity}. 
 The next lemma controls the effect of $u_i^2$ to the score at $t=i,i+1,\ldots$.  Define 
$A_i=\sum_{j\ge0}u_i^2(\partial s_{i+j}(\theta_0)/\partial (u_{i}^2))$.
Although $A_i$ depends on future innovations, the effect of a
distant innovation decreases geometrically. 

\begin{lemma}
\label{lem:loading-local}
Under the conditions of Lemma~\ref{lem:score-regularity},
$A_i\in L^p$, where $p=q/2>2$, $EA_i=C_\tau$, and
\[
\|A_k-A_k^*\|_2\le C\rho^{|k|},\qquad k\in\mathbb Z,
\]
for some $C<\infty$ and $\rho\in(0,1)$.
Here $A_k^*$ is obtained by replacing only $\epsilon_0$ by an
independent copy.
\end{lemma}

\begin{proof}
Let $ W_{i,j}=u_i^2(\partial s_{i+j}(\theta_0)/\partial (u_{i}^2))$.
Lemma~\ref{lem:score-regularity} gives
$\|W_{i,j}\|_p\le C\varrho^j$, so its summation is also  $L^p$.
By stationarity,
$EW_{i,j}=E(u_{t-j}^2(\partial s_{t}(\theta_0)/\partial (u_{t-j}^2)))$, and summing over $j$
gives $EA_i=C_\tau$.
If $k<0$, replacing $\epsilon_0$ leaves $W_{k,j}$
unchanged for $j<-k$. Thus
\[
\|A_k-A_k^*\|_2
\le 2\sum_{j\ge-k}\|W_{k,j}\|_2
\le C\varrho^{|k|}.
\]

For $k\ge0$, we use the observation-derivative recursions \eqref{eq:dh-du2-recursion} and \eqref{eq:d2h-dtheta-du2-recursion} in the proof of Lemma~\ref{lem:score-regularity}, evaluated at $\theta=\theta_0$. By \eqref{eq:dh-du2-deterministic} and \eqref{eq:d2h-du2-moment}, for some $\gamma\in(\bar\beta,1)$,
\begin{equation}\label{eq:obs-deriv-marginal-bounds}
\left|\frac{\partial h_{k+j}(\theta_0)}{\partial(u_k^2)}\right|\le C\gamma^j,\qquad
\left\|\frac{\partial^2 h_{k+j}(\theta_0)}{\partial\theta\,\partial(u_k^2)}\right\|_q\le C\gamma^j,\qquad j\ge0.
\end{equation}
Here the second bound follows directly from iterating \eqref{eq:d2h-dtheta-du2-recursion} in $L^q$ using $\|\partial_\theta\{\beta(\partial g/\partial b)\}\|_q\le C\|U_{k+j}\|_q\le C$. Let a superscript $*$ denote a quantity evaluated using the single-shock coupled process (where only $\epsilon_0$ is replaced by $\epsilon_0^\circ$). By Lemma~\ref{lem:augmented-gmc} and \eqref{eq:gmc-interpolation-bound} in Lemma~\ref{lem:contractive-master}(ii)(a), for some $\eta\in(0,1)$ and all $t\ge1$,
\begin{equation}\label{eq:single-shock-process-bounds}
\|u_t^2-(u_t^*)^2\|_q+\|h_t(\theta_0)-h_t^*(\theta_0)\|_q\le C\eta^t,\qquad
\left\|\frac{\partial h_t(\theta_0)}{\partial\theta}-\frac{\partial h_t^*(\theta_0)}{\partial\theta}\right\|_{q/2}\le C\eta^t.
\end{equation}
Applying the Lipschitz bounds \eqref{eq:g-deriv-lipschitz} to \eqref{eq:dtheta-beta-dgb} and using \eqref{eq:single-shock-process-bounds} with H\"older's inequality yields, for all $t\ge1$,
\begin{equation}\label{eq:beta-dgb-coupling}
\left\|\beta_0\frac{\partial g}{\partial b}-\beta_0\left(\frac{\partial g}{\partial b}\right)^*\right\|_q\le C\eta^t,\qquad
\left\|\frac{\partial}{\partial\theta}\left(\beta\frac{\partial g}{\partial b}\right)-\left[\frac{\partial}{\partial\theta}\left(\beta\frac{\partial g}{\partial b}\right)\right]^*\right\|_{q/2}\le C\eta^t.
\end{equation}
For $t=k+j\ge k+2$, subtracting the coupled versions from \eqref{eq:dh-du2-recursion} and \eqref{eq:d2h-dtheta-du2-recursion} gives
\begin{align*}
\frac{\partial h_t}{\partial(u_k^2)}-\left(\frac{\partial h_t}{\partial(u_k^2)}\right)^*
={}&\beta_0\frac{\partial g}{\partial b}\left\{\frac{\partial h_{t-1}}{\partial(u_k^2)}-\left(\frac{\partial h_{t-1}}{\partial(u_k^2)}\right)^*\right\}
+\beta_0\left\{\frac{\partial g}{\partial b}-\left(\frac{\partial g}{\partial b}\right)^*\right\}\left(\frac{\partial h_{t-1}}{\partial(u_k^2)}\right)^*,\\[3pt]
\frac{\partial^2 h_t}{\partial\theta\,\partial(u_k^2)}-\left(\frac{\partial^2 h_t}{\partial\theta\,\partial(u_k^2)}\right)^*
={}&\beta_0\frac{\partial g}{\partial b}\left\{\frac{\partial^2 h_{t-1}}{\partial\theta\,\partial(u_k^2)}-\left(\frac{\partial^2 h_{t-1}}{\partial\theta\,\partial(u_k^2)}\right)^*\right\}\\
&+\beta_0\left\{\frac{\partial g}{\partial b}-\left(\frac{\partial g}{\partial b}\right)^*\right\}\left(\frac{\partial^2 h_{t-1}}{\partial\theta\,\partial(u_k^2)}\right)^*\\
&+\frac{\partial}{\partial\theta}\left(\beta\frac{\partial g}{\partial b}\right)\left\{\frac{\partial h_{t-1}}{\partial(u_k^2)}-\left(\frac{\partial h_{t-1}}{\partial(u_k^2)}\right)^*\right\}\\
&+\left\{\frac{\partial}{\partial\theta}\left(\beta\frac{\partial g}{\partial b}\right)-\left[\frac{\partial}{\partial\theta}\left(\beta\frac{\partial g}{\partial b}\right)\right]^*\right\}\left(\frac{\partial h_{t-1}}{\partial(u_k^2)}\right)^*.
\end{align*}
Taking the $L^q$ and $L^{q/2}$ norms of these two identities and applying \eqref{eq:obs-deriv-marginal-bounds} and \eqref{eq:beta-dgb-coupling} with H\"older's inequality gives
\begin{align*}
\left\|\frac{\partial h_{k+j}}{\partial(u_k^2)}-\left(\frac{\partial h_{k+j}}{\partial(u_k^2)}\right)^*\right\|_q
&\le\bar\beta\left\|\frac{\partial h_{k+j-1}}{\partial(u_k^2)}-\left(\frac{\partial h_{k+j-1}}{\partial(u_k^2)}\right)^*\right\|_q
+C\gamma^j\eta^{k+j},\\
\left\|\frac{\partial^2 h_{k+j}}{\partial\theta\,\partial(u_k^2)}-\left(\frac{\partial^2 h_{k+j}}{\partial\theta\,\partial(u_k^2)}\right)^*\right\|_{q/2}
&\le\bar\beta\left\|\frac{\partial^2 h_{k+j-1}}{\partial\theta\,\partial(u_k^2)}-\left(\frac{\partial^2 h_{k+j-1}}{\partial\theta\,\partial(u_k^2)}\right)^*\right\|_{q/2}\\
&\quad+C\|U_{k+j}\|_q\left\|\frac{\partial h_{k+j-1}}{\partial(u_k^2)}-\left(\frac{\partial h_{k+j-1}}{\partial(u_k^2)}\right)^*\right\|_q
+C\gamma^j\eta^{k+j}.
\end{align*}
Checking the initial values at $j=0,1$ directly from \eqref{eq:dh-du2-recursion} and \eqref{eq:g-deriv-lipschitz}, and iterating in $j\ge0$ via Lemma~\ref{lem:contractive-master}(i)(a) (after enlarging $\gamma\in(\bar\beta,1)$ if necessary), yields
\begin{equation}\label{eq:obs-deriv-coupled-bounds}
\left\|\frac{\partial h_{k+j}}{\partial(u_k^2)}-\left(\frac{\partial h_{k+j}}{\partial(u_k^2)}\right)^*\right\|_q
+\left\|\frac{\partial^2 h_{k+j}}{\partial\theta\,\partial(u_k^2)}-\left(\frac{\partial^2 h_{k+j}}{\partial\theta\,\partial(u_k^2)}\right)^*\right\|_{q/2}
\le C\gamma^j\eta^k,\qquad k,j\ge0.
\end{equation}

Next, by \eqref{eq:ds-du2-identity}, $W_{k,j}=u_k^2\{\partial s_{k+j}(\theta_0)/\partial(u_k^2)\}$ takes the explicit form
\begin{equation}\label{eq:Wkj-explicit}
\begin{aligned}
W_{k,j}=u_k^2\Bigg[&
\frac{\partial f}{\partial h}\frac{\partial^2 h_{k+j}}{\partial\theta\,\partial(u_k^2)}
+\left(\frac{\partial^2 f}{\partial h^2}\frac{\partial h_{k+j}}{\partial\theta}+\frac{\partial^2 f}{\partial\varphi\,\partial h}e_\varphi\right)\frac{\partial h_{k+j}}{\partial(u_k^2)}\\
&+\left(\frac{\partial^2 f}{\partial h\,\partial v}\frac{\partial h_{k+j}}{\partial\theta}+\frac{\partial^2 f}{\partial\varphi\,\partial v}e_\varphi\right)\mathbf1\{j=0\}
\Bigg],
\end{aligned}
\end{equation}
where the partial derivatives of $f$ are evaluated at $(u_{k+j}^2,h_{k+j}(\theta_0),\varphi_0)$. Subtract the coupled version $W_{k,j}^*$ from \eqref{eq:Wkj-explicit} by changing one factor at a time. By \eqref{eq:f-deriv-lipschitz} and \eqref{eq:single-shock-process-bounds}, the differences of the partial derivatives of $f$ at time $k+j$ satisfy
\[
\left\|\frac{\partial f}{\partial h}-\left(\frac{\partial f}{\partial h}\right)^*\right\|_{q/2}
+\left\|\frac{\partial^2 f}{\partial h^2}-\left(\frac{\partial^2 f}{\partial h^2}\right)^*\right\|_{q/2}
+\left\|\frac{\partial^2 f}{\partial\varphi\,\partial h}-\left(\frac{\partial^2 f}{\partial\varphi\,\partial h}\right)^*\right\|_{q/2}
\le C\eta^{k+j}.
\]
Combining this with \eqref{eq:obs-deriv-marginal-bounds}, \eqref{eq:single-shock-process-bounds}, and \eqref{eq:obs-deriv-coupled-bounds} via H\"older's inequality yields
\[
\|W_{k,j}-W_{k,j}^*\|_{q/4}\le C\gamma^j\eta^k.
\]
Finally, since $\|W_{k,j}-W_{k,j}^*\|_{q/2}\le\|W_{k,j}\|_{q/2}+\|W_{k,j}^*\|_{q/2}\le C\gamma^j$ by \eqref{eq:path-derivative-decay}, applying the moment interpolation \eqref{eq:gmc-interpolation-bound} in Lemma~\ref{lem:contractive-master}(ii)(a) with $(r,p,a)=(q/4,2,q/2)$ when $4<q<8$ (and norm monotonicity $\|W_{k,j}-W_{k,j}^*\|_2\le\|W_{k,j}-W_{k,j}^*\|_{q/4}$ when $q\ge8$) gives $\|W_{k,j}-W_{k,j}^*\|_2\le C\gamma^j\widetilde\eta^k$ for some $\widetilde\eta\in(0,1)$. Minkowski's inequality then yields $\|A_k-A_k^*\|_2\le\sum_{j=0}^\infty\|W_{k,j}-W_{k,j}^*\|_2\le C\widetilde\eta^k\sum_{j=0}^\infty\gamma^j\le C\widetilde\eta^k$ for all $k\ge0$.
\end{proof}

The coefficient $C_\tau$ can also be written as an expectation of
likelihood derivatives. This representation avoids the classical HAC estimator   when estimating $C_\tau$.
\begin{lemma}
\label{lem:Ctau-estimation}
Suppose the innovations are standard Gaussian,
Assumptions~\ref{Asm1}--\ref{Asm6} hold, and $\theta_0$
is interior to $\Theta$. In derivatives with respect to $\omega$
below, treat the intercept as independent of
$(\alpha,\beta,\varphi)$, hold the observations fixed, and
evaluate at $\omega=\omega(\theta)$ afterwards.
Derivatives with respect to $\theta$ and $\varphi$ retain the
normalization $\omega(\theta)=1-\alpha-\beta-3\varphi+2\beta\varphi$.
Then
\begin{equation}
C_\tau=-E\left[
\frac{\partial\ell_t(\theta_0)}{\partial\theta}
\left\{
\varphi_0\frac{\partial\ell_t(\theta_0)}{\partial\varphi}
+(1-\alpha_0-\beta_0)
\frac{\partial\ell_t(\theta_0)}{\partial\omega}
\right\}\right].
\label{eq:Ctau-score-product}
\end{equation}
The additional derivative is obtained from
\begin{align}
\frac{\partial h_t(\theta)}{\partial\omega}
&=\frac{\partial g}{\partial b}
\left\{1+\beta\frac{\partial h_{t-1}(\theta)}{\partial\omega}\right\},
\label{eq:intercept-process-derivative}\\
\frac{\partial\ell_t(\theta)}{\partial\omega}
&=\left\{-\frac{u_t^2}{2h_t^2(\theta)}
+\frac{2h_t(\theta)}{h_t^2(\theta)+\varphi u_t^2}
-\frac{3}{2h_t(\theta)}\right\}
\frac{\partial h_t(\theta)}{\partial\omega}.
\label{eq:intercept-likelihood-derivative}
\end{align}
In \eqref{eq:intercept-process-derivative}, $\partial g/\partial b$
is evaluated at $(b_{t-1}(\theta),u_t^2;\varphi)$.
\end{lemma}

\begin{proof}
We first derive \eqref{eq:Ctau-score-product}.
For $c>0$ near one, replace every squared input $u_i^2$,
$i\le t$, by $u_i^2/c$, and denote the resulting stationary
volatility process by $h_t(c,\theta)$. Multiplying the long-run scale by $c$
gives the conditional negative log-likelihood
\[
\ell_t(c,\theta)=\frac12\log c+
f\{u_t^2/c,h_t(c,\theta),\varphi\},
\]
where $f$ is defined in \eqref{eq:g-and-f-def}.
At $c=1$, this is $\ell_t(\theta)$. The chain rule gives
\begin{equation}
\left.
\frac{\partial^2\ell_t(c,\theta_0)}
{\partial\theta\,\partial\log c}\right|_{c=1}
=-\sum_{j=0}^{\infty}
u_{t-j}^2\frac{\partial s_t(\theta_0)}{\partial(u_{t-j}^2)}.
\label{eq:Ctau-scale-cross-derivative}
\end{equation}
The geometric bound in Lemma~\ref{lem:score-regularity}
makes this series summable in $L^{q/2}$ and justifies taking
expectations. Hence the expectation in
\eqref{eq:Ctau-scale-cross-derivative} is $-C_\tau$.

To express the scale derivative using the parameter derivatives,
observe that
\[
g(cb,cu^2;c\varphi)=c\,g(b,u^2;\varphi),
\qquad
f(cu^2,ch,c\varphi)=f(u^2,h,\varphi)+\tfrac12\log c.
\]
It follows that $\ell_t(c,\theta)$ is also the likelihood
evaluated at the original observations with intercept
$c\omega(\theta)$ and contemporaneous coefficient $c\varphi$,
while $\alpha$ and $\beta$ remain fixed.
Since the normalized derivative with respect to $\varphi$
already includes the change in $\omega(\theta)$, the chain rule yields
\[
\begin{aligned}
\left.\frac{\partial\ell_t(c,\theta)}{\partial\log c}\right|_{c=1}
&=\varphi\frac{\partial\ell_t(\theta)}{\partial\varphi}
+\left\{\omega(\theta)
-\varphi\frac{\partial\omega(\theta)}{\partial\varphi}\right\}
\frac{\partial\ell_t(\theta)}{\partial\omega}\\
&=\varphi\frac{\partial\ell_t(\theta)}{\partial\varphi}
+(1-\alpha-\beta)
\frac{\partial\ell_t(\theta)}{\partial\omega},
\end{aligned}
\]
where $\partial\omega(\theta)/\partial\varphi=-3+2\beta$.
Under the correctly specified Gaussian conditional density,
differentiating its integral, which equals one, gives the
mixed information identity
\[
E\left[
\left.\frac{\partial^2\ell_t(c,\theta_0)}
{\partial\theta\,\partial\log c}\right|_{c=1}
\right]
=
E\left[
\frac{\partial\ell_t(\theta_0)}{\partial\theta}
\left.\frac{\partial\ell_t(c,\theta_0)}{\partial\log c}
\right|_{c=1}
\right].
\]
The moment bounds in Lemmas~\ref{lem:process-moment-bounds}
and~\ref{lem:score-regularity} justify differentiation under
the conditional integral. Combining the last identity with
\eqref{eq:Ctau-scale-cross-derivative} proves
\eqref{eq:Ctau-score-product}.

\end{proof}

Finally, define
\begin{equation}
 \psi_t=-C_\tau z_t,
 \qquad
 \xi_t=s_t(\theta_0)+\psi_t
       =s_t(\theta_0)-C_\tau z_t,
 \label{eq:xi-definition}
\end{equation}
and
\begin{equation*}
 V=\sum_{j=-\infty}^{\infty}
 \operatorname{Cov}(\xi_t,\xi_{t-j}).
\end{equation*}
We then give a simplified expression for $V.$
\begin{lemma}
\label{lem:covariance-simplification}
Suppose the conditions of Theorem~\ref{thm:asymptotic-normality} hold. 
Define
\[
\zeta_t=u_t^2-b_{t-1}-3\varphi,
\qquad
\chi_t=
\frac{(1-\beta)\zeta_t+\beta\varphi(\epsilon_t^2-1)}
{1-\alpha-\beta}.
\]
Then $\{\zeta_t\}$ and $\{\chi_t\}$ are martingale differences
with respect to $\{\mathcal F_t\}$, and
\begin{equation}
z_t=(\alpha+\beta)z_{t-1}
+\zeta_t-\beta\zeta_{t-1}
+\beta\varphi(\epsilon_{t-1}^2-1).
\label{eq:z-arma}
\end{equation}
Moreover,
\begin{equation}
\frac1{\sqrt T}\sum_{t=1}^T(z_t-\chi_t)
\to0
\quad\text{in }L^2,
\qquad
E\{s_t(\theta_0)\chi_t\}=0.
\label{eq:z-martingale-equivalence}
\end{equation}
Consequently,
\begin{equation}
V=H+\Omega_z C_\tau C_\tau^\top,
\qquad
\Sigma=H^{-1}
+\Omega_z H^{-1}C_\tau C_\tau^\top H^{-1},
\label{eq:covariance-simplification}
\end{equation}
where $\Omega_z=E(\chi_t^2)$. In the Gaussian case, we further have that 
\begin{align}
\operatorname{Var}(b_t)
&=
\frac{
2\{\alpha+(3\alpha+\beta)\varphi\}^2
+24\alpha^2\varphi^2
}{
1-3\alpha^2-2\alpha\beta-\beta^2
},
\label{eq:b-variance}\\
\Omega_z
&=
\frac{
2(1-\beta)^2\operatorname{Var}(b_t)
+2\{1-\beta+(3-2\beta)\varphi\}^2
+24(1-\beta)^2\varphi^2
}{
(1-\alpha-\beta)^2
}.
\label{eq:Omega-z-explicit}
\end{align}
 
\end{lemma}

\begin{proof}
We divide the proof into four steps: (1)~deriving the ARMA representation \eqref{eq:z-arma} and the  decomposition for $z_t-\chi_t$; (2)~establishing the orthogonality $E\{s_t(\theta_0)\chi_t\}=0$ via a general conditional score identity; (3)~deriving the rank-one covariance update \eqref{eq:covariance-simplification}; and (4)~computing the closed-form Gaussian variances \eqref{eq:b-variance}--\eqref{eq:Omega-z-explicit} via orthogonal polynomial innovations.

\medskip
\noindent\emph{Step 1: ARMA representation}
Since $u_t^2=b_{t-1}\epsilon_t^2+\varphi\epsilon_t^4$ and $\epsilon_t\sim\mathcal N(0,1)$ is independent of $\mathcal F_{t-1}$, we have $E(u_t^2\mid\mathcal F_{t-1})=b_{t-1}+3\varphi$ and $E(\epsilon_t^2-1\mid\mathcal F_{t-1})=0$. Thus $\zeta_t=u_t^2-E(u_t^2\mid\mathcal F_{t-1})$ and $\chi_t$ are martingale differences with respect to $\{\mathcal F_t\}$.

From $z_t=u_t^2-1$ and $\zeta_t=u_t^2-b_{t-1}-3\varphi$, we have $z_t-\zeta_t=b_{t-1}+3\varphi-1$. Moreover, eliminating $b_{t-2}$ between $h_{t-1}=b_{t-2}+\varphi\epsilon_{t-1}^2$ and $u_{t-1}^2=b_{t-2}+3\varphi+\zeta_{t-1}$ gives
\[
h_{t-1}=u_{t-1}^2-\zeta_{t-1}+\varphi(\epsilon_{t-1}^2-1)-2\varphi.
\]
Substituting this expression and the normalization $\omega=1-\alpha-\beta-3\varphi+2\beta\varphi$ into $b_{t-1}=\omega+\alpha u_{t-1}^2+\beta h_{t-1}$ yields
\[
z_t-\zeta_t=b_{t-1}+3\varphi-1=(\alpha+\beta)z_{t-1}-\beta\zeta_{t-1}+\beta\varphi(\epsilon_{t-1}^2-1),
\]
which is \eqref{eq:z-arma}. Subtracting $(\alpha+\beta)z_t-\beta\zeta_t+\beta\varphi(\epsilon_t^2-1)$ from both sides of \eqref{eq:z-arma} and dividing by $1-\alpha-\beta>0$ gives 
\[
z_t-\chi_t=-\frac{1}{1-\alpha-\beta}\,\Delta\bigl\{(\alpha+\beta)z_t-\beta\zeta_t+\beta\varphi(\epsilon_t^2-1)\bigr\},
\qquad \Delta x_t:=x_t-x_{t-1}.
\]
Summing over $t=1,\ldots,T$ yields
\[
(1-\alpha-\beta)\sum_{t=1}^T(z_t-\chi_t)
=(\alpha+\beta)(z_0-z_T)-\beta(\zeta_0-\zeta_T)+\beta\varphi(\epsilon_0^2-\epsilon_T^2).
\]
Since $z_t,\zeta_t,\epsilon_t^2\in L^2$ by stationarity and Assumption~\ref{Asm6}, dividing by $\sqrt T$ proves $T^{-1/2}\sum_{t=1}^T(z_t-\chi_t)\to0$ in $L^2$, and the martingale-difference property of $\chi_t$ implies $\Omega_z=E(\chi_t^2)$.

\medskip
\noindent\emph{Step 2: Orthogonality $E\{s_t(\theta_0)\chi_t\}=0$ via the conditional score identity.}
Let $f_t(r\mid\mathcal F_{t-1};\theta)$ be the conditional density of $r_t$, so that $s_t(\theta)=-\partial_\theta\log f_t(r_t\mid\mathcal F_{t-1};\theta)$. For any smooth function $\Psi(u_t,\theta)$ whose value and $\theta$-derivative (holding $u_t$ fixed) are dominated on $\mathcal N$ by an integrable envelope, differentiating $\int\Psi(r/\sqrt{\tau_t},\theta)f_t(r\mid\mathcal F_{t-1};\theta)\,dr=E_\theta\{\Psi(u_t,\theta)\mid\mathcal F_{t-1}\}$ at $\theta=\theta_0$ gives the general conditional score identity
\begin{equation}\label{eq:general-bartlett}
E\bigl\{s_t(\theta_0)\Psi(u_t,\theta_0)\;\big|\;\mathcal F_{t-1}\bigr\}
=E\left\{\frac{\partial\Psi(u_t,\theta_0)}{\partial\theta}\;\middle|\;\mathcal F_{t-1}\right\}
-\left.\frac{\partial}{\partial\theta}E_\theta\bigl\{\Psi(u_t,\theta)\;\big|\;\mathcal F_{t-1}\bigr\}\right|_{\theta=\theta_0},
\end{equation}
where differentiation under the integral is justified by Lemmas~\ref{lem:process-moment-bounds} and~\ref{lem:score-regularity}. Applying \eqref{eq:general-bartlett} to three choices of $\Psi(u_t,\theta)$ yields:
\begin{enumerate}[1.]
\item Taking $\Psi(u_t,\theta)\equiv1$ gives $E\{s_t(\theta_0)\mid\mathcal F_{t-1}\}=0$, so $\{s_t(\theta_0)\}$ is a martingale difference sequence.
\item Taking $\Psi(u_t,\theta)=u_t^2-b_{t-1}(\theta)-3\varphi$ (which equals $\zeta_t$ at $\theta_0$ and satisfies $E_\theta\{\Psi(u_t,\theta)\mid\mathcal F_{t-1}\}\equiv0$ for all $\theta$) gives
\[
E\{s_t(\theta_0)\zeta_t\mid\mathcal F_{t-1}\}=-\frac{\partial b_{t-1}(\theta_0)}{\partial\theta}-3e_\varphi.
\]
\item Taking $\Psi(u_t,\theta)=u_t^2/h_t(\theta)-1$ (which equals $\epsilon_t^2-1$ at $\theta_0$ and satisfies $E_\theta\{\Psi(u_t,\theta)\mid\mathcal F_{t-1}\}\equiv0$ for all $\theta$) gives
\[
E\{s_t(\theta_0)(\epsilon_t^2-1)\mid\mathcal F_{t-1}\}
=-E\left\{\frac{\epsilon_t^2}{h_t(\theta_0)}\frac{\partial h_t(\theta_0)}{\partial\theta}\;\middle|\;\mathcal F_{t-1}\right\}.
\]
\end{enumerate}
To relate these two expectations, differentiate the quadratic identity $h_t(\theta)=b_{t-1}(\theta)+\varphi u_t^2/h_t(\theta)$ with respect to $\theta$ at $\theta_0$ (holding $u_t^2$ fixed) to obtain
\[
-\varphi\frac{\epsilon_t^2}{h_t(\theta_0)}\frac{\partial h_t(\theta_0)}{\partial\theta}
=\frac{\partial h_t(\theta_0)}{\partial\theta}-\frac{\partial b_{t-1}(\theta_0)}{\partial\theta}-\epsilon_t^2 e_\varphi.
\]
Meanwhile, taking unconditional expectations in $\partial_\theta b_{t-1}(\theta_0)=(-1,-1,-3+2\beta)^\top+e_\alpha u_{t-1}^2+e_\beta h_{t-1}+\beta\partial_\theta h_{t-1}(\theta_0)$ and using $E(u_{t-1}^2)=1$ and $E(h_{t-1})=1-2\varphi$ gives
\[
E\left\{\frac{\partial b_{t-1}(\theta_0)}{\partial\theta}\right\}
=(-3+2\beta)e_\varphi+\beta E\left\{\frac{\partial h_t(\theta_0)}{\partial\theta}\right\}.
\]
Substituting this into the unconditional expectations of items~2 and~3 above shows that both covariances are collinear with $D_h:=E\{\partial_\theta h_t(\theta_0)\}+2e_\varphi$:
\[
E\{s_t(\theta_0)\zeta_t\}=-\beta D_h,\qquad
\varphi E\{s_t(\theta_0)(\epsilon_t^2-1)\}=(1-\beta)D_h.
\]
Therefore, $(1-\beta)E\{s_t(\theta_0)\zeta_t\}+\beta\varphi E\{s_t(\theta_0)(\epsilon_t^2-1)\}=0$, which proves $E\{s_t(\theta_0)\chi_t\}=0$.

\medskip
\noindent\emph{Step 3: Long-run covariance simplification.}
Since both $\{s_t(\theta_0)\}$ and $\{\chi_t\}$ are stationary martingale difference sequences with respect to $\{\mathcal F_t\}$ and are mutually orthogonal at lag zero ($E\{s_t(\theta_0)\chi_t\}=0$), all auto- and cross-covariances of $s_t(\theta_0)$ and $\chi_t$ at nonzero lags vanish. By Step~1, $T^{-1/2}\sum_{t=1}^T\xi_t=T^{-1/2}\sum_{t=1}^T\{s_t(\theta_0)-C_\tau\chi_t\}+o_{L^2}(1)$. Moreover, since $q>4$, Proposition~\ref{prop:rt-gmc} and Lemma~\ref{lem:augmented-gmc}(iii) imply that $\{\xi_t\}$ is $L^2$ GMC, so Lemma~\ref{lem:contractive-master}(ii)(b) gives
\[
\begin{aligned}
V=\sum_{j=-\infty}^{\infty}\operatorname{Cov}(\xi_t,\xi_{t-j})
&=\lim_{T\to\infty}\operatorname{Var}\!\left(\frac1{\sqrt T}\sum_{t=1}^T\xi_t\right)\\
&=E\bigl[\{s_t(\theta_0)-C_\tau\chi_t\}\{s_t(\theta_0)-C_\tau\chi_t\}^\top\bigr]
=E\{s_t(\theta_0)s_t(\theta_0)^\top\}+\Omega_z C_\tau C_\tau^\top.
\end{aligned}
\]
Differentiating $\int f_t(r\mid\mathcal F_{t-1};\theta)\,dr=1$ twice at $\theta=\theta_0$ gives the conditional information equality
\[
E\bigl\{s_t(\theta_0)s_t(\theta_0)^\top\;\big|\;\mathcal F_{t-1}\bigr\}
=E\left\{\frac{\partial^2\ell_t(\theta_0)}{\partial\theta\,\partial\theta^\top}\;\middle|\;\mathcal F_{t-1}\right\},
\]
and taking expectations yields $E\{s_t(\theta_0)s_t(\theta_0)^\top\}=H$. Substituting this into $\Sigma=H^{-1}VH^{-\top}$ proves \eqref{eq:covariance-simplification}.

\medskip
\noindent\emph{Step 4: Explicit Gaussian variances via orthogonal polynomial innovations.}
For $\epsilon_t\sim\mathcal N(0,1)$, define the centered Hermite-type polynomials
\[
H_{1,t}:=\epsilon_t^2-1,\qquad H_{2,t}:=\epsilon_t^4-6\epsilon_t^2+3.
\]
Direct Gaussian moment calculation gives $E(H_{1,t})=E(H_{2,t})=E(H_{1,t}H_{2,t})=0$, $E(H_{1,t}^2)=2$, and $E(H_{2,t}^2)=24$. Using $\bar b:=E(b_t)=1-3\varphi$ and $\epsilon_t^4=H_{2,t}+6H_{1,t}+3$, we can expand both the centered state $b_t-\bar b=(\alpha\epsilon_t^2+\beta)b_{t-1}+\omega+\varphi(\alpha\epsilon_t^4+\beta\epsilon_t^2)-\bar b$ and the martingale difference $(1-\alpha-\beta)\chi_t=(1-\beta)\{(b_{t-1}+\varphi\epsilon_t^2)\epsilon_t^2-b_{t-1}-3\varphi\}+\beta\varphi(\epsilon_t^2-1)$ in the mutually orthogonal basis $\{b_{t-1}-\bar b,\,H_{1,t},\,H_{2,t}\}$:
\begin{align*}
b_t-\bar b
&=(\alpha\epsilon_t^2+\beta)(b_{t-1}-\bar b)
+\{\alpha+(3\alpha+\beta)\varphi\}H_{1,t}
+\alpha\varphi H_{2,t},\\
(1-\alpha-\beta)\chi_t
&=(1-\beta)(b_{t-1}-\bar b)H_{1,t}
+\{1-\beta+(3-2\beta)\varphi\}H_{1,t}
+(1-\beta)\varphi H_{2,t}.
\end{align*}
Since $b_{t-1}-\bar b$ has mean zero and is independent of $\epsilon_t$, while $E(\alpha\epsilon_t^2+\beta)^2=3\alpha^2+2\alpha\beta+\beta^2<1$ (by Proposition~\ref{prop:rt-gmc} with $p=2$), taking variances in the first equation yields
\[
\operatorname{Var}(b_t)
=(3\alpha^2+2\alpha\beta+\beta^2)\operatorname{Var}(b_t)
+2\{\alpha+(3\alpha+\beta)\varphi\}^2+24\alpha^2\varphi^2,
\]
which solves to \eqref{eq:b-variance}. Similarly, since $(b_{t-1}-\bar b)H_{1,t}$, $H_{1,t}$, and $H_{2,t}$ are mutually orthogonal with variances $2\operatorname{Var}(b_t)$, $2$, and $24$, taking the second moment of the second equation gives
\[
(1-\alpha-\beta)^2\Omega_z
=(1-\alpha-\beta)^2 E(\chi_t^2)
=2(1-\beta)^2\operatorname{Var}(b_t)+2\{1-\beta+(3-2\beta)\varphi\}^2+24(1-\beta)^2\varphi^2,
\]
proving \eqref{eq:Omega-z-explicit}. This generalizes the ARMA reduction in equations~(A.8)--(A.9) of \citet{jiang2021adaptive} by accounting for the contemporaneous feedback term $\beta\varphi(\epsilon_{t-1}^2-1)$ in \eqref{eq:z-arma}.
\end{proof}
\begin{proof}[Proof of Theorem~\ref{thm:asymptotic-normality}]
The proof consists of three steps. We first show that the sample Hessian matrix converges
to $H$. We then derive the asymptotic expansion of the score function by replacing the long-run component with its estimates in the first step. Finally, we apply the central limit theorem for the martingale difference sequence. 

\paragraph{Step 1.}
Because $\theta_0$ is an interior point, by the first order condition, we have that 
\[
 0=\frac{\partial\widehat L_T(\widehat\theta_T)}{\partial\theta}.
\]
By the Taylor expansion, we have that
\[
0=\frac1{\sqrt T}\frac{\partial \widehat L_T(\theta_0)}{\partial \theta}
+ H_T^*\sqrt T(\widehat\theta_T-\theta_0),
\]
where  
\[
 H_T^*
=
\int_0^1\frac1T\frac{\partial^2\widehat L_T\{\theta_0+s(\widehat\theta_T-\theta_0)\}}{\partial\theta\,\partial\theta^\top}\,ds.\]
Let $\mathcal N(\theta_0)\subset\mathcal N$ be a compact neighborhood of $\theta_0$ contained in the interior of $\Theta$. By   
Lemma \ref{lem:initial-values} and \ref{lem:criterion-differences}, we can show that 
\[
 \sup_{\theta\in\mathcal N(\theta_0)}
 \left\|
 \frac1T\sum_{t=1}^T
 \left\{
 \frac{\partial^2\widehat\ell_t(\theta)}
 {\partial\theta\partial\theta^\top}
 -
 \frac{\partial^2\ell_t(\theta)}
 {\partial\theta\partial\theta^\top}
 \right\}
 \right\|=o_p(1)
\]
for a neighborhood $\mathcal N(\theta_0)$ of $\theta_0$.
Moreover,  Lemma~\ref{lem:score-regularity} and  Lemma~2.4 of \citet{newey1994large}, we can show that \[
\sup_{\theta\in\mathcal N(\theta_0)}
\left\|\frac1T\sum_{t=1}^T\frac{\partial^2\ell_t(\theta)}{\partial\theta\,\partial\theta^\top}
-E\frac{\partial^2\ell_t(\theta)}{\partial\theta\,\partial\theta^\top}\right\|=o_p(1).
\]
Since $E[\partial^2\ell_t(\theta)/(\partial\theta\,\partial\theta^\top)]$ is continuous on $\mathcal N(\theta_0)$ by dominated convergence theorem, and that $\widehat\theta_T\to_p\theta_0$, the line segment $\{\theta_0+s(\widehat\theta_T-\theta_0):s\in[0,1]\}$ lies in $\mathcal N(\theta_0)$ with probability approaching one, which yields
\begin{equation}\label{eq:hessian-limit-A6}
H_T^*\to_pE\left[\frac{\partial^2\ell_t(\theta_0)}{\partial\theta\,\partial\theta^\top}\right]=H.
\end{equation}
Since $H$ is nonsingular, $H_T^*$ is invertible with
probability approaching one. Hence, we can write
\begin{equation}
 \sqrt T(\widehat\theta_T-\theta_0)
 =- H_T^{*-1}
 \frac1{\sqrt T}\frac{\partial \widehat L_T(\theta_0)}{\partial \theta}=-[H^{-1} +o_p(1)]\frac1{\sqrt T}\frac{\partial \widehat L_T(\theta_0)}{\partial \theta}.
 \label{eq:taylor-estimator-A6}
\end{equation}

\paragraph{Step 2}
Throughout Step 2, we work on the high-probability event $\delta_T:=\max_{1\le i\le T}|\tau_i/\widehat\tau_i-1|\le1/2$. Recall the linear interpolation of squared returns $u_i^2(a)=u_i^2+a(\widehat u_i^2-u_i^2)$ for $a\in[0,1]$ and $1\le i\le T$, and let $s_t(a,\theta_0)$ denote the score at $\theta_0$ evaluated along the sample inputs $\{u_i^2(a)\}_{i=1}^t$ while holding the stationary presample state $(u_0^2,h_0(\theta_0))$ fixed. We decompose $T^{-1/2}\partial_\theta\widehat L_T(\theta_0)$ in four sub-steps.

\medskip
\noindent\emph{Step 2(a): Taylor expansion.}
At $a=1$, the feasible score $\partial_\theta\widehat\ell_t(\theta_0)$ and the path endpoint $s_t(1,\theta_0)$ share identical sample inputs $\{\widehat u_i^2\}_{i=1}^t$ and differ only in their initial values at $t=0$. By Lemma~\ref{lem:initial-values} ($r=1$ at $a=1$),
\[
R_{1,T}:=\frac1{\sqrt T}\sum_{t=1}^T\left\{\frac{\partial\widehat\ell_t(\theta_0)}{\partial\theta}-s_t(1,\theta_0)\right\}=O_p(T^{-1/2})=o_p(1).
\]
At $a=0$, $s_t(0,\theta_0)=s_t(\theta_0)$ is the true stationary infeasible score. Applying the second-order Taylor  to $a\mapsto\sum_{t=1}^T s_t(a,\theta_0)$ around $a=0$ gives
\[
\frac1{\sqrt T}\sum_{t=1}^T s_t(1,\theta_0)
=\frac1{\sqrt T}\sum_{t=1}^T s_t(\theta_0)
+\frac1{\sqrt T}\sum_{t=1}^T\left.\frac{\partial s_t(a,\theta_0)}{\partial a}\right|_{a=0}
+R_{2,T},
\]
where $R_{2,T}:=T^{-1/2}\int_0^1(1-a)\sum_{t=1}^T\{\partial^2 s_t(a,\theta_0)/\partial a^2\}\,da$. By \eqref{eq:score-directional-bound} in Lemma~\ref{lem:score-regularity}, the ergodic theorem for $J_t\in L^1$, and $\sqrt T\kappa_T^2\le 2\log T/(\sqrt Th)+2\sqrt Th^2\to0$ under Assumption~\ref{Asm5},
\[
\|R_{2,T}\|
\le\frac1{\sqrt T}\sum_{t=1}^T\sup_{0\le a\le1}\left\|\frac{\partial^2 s_t(a,\theta_0)}{\partial a^2}\right\|
\le C\sqrt T\,\delta_T^2\left(\frac1T\sum_{t=1}^T J_t\right)
=O_p(\sqrt T\kappa_T^2)=o_p(1).
\]

\medskip
\noindent\emph{Step 2(b): Linearization in the first-step relative error.}
By the chain rule, $\left.\partial s_t(a,\theta_0)/\partial a\right|_{a=0}=\sum_{i=1}^t\{\partial s_t(\theta_0)/\partial(u_i^2)\}(\widehat u_i^2-u_i^2)$. Interchanging the order of summation $\sum_{t=1}^T\sum_{i=1}^t=\sum_{i=1}^T\sum_{t=i}^T$ and substituting $\widehat u_i^2-u_i^2=-u_i^2\delta_{i,T}+u_i^2 R_{iT}$ from \eqref{eq:uhat-expansion} in Lemma~\ref{lem:first-step-expansion} (where $\delta_{i,T}=(\widehat\tau_i-\tau_i)/\tau_i$ and $\max_{1\le i\le T}|R_{iT}|=O_p(\kappa_T^2)$) gives
\[
\frac1{\sqrt T}\sum_{t=1}^T\left.\frac{\partial s_t(a,\theta_0)}{\partial a}\right|_{a=0}
=-\frac1{\sqrt T}\sum_{i=1}^T A_{i,T}\delta_{i,T}+R_{3,T},
\qquad
A_{i,T}:=u_i^2\sum_{t=i}^T\frac{\partial s_t(\theta_0)}{\partial(u_i^2)},
\]
where $R_{3,T}:=T^{-1/2}\sum_{i=1}^T A_{i,T}R_{iT}$. By \eqref{eq:path-derivative-decay} in Lemma~\ref{lem:score-regularity}, $\sup_{1\le i\le T}\|A_{i,T}\|_p\le\sum_{j=0}^\infty\|W_{i,j}\|_p\le C<\infty$ ($p=q/2>2$), so Markov's inequality implies $T^{-1}\sum_{i=1}^T\|A_{i,T}\|=O_p(1)$ and hence
\[
\|R_{3,T}\|\le\sqrt T\left(\max_{1\le i\le T}|R_{iT}|\right)\frac1T\sum_{i=1}^T\|A_{i,T}\|=O_p(\sqrt T\kappa_T^2)=o_p(1).
\]
Combining Steps~2(a) and~2(b) yields the intermediate score expansion
\begin{equation}\label{eq:feasible-score-intermediate}
\frac1{\sqrt T}\frac{\partial\widehat L_T(\theta_0)}{\partial\theta}
=\frac1{\sqrt T}\sum_{t=1}^T s_t(\theta_0)
-\frac1{\sqrt T}\sum_{i=1}^T A_{i,T}\delta_{i,T}+o_p(1).
\end{equation}

\medskip
\noindent\emph{Step 2(c): Split boundary and interior observations.}
We partition the input index $i\in\{1,\ldots,T\}$ into the boundary set $\mathcal B_T$ ($|\mathcal B_T|=2m_T=O(Th)$) and the interior set $\mathcal I_T=\{m_T+1,\ldots,T-m_T\}$.
\begin{enumerate}[1.]
\item \emph{Boundary terms ($i\in\mathcal B_T$):} Since $\sup_{i,T}E\|A_{i,T}\|\le C$, Markov's inequality gives $\sum_{i\in\mathcal B_T}\|A_{i,T}\|=O_p(Th)$. Combining this with $\max_{1\le i\le T}|\delta_{i,T}|=O_p(\kappa_T)$ from Proposition~\ref{prop:first-step-uniform} and Assumption~\ref{Asm5} yields
\begin{equation}\label{eq:boundary-loading}
\frac1{\sqrt T}\left\|\sum_{i\in\mathcal B_T}A_{i,T}\delta_{i,T}\right\|
\le\frac{\max_{1\le i\le T}|\delta_{i,T}|}{\sqrt T}\sum_{i\in\mathcal B_T}\|A_{i,T}\|
=O_p\bigl(\sqrt{T}h\,\kappa_T\bigr)
=o_p(1).
\end{equation}
\item \emph{Replacement of $A_{i,T}$ by $A_i$ on $\mathcal I_T$:} Recall  $A_i=u_i^2\sum_{t=i}^\infty\{\partial s_t(\theta_0)/\partial(u_i^2)\}$ from Lemma~\ref{lem:loading-local}. By \eqref{eq:path-derivative-decay}, $\|A_i-A_{i,T}\|_p\le\sum_{j=T-i+1}^\infty\|W_{i,j}\|_p\le C\varrho^{T-i}$ for some $\varrho\in(0,1)$. Since $T-i\ge m_T=\lfloor Th\rfloor$ for $i\in\mathcal I_T$, we have $E\sum_{i\in\mathcal I_T}\|A_{i,T}-A_i\|\le C\sum_{k=m_T}^\infty\varrho^k=O(\varrho^{Th})$, and therefore
\[
\frac1{\sqrt T}\left\|\sum_{i\in\mathcal I_T}(A_{i,T}-A_i)\delta_{i,T}\right\|
\le\frac{\max_{1\le i\le T}|\delta_{i,T}|}{\sqrt T}\sum_{i\in\mathcal I_T}\|A_{i,T}-A_i\|
=O_p\bigl(T^{-1/2}\kappa_T\varrho^{Th}\bigr)=o_p(1).
\]
\item \emph{Substitution of the interior kernel decomposition:} Note that for $t\in\mathcal I_T$,  we have 
\begin{equation}
\delta_{t,T}
=\frac1{Th}\sum_{s=1}^T
K\!\left(\frac{t-s}{Th}\right)z_s
+d_{t,T}+r_{t,T},
\label{eq:first-step-expansion}
\end{equation}
where  $d_{t,T}=h^2C_2\tau''(t/T)/(2\tau_t)$, and 
$
\max_{t\in\mathcal I_T}|d_{t,T}|=O(h^2), 
\sqrt T\max_{t\in\mathcal I_T}|r_{t,T}|=O(T^{-1/2}h^{-1})+o(\sqrt{T}h^2)+O_p(h\sqrt{\log T})=o_p(1).
$
Then let $M:=Th$, we obtain that 
\[
\frac1{\sqrt T}\sum_{i\in\mathcal I_T}A_i\delta_{i,T}
=\frac1{\sqrt T M}\sum_{i\in\mathcal I_T}\sum_{s=1}^T K\!\left(\frac{i-s}{M}\right)A_i z_s+R_{4,T},
\]
where
\[
\|R_{4,T}\|\le\sqrt T\left(\max_{i\in\mathcal I_T}|d_{i,T}|+\max_{i\in\mathcal I_T}|r_{i,T}|\right)\frac1T\sum_{i=1}^T\|A_i\|=O_p(\sqrt Th^2)+o_p(1)=o_p(1).
\]
\end{enumerate}
\medskip
\noindent\emph{Step 2(d): Evaluation of the kernel projection via centering $A_i=C_\tau+(A_i-C_\tau)$.}
Define the local kernel average $\bar z_{i,M}:=M^{-1}\sum_{s=1}^T K\{(i-s)/M\}z_s$. Since $\{z_s\}$ is stationary with $\sum_{k\in\mathbb Z}|\operatorname{Cov}(z_0,z_k)|<\infty$ and $K$ is bounded and supported on $[-1,1]$,
\[
\begin{aligned}
\sup_{1\le i\le T}\|\bar z_{i,M}\|_2^2
&=\sup_{1\le i\le T}\frac1{M^2}\sum_{s,s'=1}^T K\!\left(\frac{i-s}{M}\right)K\!\left(\frac{i-s'}{M}\right)\operatorname{Cov}(z_s,z_{s'})\\
&\le\frac{C}{M}\sum_{k\in\mathbb Z}|\operatorname{Cov}(z_0,z_k)|
=O(M^{-1}),
\end{aligned}
\]
so $\sup_{1\le i\le T}\|\bar z_{i,M}\|_2\le CM^{-1/2}$. We now evaluate $T^{-1/2}\sum_{i\in\mathcal I_T}A_i\bar z_{i,M}$ in three steps:
\begin{enumerate}[1.]
\item \emph{Extending the outer sum from $\mathcal I_T$ to $\{1,\ldots,T\}$:} By Cauchy--Schwarz inequality, $E\|A_i\bar z_{i,M}\|\le\|A_i\|_2\|\bar z_{i,M}\|_2\le CM^{-1/2}$. Since $|\mathcal B_T|=O(M)$,
\[
E\left\|\frac1{\sqrt T}\sum_{i\in\mathcal B_T}A_i\bar z_{i,M}\right\|
\le\frac1{\sqrt T}\sum_{i\in\mathcal B_T}\|A_i\|_2\|\bar z_{i,M}\|_2
\le C\frac{M}{\sqrt T}M^{-1/2}=O(\sqrt h)=o(1).
\]
Thus we may replace $\sum_{i\in\mathcal I_T}$ by $\sum_{i=1}^T$ at cost $o_p(1)$, and decompose $A_i=C_\tau+\widetilde A_i$ with $\widetilde A_i:=A_i-C_\tau$.
\item \emph{The mean-loading term $C_\tau$:} Put $\omega_{s,T}:=M^{-1}\sum_{i=1}^T K\{(i-s)/M\}$. Since $K$ is Lipschitz and $\int_{-1}^1 K(v)\,dv=1$, Riemann-sum approximation gives $\max_{s\in\mathcal I_T}|\omega_{s,T}-1|=O(M^{-1})$, while $\max_{s\in\mathcal B_T}|\omega_{s,T}-1|\le C$ on the $O(M)$ boundary indices. By covariance summability of $\{z_s\}$,
\[
\left\|\frac1{\sqrt T}\sum_{s=1}^T(\omega_{s,T}-1)z_s\right\|_2^2
\le\frac{C}{T}\sum_{s=1}^T|\omega_{s,T}-1|^2\sum_{k\in\mathbb Z}|\operatorname{Cov}(z_0,z_k)|
=O(h+M^{-2})=o(1).
\]
Consequently, $$C_\tau T^{-1/2}\sum_{i=1}^T\bar z_{i,M}=C_\tau T^{-1/2}\sum_{s=1}^T\omega_{s,T}z_s=C_\tau T^{-1/2}\sum_{s=1}^T z_s+o_p(1).$$
\item \emph{The centered cross-term $\widetilde A_i=A_i-C_\tau$:} It remains to show that
\[
\Delta_{A,T}:=\frac1{\sqrt T M}\sum_{i,s=1}^T K\!\left(\frac{i-s}{M}\right)\widetilde A_i z_s
=\frac1{\sqrt Th}\left\{\frac1T\sum_{i,s=1}^T K\!\left(\frac{i-s}{M}\right)\widetilde A_i z_s\right\}=o_p(1).
\]
For each coordinate $a_i$ of $\widetilde A_i$, Lemma~\ref{lem:loading-local} and Proposition~\ref{prop:rt-gmc} give summable $L^2$ physical dependence coefficients $\sum_{k\in\mathbb Z}\|Y_k-Y_k^*\|_2<\infty$ for the two-sided centered stationary scalar processes $Y_i=a_i+z_i$ and $Y_i=a_i-z_i$. Under Assumption~\ref{Asm3}, the compactly supported Lipschitz kernel $K$ has an absolutely integrable Fourier transform; if $K(0)\ne0$ we rescale by $K(0)$, and if $K(0)=0$ we write $K=(K+K_0)-K_0$ with $K_0(x)=(1-|x|)\mathbf1\{|x|\le1\}$. Applying Theorem~1 and Section~2 of \citet{liu2010asymptotics} at frequency zero (with $n=T$, $B_n=M=Th\to\infty$, $B_n/n=h\to0$, and $p=2$) yields $T^{-1}\sum_{i,s=1}^T K\{(i-s)/M\}Y_iY_s=O_p(1)$ for both $Y_i=a_i+z_i$ and $Y_i=a_i-z_i$. Using the symmetry of $K$, we then obtain that 
\[
\frac1T\sum_{i,s=1}^T K\!\left(\frac{i-s}{M}\right)\widetilde A_i z_s=O_p(1)
\quad\to\quad
\Delta_{A,T}=O_p\bigl((\sqrt Th)^{-1}\bigr)=o_p(1),
\]
since $\sqrt Th=c_h T^{1/2-\lambda_h}\to\infty$ under Assumption~\ref{Asm5}.
\end{enumerate}
Combining Steps~2(c) and~2(d) establishes that
\begin{equation}\label{eq:first-step-score-projection}
\frac1{\sqrt T}\sum_{i=1}^T A_{i,T}\delta_{i,T}
=C_\tau\frac1{\sqrt T}\sum_{s=1}^T z_s+o_p(1).
\end{equation}
Substituting \eqref{eq:first-step-score-projection} into \eqref{eq:feasible-score-intermediate} completes Step~2:
\[
\frac1{\sqrt T}\frac{\partial\widehat L_T(\theta_0)}{\partial\theta}
=\frac1{\sqrt T}\sum_{t=1}^T\{s_t(\theta_0)-C_\tau z_t\}+o_p(1)
=\frac1{\sqrt T}\sum_{t=1}^T\xi_t+o_p(1).
\]

\paragraph{Step 3}
By \eqref{eq:z-martingale-equivalence} in Lemma~\ref{lem:covariance-simplification}, replacing $z_t$ by the martingale difference $\chi_t$ costs $o_{L^2}(1)$:
\[
\frac1{\sqrt T}\sum_{t=1}^T\xi_t
=\frac1{\sqrt T}\sum_{t=1}^T\{s_t(\theta_0)-C_\tau\chi_t\}+o_{L^2}(1).
\]
The summands $\tilde\xi_t:=s_t(\theta_0)-C_\tau\chi_t$ form a stationary, ergodic martingale difference sequence with respect to $\{\mathcal F_t\}$, with covariance matrix $V=E(\tilde\xi_t\tilde\xi_t^\top)=H+\Omega_z C_\tau C_\tau^\top$ and finite $p$th moment ($p=q/2>2$). By the ergodic theorem,
\[
\frac1T\sum_{t=1}^T E\bigl(\tilde\xi_t\tilde\xi_t^\top\mid\mathcal F_{t-1}\bigr)\to V\quad\text{a.s.},
\]
and for any $\varepsilon>0$, the conditional Lindeberg term satisfies
\[
\frac{(\varepsilon\sqrt T)^{-(p-2)}}{T}\sum_{t=1}^T E\bigl(\|\tilde\xi_t\|^p\mid\mathcal F_{t-1}\bigr)=O_p\bigl(T^{-(p-2)/2}\bigr)=o_p(1).
\]
By Theorem~3(i) of \citet{wu2005nonlinear}, we have that  $T^{-1/2}\sum_{t=1}^T\xi_t\to_dN(0,V)$. Combining this  with \eqref{eq:hessian-limit-A6}, \eqref{eq:taylor-estimator-A6}, and Slutsky's theorem gives
\[
\sqrt T(\widehat\theta_T-\theta_0)\to_dN(0,H^{-1}VH^{-\top}),
\]
with $V$ and $\Sigma=H^{-1}VH^{-\top}$ given by \eqref{eq:covariance-simplification}.
\end{proof}

\subsection{\texorpdfstring{Multi-Step Volatility Forecasts}{Multi-Step Volatility Forecasts}}\label{app:forecast-proof}

Write $a=\alpha+\beta$ and
$b_t=\omega+\alpha u_t^2+\beta h_t$. The true-state representation
used in the stationarity proof gives
\[
b_t=\omega+(\alpha\epsilon_t^2+\beta)b_{t-1}
+\varphi(\alpha\epsilon_t^4+\beta\epsilon_t^2).
\]
Consequently,
$E(b_t\mid\mathcal F_{t-1})=a b_{t-1}
+\omega+\varphi(\alpha\mu_4+\beta)$.
By normalization, $\bar b=E(b_t)=1-\varphi\mu_4$, so the constant
term equals $(1-a)\bar b$. Iteration from the current state $b_T$
yields
\[
E(b_{T+k-1}\mid\mathcal F_T)
=\bar b+a^{k-1}(b_T-\bar b),\qquad k\ge1.
\]
Since
$E(u_{T+k}^2\mid\mathcal F_{T+k-1})=b_{T+k-1}+\varphi\mu_4$,
the tower property then gives
\begin{equation}
E(r_{T+k}^2\mid\mathcal F_T)
=\tau_{T+k}\left[1+a^{k-1}\{b_T+\varphi\mu_4-1\}\right].
\label{eq:return-multistep-forecast}
\end{equation}
This identity requires only the innovation moments in
Assumption~\ref{Asm1}, not Gaussianity. The short-run forecast
converges to one as $k\to\infty$ because $a<1$.

Under Gaussian innovations, $\mu_4=3$ and
$b_T+3\varphi=1-\alpha-\beta+2\beta\varphi+\alpha u_T^2+\beta h_T$.
Thus the feasible multi-step forecast is
\[
\widehat v_{T+k\mid T}
=\widehat\tau_{T+k\mid T}
\left[1+(\widehat\alpha+\widehat\beta)^{k-1}
\{\widehat b_T+3\widehat\varphi-1\}\right],
\]
where
$\widehat b_T=\omega(\widehat\theta)+\widehat\alpha\widehat u_T^2
+\widehat\beta\widehat h_T$.
For $k=1$, this is the expression in
Section~\ref{sec:volatility-forecasting}.

\subsection{Randomly Weighted Bootstrap}
\label{app:weighted-bootstrap}

As a finite-sample robustness check, we also consider a randomly weighted bootstrap that perturbs both stages of the estimation procedure. The procedure is not used for formal inference, since its validity for reproducing the long-run covariance matrix in Theorem~\ref{thm:asymptotic-normality} is not established here.

Let $\{r_t\}_{t=1}^T$ denote the observed return series and let $\widehat\theta_T$ be the corresponding two-step estimator. For bootstrap replication $b$, proceed as follows.

\begin{algorithm}[Randomly weighted bootstrap]
\begin{enumerate}
   \item
Generate i.i.d.\ weights
\[
w_t^{(b)}\sim \operatorname{Exp}(1),
\qquad t=1,\ldots,T,
\]
so that
\[
E(w_t^{(b)})=1,
\qquad
\operatorname{Var}(w_t^{(b)})=1.
\]

\item
Using the bandwidth selected from the original sample, compute the weighted kernel estimator
\begin{equation*}
\widehat\tau^{(b)}(x)
=
\frac{
\displaystyle
\sum_{s=1}^{T}
w_s^{(b)}
K_h\left(x-\frac{s}{T}\right)r_s^2
}{
\displaystyle
\sum_{s=1}^{T}
w_s^{(b)}
K_h\left(x-\frac{s}{T}\right)
}.
\end{equation*}
At boundary points, we apply the same reflection scheme as in the original first-step estimator. Set
\[
\widehat\tau_t^{(b)}
=
\widehat\tau^{(b)}(t/T),
\qquad
\widehat u_t^{(b)}
=
\frac{r_t}{\sqrt{\widehat\tau_t^{(b)}}}.
\]

\item
For each $\theta\in\Theta$, construct
$\widehat h_t^{(b)}(\theta)$ recursively from
$\{\widehat u_t^{(b)}\}_{t=1}^T$ using the same initialization rule as in the original estimation, and define
\begin{equation*}
\widehat L_T^{(b)}(\theta)
=
\sum_{t=1}^{T}
w_t^{(b)}
\widehat\ell_t^{(b)}(\theta),
\end{equation*}
where $\widehat\ell_t^{(b)}(\theta)$ is the negative log-likelihood contribution in \eqref{eq:log-likelihood} evaluated at
$\widehat u_t^{(b)}$ and
$\widehat h_t^{(b)}(\theta)$.

\item
Compute
\begin{equation*}
\widehat\theta_T^{(b)}
=
\arg\min_{\theta\in\Theta}
\widehat L_T^{(b)}(\theta).
\end{equation*}

\item
Repeat Steps 1--4 for $b=1,\ldots,B$.

\end{enumerate}
\end{algorithm}

The bootstrap standard deviation for parameter component $j$ is computed as
\begin{equation*}
\widehat{\operatorname{se}}_{\mathrm{boot}}
\left(
\widehat\theta_{j,T}
\right)
=
\left[
\frac{1}{B-1}
\sum_{b=1}^{B}
\left(
\widehat\theta_{j,T}^{(b)}
-
\overline{\widehat\theta}_{j,T}^{\,*}
\right)^2
\right]^{1/2},
\end{equation*}
where
\begin{equation*}
\overline{\widehat\theta}_{j,T}^{\,*}
=
\frac{1}{B}
\sum_{b=1}^{B}
\widehat\theta_{j,T}^{(b)}.
\end{equation*}

\end{document}